\documentclass{aa}  
\usepackage[super]{nth}
\usepackage{graphicx}
\usepackage{txfonts}
\usepackage{lscape}
\usepackage{float}
\usepackage{placeins}

\usepackage{xcolor}

\usepackage{natbib}
\usepackage{hyperref} 

\makeatletter
\renewcommand*\aa@pageof{, page \thepage{} of \pageref*{LastPage}}
\makeatother

\bibpunct{(}{)}{;}{a}{}{,} 

\usepackage{tikz}
\definecolor{lime}{HTML}{A6CE39}
\DeclareRobustCommand{\orcidicon}{
	\begin{tikzpicture}
	\draw[lime, fill=lime] (0,0) 
	circle [radius=0.16] 
	node[white] {{\fontfamily{qag}\selectfont \tiny ID}};
	\draw[white, fill=white] (-0.0625,0.095) 
	circle [radius=0.007];
	\end{tikzpicture}
	\hspace{-2mm}}

\foreach \x in {A, ..., Z}{\expandafter\xdef\csname orcid\x
\endcsname{\noexpand\href{https://orcid.org/\csname orcidauthor\x\endcsname}{\noexpand\orcidicon}}}

\newcommand{\kms}{\mbox{km\,s$^{-1}$}}
\newcommand{\vsini}{\mbox{$v\sin i$}}

\newcommand{\Teff}{\mbox{$T_{\rm eff}$}}
\newcommand{\micro}{$\xi$}
\newcommand{\MSol}{\mbox{M$_\odot$}}

\newcommand{\logg}{\mbox{$\log g$}}

\newcommand{\logq}{\mbox{$\log Q$}}
\newcommand{\logLs}{$\log (\mathcal{L}/\mathcal{L}_{\odot})$}

\newcommand{\ls}{\mbox{$\lesssim$}}
\newcommand{\gs}{\mbox{$\gtrsim$}}
\newcommand{\fwhb}{\textit{FW3414(H$\beta$)}}

\begin{document}

\title{The IACOB project}
\subtitle{IX. Building a modern empirical database of Galactic O9\,--\,B9 supergiants:\\sample selection, description, and completeness\thanks{Tables~\ref{tab:observations}, \ref{tab:phot_kinem_Gaia} and \ref{tab:phot_kinem_Hipp} are only available in electronic form at the CDS via anonymous ftp to {\tt cdsarc.cds.unistra.fr} (130.79.128.5) or via \url{https://cdsarc.cds.unistra.fr/cgi-bin/qcat?J/A+A/}}}
\titlerunning{Sample selection, description, and completeness of Galactic O9\,--\,B9 supergiants}
\author{A. de~Burgos\inst{1,2\orcidA{}}, S. Sim\'on-D\'iaz\inst{1,2\orcidB{}} , M.~A. Urbaneja\inst{3\orcidC{}}, I. Negueruela\inst{4\orcidD{}}}
\authorrunning{A. de~Burgos et al.}
\institute{
Universidad de La Laguna, Departamento de Astrof\'isica, 38206 La Laguna, Tenerife, Spain
\and
Instituto de Astrof\'isica de Canarias, Avenida V\'ia L\'actea s/n, 38205 La Laguna, Tenerife, Spain
\and
Universität Innsbruck, Institut für Astro- und Teilchenphysik, Technikerstr. 25/8, 6020 Innsbruck, Austria
\and
Departamento de F\'{\i}sica Aplicada, Facultad de Ciencias, Universidad de Alicante, Carretera de San Vicente s/n, 03690 San Vicente del Raspeig, Spain
}
\date{Received 20 February 2023 / Accepted 2 May 2023}
\abstract 
{Blue supergiants (BSGs) are key objects to study the intermediate phases of massive star evolution, helping to constrain evolutionary models. However, the lack of a holistic study of a statistically significant and unbiased sample of these objects makes several long-standing questions about their physical properties and evolutionary nature remain unsolved.
}
{The present and other upcoming papers of the IACOB series are focused in studying -- from a pure empirical point of view -- a sample of about 500 Galactic O9\,--\,B9 stars with luminosity classes I and II (plus 250 late O- and early B-type stars with luminosity classes III, IV and V) and covering distances up to $\sim$4\,kpc from the Sun.
}
{We compile an initial set of $\approx$11\,000 high-resolution spectra from $\approx$1\,600 Galactic late O- and B-type stars. We use a novel spectroscopic strategy based on a simple fitting of the H$\beta$ line to select stars in a specific region of the spectroscopic HR diagram. We evaluate the completeness of our sample using the Alma Luminous Star catalog (ALS~III) and {\em Gaia-DR3} data.
}
{We show the benefits of the proposed strategy for identifying BSGs descending, in the context of single star evolution, from stellar objects born as O-type stars. The resulting sample reaches a high level of completeness with respect to the ALS~III catalog, gathering $\approx$80\% of all-sky targets brighter than $B_{mag}$\,<\,9 located within 2\,kpc. However, we identify the need for new observations in specific regions of the Southern hemisphere.
}
{We have explored a very fast and robust method to select BSGs, providing a valuable tool for large spectroscopic surveys as WEAVE-SCIP or 4MIDABLE-LR, and highlighting the risk of using spectral classifications from the literature. Upcoming studies will make use of this large and homogeneous spectroscopic sample to study specific properties of these stars in detail. We initially provide first results about their rotational properties (in terms of projected rotational velocities, \vsini). 
}
\maketitle


\section{Introduction}
\label{section:1_intro}

Massive stars are very interesting and important objects for several branches of astrophysics. From their feedback to their circumstellar and interstellar medium \citep{1996A&A...305..229G,2013A&A...550A..49K}, stellar clusters \citep{2013MNRAS.431.1337R} and larger structures such as giant molecular clouds \citep{2002ApJ...566..302M} with a global importance in the chemo-dynamical evolution of galaxies \citep{2008IAUS..250..391M}, to their use as extragalactic tools \citep{2000ARA&A..38..613K,2003LNP...635..123K,2003ApJ...582L..83K,2008ApJ...681..269K}, and main role as progenitors of gravitational wave emitters \citep{2016Natur.534..512B,2016A&A...588A..50M}. 

Their different evolutionary phases are of particular interest to study the different physical processes there involved. While the progress into understanding each of these phases has been notorious over the last decades, they still gather many questions yet unresolved \citep{2012ARA&A..50..107L}. In particular, the so-called blue supergiants have always been among the less constrained ones from an evolutionary perspective \citep[e.g.][and many others]{1990ApJ...363..119F,1992A&AS...94..569L,1993A&AS...97..559L,2000A&A...361..101M}. 

These objects can have two very different fates according to single evolutionary models, which essentially depends on their masses \citep[e.g.][]{2011A&A...530A.115B,2012A&A...537A.146E}. Either they transition the post-main sequence (MS) likely becoming luminous blue variables \citep[LBVs,][]{1994PASP..106.1025H,2011MNRAS.415..773S,2020Galax...8...20W} and ultimately Wolf-Rayet (WR) objects \citep{1987ARA&A..25..113A,1991IAUS..143..485H,2000A&A...360..227N}, or they go through it to become red supergiants \citep[RSgs,][]{2005ApJ...628..973L} and, in the event of undergoing one or more of the so called ``blue loops" \citep{1975ApJ...198..407S,2012A&A...537A.146E}, turning back into a bluer and hotter state. In both cases, they will end their lives as supernovae \citep{2002RvMP...74.1015W,2009ARA&A..47...63S} or faded supernovae \citep{2017MNRAS.468.4968A}.

One of the still unresolved but most important questions arises from the existence of an overdensity of blue supergiants in the region of the Hertzsprung–Russell diagram where evolutionary models predict a gap of stars, as they are expected to evolve rapidly towards the RSg phase after leaving the MS. This overdensity was first shown in \citet{1990ApJ...363..119F} using photometric data, and has been observed in more recent spectroscopic studies of large samples of massive stars such in \citet{2014A&A...570L..13C}. Several scenarios have been proposed to explain this overdensity. One of them is the overlapping of blue supergiants going to and coming from the RSg phase through the above-mentioned assumption of the \textit{blue loops}. Additionally, this region can also be populated by stars that have undergone binary interaction, resulting in additional hydrogen in their cores and therefore allowing them to extend their MS \citep{2012Sci...337..444S,2013ApJ...764..166D}. Lastly, it can also be an issue with the models themselves, being corrected by the fine-tuning of the parameters used in the evolutionary models such the overshooting or inflation \citep{2014A&A...570L..13C,2021A&A...648A.126M}.

Theoretical models have highlighted the implications that rotation may have in the evolution and properties of massive stars \citep{2000ARA&A..38..143M,2012ARA&A..50..107L} as it can alter their surface composition \citep{2000ApJ...544.1016H,2005A&A...440.1041M,2009A&A...496..841H}, mass-loss rate \citep{2008A&ARv..16..209P} and the final fate \citep{2005A&A...429..581M}. As a continuation of the study performed by Holgado et al. (2020, 2022) for the case of Galactic O-type stars, gathering empirical information for large sample of blue supergiants is a key to constraint state-of-the-art evolutionary models. In this matter, another important question arises from the observational evidence of a drop in the number fast-rotating stars at approximately 22\,000\,K \citep{2010A&A...512L...7V}. This drop has been proposed to be caused by enhanced mass loss (stronger angular momentum loss) in the vicinity of the theoretical bi-stability jump \citep{2000A&A...362..295V}. However, as seen in Fig.~1 of \citet{2014A&A...570L..13C}, the presence of this drop might be located at the same position as a new empirical Terminal Age Main Sequence for stars with masses above 20\,\MSol, which may indicate that the drop could also be cause by stars leaving the MS as they decrease their rotational velocities.

To try to unlock this situation, the present and upcoming set of papers are focused on the study of the physical and evolutionary nature of Galactic blue supergiants from a pure empirical point of view, and more specifically, those Galactic O9\,--\,B9-type stars evolving from stars born with masses in the range (\gs20\,--\,80\,\MSol) (i.e. those born as O-type stars). For this endeavor, we benefit from a significantly large and homogeneous sample of such stars observed with high-resolution spectrographs. In this first paper, we have developed a methodology based on the H$\beta$ line profile for selecting them independently from previous spectral classifications. We have carried out a thorough description of the selected sample, identifying miss-classified stars from the literature, and also identifying double line spectroscopic binaries. We have used the {\em Gaia} DR3 \citep{2016A&A...595A...1G,2022arXiv220800211G,2022arXiv220605989B} and the Alma Luminous Star catalog \citep[ALS~III, Pantaleoni González, et al., in prep.;][]{2021MNRAS.504.2968P} to evaluate the completeness of our sample in the Solar neighborhood. This is of particular importance not to add potential observational biases when constraining evolutionary models that can subsequently introduce further biases in population synthesis of massive stars \citep[e.g.][]{2009A&A...504..531V}. Lastly, we have made use of {\tt IACOB-BROAD} \citep{2014A&A...562A.135S} tool to obtain the \vsini\ distribution as a preliminary approach to constraint the MS in this high range of masses. For the stars selected in this work, upcoming papers in preparation cover their stellar parameters and abundances derived from the quantitative spectroscopic analysis, the analysis of their variability and full identification of binaries using multi-epoch data, and the analyses of their pulsational properties, among other topics all connected to answer those still open questions.

The paper is organized as follows: Sect.~\ref{section:2_XXXXX} summarizes the spectroscopic data of the input sample of stars, and also the data retrieved from the {\em Gaia} mission. Sect.~\ref{section:3_XXXXX} describes the initial motivation and methodology used to build the sample of stars, including the visual inspection and line-broadening analyses of the selected spectra. Sect.~\ref{section:4_ZZZZZ} first analyzes the utility of the used methodology and provides a description of the resulting sample in terms of spectral classifications, photometric magnitudes and distances, evaluating also the completeness of the sample respect to the ALS~III catalog. There is also description of the sample in terms of binaries and the different H$\beta$ and H$\alpha$ profiles as they are affected by stellar winds. First results from the \vsini\ distribution are also discussed. Lastly, Sect.~\ref{section:5_XXXXXX} includes the summary, conclusions and  the future work to be carried out as a continuation of this one.


\section{Observations}
\label{section:2_XXXXX}


\subsection{Ground-based spectroscopy}
\label{subsection:21_YYYYY}


The starting point of our study was the gathering of high-resolution optical spectroscopy for the largest possible sample of Galactic O9\,--\,B9 stars with luminosity classes (LCs) I\,--\,II and O9\,--\,B3 stars with LCs III\,--\,IV\,--\,V, reachable with any of the facilities indicated below using reasonable exposure times.

To this aim, we built an initial list of stars fulfilling these criteria based on the recommended spectral classifications quoted in the \textit{Simbad} astronomical database \citep{2000A&AS..143....9W}. This initial list was then complemented with a few other sources known to be B-type stars, but wrongly quoted as A-type stars in \textit{Simbad}, plus a small sample of stars for which there were spectra available in the IACOB spectroscopic database (see below) but either no information of the corresponding LC is quoted in \textit{Simbad}, or they are simply labeled as ``OB" or ``O" (see Sect~\ref{subsection:42_YYYYY} and Tables~\ref{tab:observations}, \ref{tab:newclass_noLC} from Appendix~\ref{apen.newclass}).

We then focused on collecting spectra for as many stars as possible from the above-mentioned list using three main instruments: the FIbre-fed Echelle Spectrograph \citep[FIES,][]{2014AN....335...41T} mounted at the 2.56\,m Nordic Optical Telescope (NOT) located at the Observatorio del Roque de los Muchachos in La Palma, Canary Islands, Spain; the High Efficiency and Resolution Mercator Echelle Spectrograph \citep[HERMES,][]{2011A&A...526A..69R} at the 1.2\,m Mercator semi-robotic telescope, also located at the Observatorio del Roque de los Muchachos; and the Fiber-fed Extended Range Optical Spectrograph \citep[FEROS,][]{1997Msngr..89....1K} at the 2.2\,m MPG/ESO telescope located at ESO's La Silla Observatory, Chile. All these instruments provide high resolution spectra with a resolving power ranging from 25\,000 to 85\,000, and a common wavelength coverage in the 3\,800\,--\,9\,000\,\AA\ range\footnote{Except for those FIES spectra taken before the installation of the current CCD (30/9/2016), which only cover up to 7\,200\,\AA. In addition, FEROS spectra have a wider coverage, from 3\,500\,\AA\ to 9\,200\,\AA.}.

We initially benefited from those FIES and HERMES spectra\footnote{Some of them were obtained between 2019 and 2020 with the NOT during Technical and Nordic Service nights within the scope of the NOT studentship program of the main author.} of Northern Galactic OB stars (with declination \gs\,-20\,$\degr$) compiled by the IACOB project until 2020 \citep[see][for the latest review of the so-called IACOB spectroscopic database]{2020sea..confE.187S}. Pursuing the objectives described above, several additional observing campaigns with NOT and Mercator were planned and executed in the framework of the IACOB project between 2020 and 2023. In addition, we searched into the ESO-archive to retrieve all public available spectra of those Southern stars observable from La Silla (with declination \ls\,30\,$\degr$) and matching the above-mentioned criteria.

By the time of submitting this paper, we were able to gather a total of $\approx$11\,000 spectra of $\approx$1\,600 stars, of which $\approx$610 have been observed within the scope of this paper. Of all of them, $\approx$670 correspond to O9\,--\,B9 type stars with LCs I and II, and $\approx$710 to O9\,--\,B3 type with LCs III\,--\,IV\,--\,V. From the ESO-archive, we downloaded spectra of $\approx$330 stars. The list of programs from where these stars were taken are included in the acknowledgments.

In all cases, we directly considered the reduced spectra provided by the pipelines installed at the telescopes. In addition, we applied our own routines to produce a normalized version of each individual spectrum, and used \textit{pyIACOB}\footnote{The \textit{pyIACOB} package is described in Appendix~\ref{apen.pyiacob}.} to correct all the spectra from heliocentric velocity, cosmic rays and some systematic cosmetic defects mainly present in the FEROS spectra.

For the purposes of this paper, we selected for each star the best available spectrum with signal-to-noise (S/N) \gs\,30 in the 4\,000\,--\,5\,000\,\AA\ range, taking also into account any cosmetic or potential issues (e.g. normalization issues) that could lead to unreliable results in this or later works. 
The median of the S/N ratio in the 4\,000\,--\,5\,000\,\AA\ spectral range of the above-mentioned best spectra is $\sim$170, reaching up to 300 or more in several bright stars. In addition, we considered all available multi-epoch spectra to search for spectroscopic binaries.



\subsection{{\em Gaia} data}
\label{subsection:22_YYYYY}

We complemented the spectroscopic data with data from {\em Gaia} DR3 \citep{2016A&A...595A...1G,2022arXiv220800211G,2022arXiv220605989B}. In general terms, the {\em Gaia} DR3 release provides reliable astrometric and photometric data for most of the Galactic OB stars near the Sun with the exception of some too bright sources ($G_{mag}$ \ls 6) for which either the data is unavailable or the astrometric solution is less reliable due to uncalibrated CCD saturation \citep{2018A&A...616A...2L}, also affecting the photometric data. For the latter case we used \textit{Hipparcos} \citep{1997A&A...323L..49P} astrometric data if available, to have at least a rough estimation. In addition to this, the inclusion of Radial Velocity Spectrometer (RVS) allows for determination of line broadening parameters \citep{2022arXiv220610986F}, which we used to compare with our results, for curiosity, obtaining a poor agreement between both. Such comparison is included in Appendix~\ref{apen.vsini}

In particular and for the stars in the sample (see Sect.~\ref{section:31_XXXXX}) we downloaded available information of the astrometric parallaxes ($\varpi$) and proper motions ($\mu_{\alpha}\cos{\delta}$, $\mu_{\delta}$), line broadening measurements, \textit{G, G$_{RP}$, G$_{BP}$} bands photometry and finally associated RUWE values, an estimate for the reliability of the astrometric solution \citep{2021A&A...649A...2L}. Despite in this work we are not filtering our sample by this value, we note here that given the relatively close distance (<\,2\,500\,pc) of most of the targets in the IACOB sample, most of the stars that are not too bright have a RUWE < 1.4. Still the stars with RUWE values up to 8 can be used with considering that the errors in parallax are larger \citep[for more details see][]{2021A&A...649A..13M,2022A&A...657A.130M}. 

In order to obtain distances to the stars in the sample, we directly downloaded corrected distances from \citet{2021AJ....161..147B}, where {\em Gaia} EDR3 \citep{2021A&A...649A...1G} is used together with a direction-dependent prior and other additional information such the extinction map of the Galaxy for the determination of the distances. Alternatively and as first order approximation and comparison, we also used the inverse of the parallax (1/$\varpi$) by correcting them from zero-point offset by using the procedure described in \citet{2021A&A...649A...4L}\footnote{In particular we applied the algorithm provided by these authors at {\tt \url{https://pypi.org/project/gaiadr3-zeropoint/}}}. However, this is a poor estimate of the distance and only valid for zero-point corrected and positive parallaxes with $\sigma_{\varpi}/\varpi$ \ls 0.1 \citep[see][for more details]{2021AJ....161..147B}. Lastly, we also used preliminary results from Pantaleoni González, et al. (in prep.), who uses a different prior from \citet{2021AJ....161..147B}, more optimized for deriving distances to OB stars \citep[see][Sect.~3.3, for more details]{2021MNRAS.504.2968P}. A comparison between these distances from different methodologies is included in Appendix~\ref{apen.distances}.

\begin{figure*}[!t]
\centering
\includegraphics[width=1\textwidth]{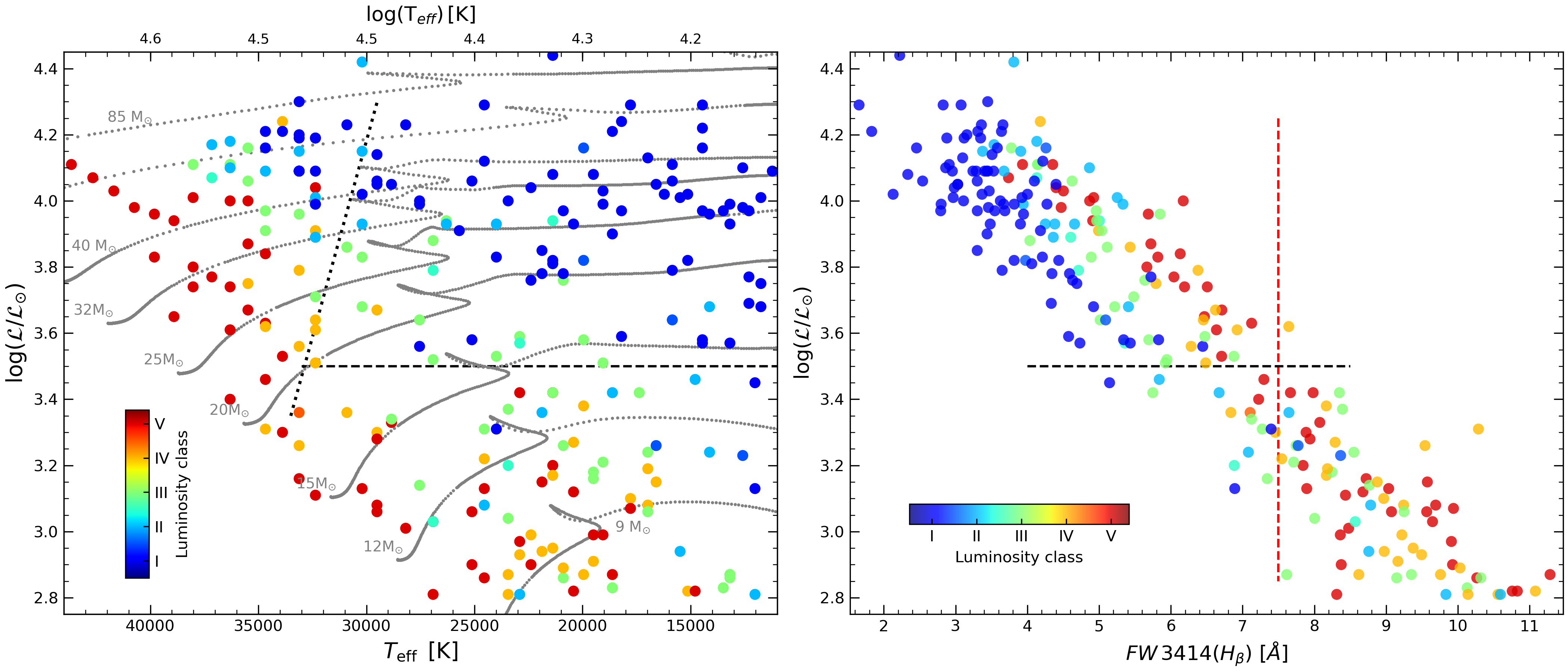}
\caption{Sub-sample of 246 Galactic O- and B-type stars investigated by \citet{2017A&A...597A..22S}, color-coded by the LC taken from \textit{Simbad}. Left panel: Location of the stars in an spectroscopic Hertzsprung-Russel diagram (see Sect.~\ref{section:311_XXXXX}). The rough boundary between the O- and B-type star domains is indicated with a black dotted diagonal line. Evolutionary tracks taken from the MESA Isochrones \& Stellar Tracks online tool for solar metallicity and no initial rotation are also included for reference purposes. Right panel: \logLs\ against \fwhb\ for the same stars. The vertical red dashed line and back dashed horizontal lines at \logLs\ = 3.5\,dex in both panels are included for reference (see Sect.~\ref{subsection:41_YYYYY}).}
\label{fig:shrd_fw_ss}
\end{figure*}


\section{Methodology}
\label{section:3_XXXXX}


\subsection{Sample selection}
\label{section:31_XXXXX}


\subsubsection{Motivation and selection criteria}
\label{section:311_XXXXX}

Left panel of Fig.~\ref{fig:shrd_fw_ss} shows the distribution of a sample of 246 O and B-type stars investigated by \citet{2017A&A...597A..22S} in the so-called spectroscopic Hertzsprung-Russel diagram \citep[sHRD, first utilized by][]{2014A&A...564A..52L}. In this figure, we separate stars by LC using different colors, and indicate with a dashed inclined line the rough boundary between the location of the O- and B-type stars in the sHRD \citep[see][]{2018A&A...613A..65Hol}. In addition, we include a set of non-rotating evolutionary tracks computed with the MESA code\footnote{Available at the MESA Isochrones \& Stellar Tracks (MIST) web-page (\url{https://waps.cfa.harvard.edu/MIST/}).} \citep[see][for references purposes]{2016ApJS..222....8D,2016ApJ...823..102C,2011ApJS..192....3P,2013ApJS..208....4P,2015ApJS..220...15P}.

As mentioned in Sect.~\ref{section:1_intro}, we are interested in continuing the efforts initiated in previous papers of the IACOB series \citep[referring to the O star domain, see][]{2020A&A...638A.157H,2022A&A...665A.150H}, and populate with the largest possible sample the region of the sHRD which corresponds to those B-type stars which are expected to be the descendants of the O-type stars.

To this aim, one possibility would be to select from the initial sample described in Sect.~\ref{subsection:21_YYYYY} those stars with spectral types in the range B0 to B9, and classified as supergiants (LC I) or bright giants (LC II). However, we have to be careful with this approach because, as will be shown in Sect.~\ref{subsection:41_YYYYY}, the spectral classifications provided by default by \textit{Simbad} for the case of B-type stars include a non negligible number of cases in which the LCs are wrong (or not even provided). In addition, as illustrated by the left panel of Fig.~\ref{fig:shrd_fw_ss}, part of the region of interest is also populated by early-B giants (and maybe also sub-giants and dwarfs). Therefore, if we follow this selection strategy we might end up in a situation in which we are excluding many stars of interest for our study.

To overcome this problem, we decided to follow a more robust (but still simple) strategy, based on the use of the quantity \fwhb\ as a proxy of the parameter log\,$\mathcal{L}$ (both described in next section).


\subsubsection{\texorpdfstring{Using the width of H$\beta$ as a proxy of \logLs}{Using the width of Hb as a proxy of logLs}}
\label{subsubsection:312_ZZZZZ}

Considering the location of the O-type stars and the behavior of the evolutionary tracks in the sHRD shown in the left panel of Fig.~\ref{fig:shrd_fw_ss}, one could easily assume that, to a first approach, we are mainly interested in stars with \logLs\,$\gtrsim$\,3.5\,dex, where the $\mathcal{L}$ parameter is defined as $T_{\rm eff}^4$\,/$g$ \citep{2014A&A...564A..52L}.

Ideally speaking, the best approach to select our working sample of evolved descendant of O-type stars would be to have estimates of the effective temperature (\Teff) and the surface gravity (log\,$g$) for each star in the list of potential targets of interest described in Sect.~\ref{subsection:21_YYYYY}. However, although viable, this strategy would be very time-consuming, since it would require going through the full quantitative spectroscopic analysis process for each target before selecting.

As an alternative, we decided to explore to what extent using a direct measurement of the profile shape of one of the Hydrogen Balmer lines -- which are known to be affected by the surface gravity of the star through the effect of the Stark broadening -- can be used to perform such a selection in a much faster way.

\begin{figure}[!t]
\centering
\includegraphics[width=0.46\textwidth]{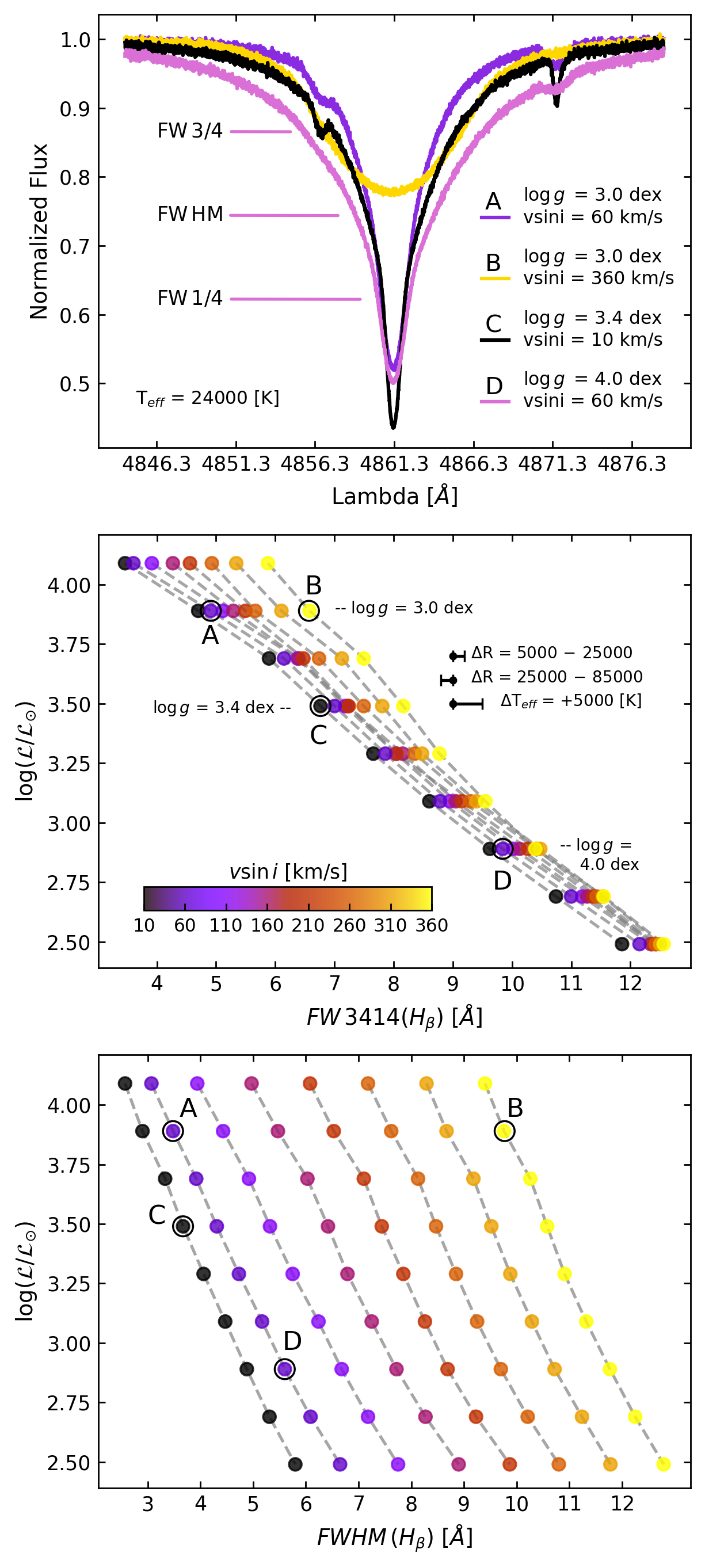}
\caption{Results from testing the methodology described in Sect.~\ref{subsubsection:312_ZZZZZ} measuring the \fwhb. Top panel: four illustrative H$\beta$ profiles named A, B, C and D for {\tt FASTWIND} models with \Teff\ = 24\,000\,K and R = 25\,000, and different pairs of \logg\ and \vsini. The horizontal pink lines indicate the position at 3/4, 1/2 and 1/4 of the line depth for the example D. Middle and bottom panels: \logLs\ against \fwhb\ and \textit{FWHM}, respectively, both measured in a grid of synthetic H$\beta$ profiles from {\tt FASTWIND} with the same \Teff\ and $R$ as in the top panel, and values of \vsini\ between 10 and 360\,\kms. The four profiles of the top panel are indicated with open circles and letters. The error-bars in the middle panel show the average shift in all measurements due to the effect of different resolutions, and the effect of increasing \Teff\ up to 30\,000\,K (approximate temperature for a B0\,I-II star).} 
\label{fig:Emul_Hb_Lspec}
\end{figure}

In particular, we chose to work with the H$\beta$ line, which is less affected by blends with other metal lines (as it is, for example, the case of the H$\gamma$ line) and is not as heavily affected by the presence of a stellar wind as the H$\alpha$ line. As illustrated in Fig.~\ref{fig:Emul_Hb_Lspec} (see also explanation below), we also identified that the quantity \fwhb\ -- defined as the difference between the width of the H$\beta$ line measured at 3/4 and 1/4 of the line-depth, respectively -- is better suited as a proxy of \logLs\ than the full-width at half-maximum (\textit{FWHM}) of the line.


To test the proposed methodology, we first used the stellar atmosphere code {\tt FASTWIND} \citep{1997A&A...323..488S,2005A&A...435..669P} to compute a grid of synthetic spectra corresponding to models covering a range in \Teff\ = 16\,000\,--\,28\,000\,K and \logLs\ = 2.5\,--\,4.1\,dex. Other parameters such as the helium abundance ($N_{\rm He}$), the microturbulence (\micro), and the wind-strength parameter (\logq)\footnote{Defined as $Q = \dot{M}/(R_{\star}v_{\infty})^{1.5}$ \citep[see][]{1996A&A...305..171P,2005A&A...435..669P}.} were kept fixed in the computation of the grid to $N_{\rm He}$ = 0.1, \micro = 10\,\kms, \logq = -14.0, respectively.

By convolving these spectra with different broadening and instrumental profiles, we were able to reproduce (to a first order) the effect that different projected rotational velocities (\vsini) and spectral resolutions ($R$) have on the line profiles. In particular, we considered values of $R$ between 2\,500 and 85\,000, and 8 values of \vsini, ranging from 10 to 400\,\kms. Some illustrative examples of the synthesized profiles can be found in the top panel of Fig.~\ref{fig:Emul_Hb_Lspec}.

We then used the corresponding \textit{pyIACOB} module to measure the full-width of the line at 3/4, 1/2 and 1/4 of its depth for each simulated H$\beta$ profile. For the same values of \Teff\ = 24\,000\,K and $R$ = 25\,000, middle and bottom panels of Fig.~\ref{fig:Emul_Hb_Lspec} shows that for a given \vsini\ value, the lower the quantity \logLs\ (i.e., the higher the surface gravity), the larger the quantities \fwhb\ and \textit{FWHM}. However, in a more general situation, the use of \fwhb\ results in a more adequate proxy of \logLs\ than \textit{FWHM}, as it mitigates the effect that \vsini\ has in the shape of the line, providing a better indirect measurement of the gravity.

To better illustrate this situation, top panel of Fig.~\ref{fig:Emul_Hb_Lspec} depicts four synthetic H$\beta$ line profiles (named A, B, C and D, respectively) corresponding to four different \logg\ and \vsini\ combinations. Their location in the other two panels is indicated with open black circles and corresponding letters.

On the one hand, the aforementioned mitigation of the \vsini\ effect can be clearly seen when comparing the resulting differences obtained in \fwhb\ ($\ls$\,2\,\AA) and \textit{FWHM} ($\gs$\,6\,\AA) when considering the profiles labeled as A and B. On the other hand, and more importantly, using the \textit{FWHM} produces a much less efficient filtering than the quantity \fwhb\ when samples of stars with a broad range of \vsini\ are considered. For example, to ensure that a star with a set of parameters similar to those considered in profile B survives the filtering process, we need to assume a filtering value of \textit{FWHM}\,$\approx$\,10\,\AA. But then, other stars with \logLs\ down to $\approx$\,2.5\,dex would also be selected, contaminating the sample with many stars which are below the initially considered limit in \logLs\ of $\approx$\,3.5\,dex (see left panel of Fig.~\ref{fig:shrd_fw_ss}). On the contrary, since profile B has a \fwhb\,$\approx$6.5\,\AA, considering this as the filtering value would only select stars down to \logLs\,$\approx$3.5\,dex, that corresponds to a star with similar parameters as those considered for profile C.

We note that the value of \Teff\ = 24\,000\,K was not randomly chosen, but rather approximates the temperature of the majority of B-type stars analyzed in this study (see Sect.~\ref{subsubsection:421_EEEEE}). Similarly, the value of $R$ = 25\,000 was selected because the spectra used in this study have the same or higher resolution. We have evaluated the effect of increasing \Teff\ up to 30\,000\,K (approximate temperature for a B0 I-II star), which shifts our results by 0.5\,\AA\ towards higher values of the quantity \fwhb. Similarly, we investigated which is the effect of considering different resolving powers. We found that within the resolutions of our spectroscopic data\,set (see Sect.~\ref{subsection:21_YYYYY}) the shift in \fwhb\ is $\sim$\,0.2\,\AA\ towards lower values. On the opposite side, we found an average shift of $\sim$\,0.2\,\AA\ compared to $R$ = 5\,000 towards higher values, being $\sim$\,0.5\,\AA\ for $R$ = 2\,500. In all cases, we obtained very similar slopes as those shown in the middle panel of Fig.~\ref{fig:Emul_Hb_Lspec}, hence demonstrating the method's ability to safely discriminate supergiants from low-luminosity counterparts at upcoming lower resolution spectroscopic surveys such as WEAVE-SCIP \citep{2020SPIE11447E..14D,2023MNRAS.tmp..715J} or 4MIDABLE-LR \citep{2019Msngr.175...30C}.

As an additional validation of our proposed strategy, we measured the \fwhb\ for all the stars included in the sHRD presented in the left panel of Fig.~\ref{fig:shrd_fw_ss}. We present the corresponding \logLs\ vs. \fwhb\ diagram in the right panel of Fig.~\ref{fig:shrd_fw_ss}, using the LC as color-code as in the left panel. We see that, despite the very wide \Teff\ span (more than 30\,000\,K) and different \vsini\ and resolutions, we are able to reproduce to a very good level what is shown in the middle panel of Fig.~\ref{fig:Emul_Hb_Lspec}. In fact, these results allow us to provide an approximate value of \logLs\ above which all stars will very likely be descendants of the O-stars. In particular, we see that if we want to include all B-type stars with LCs I and II (hereafter BSGs), a value of \logLs\,$\approx$3.5\,dex can be used and this would correspond to a \fwhb\ of 7\,--\,8\,\AA.


\subsection{Visual inspection of spectra}
\label{subsection:32_YYYYY}

The method described in Sect.~\ref{subsubsection:312_ZZZZZ} assumes a pure absorption profile for the H$\beta$ line, however, this is not always the case. Therefore, we decided to also perform a visual inspection of the used spectra in order to identify cases in which the measurement of \fwhb\ could be less reliable. 

While inspecting the spectra, we did not only concentrate on the H$\beta$ profiles, but also evaluated the global morphology of the H$\alpha$ line. As a result, we were able to provide a rough classification of the different types of H$\alpha$ and H$\beta$ profiles that we found in the full investigated sample. In addition, we benefited from the complete set of multi-epoch spectra that we had available to perform a visual identification of the double line or higher order spectroscopic binaries (hereafter SB2+).

\subsubsection{\texorpdfstring{Identification of different H$\alpha$ and H$\beta$ profiles}{Identification of different Ha and Hb profiles}}
\label{subsubsection:321_FFFFFF} 

\begin{figure}[!t]
    \centering
    \resizebox{\columnwidth}{!}{\includegraphics{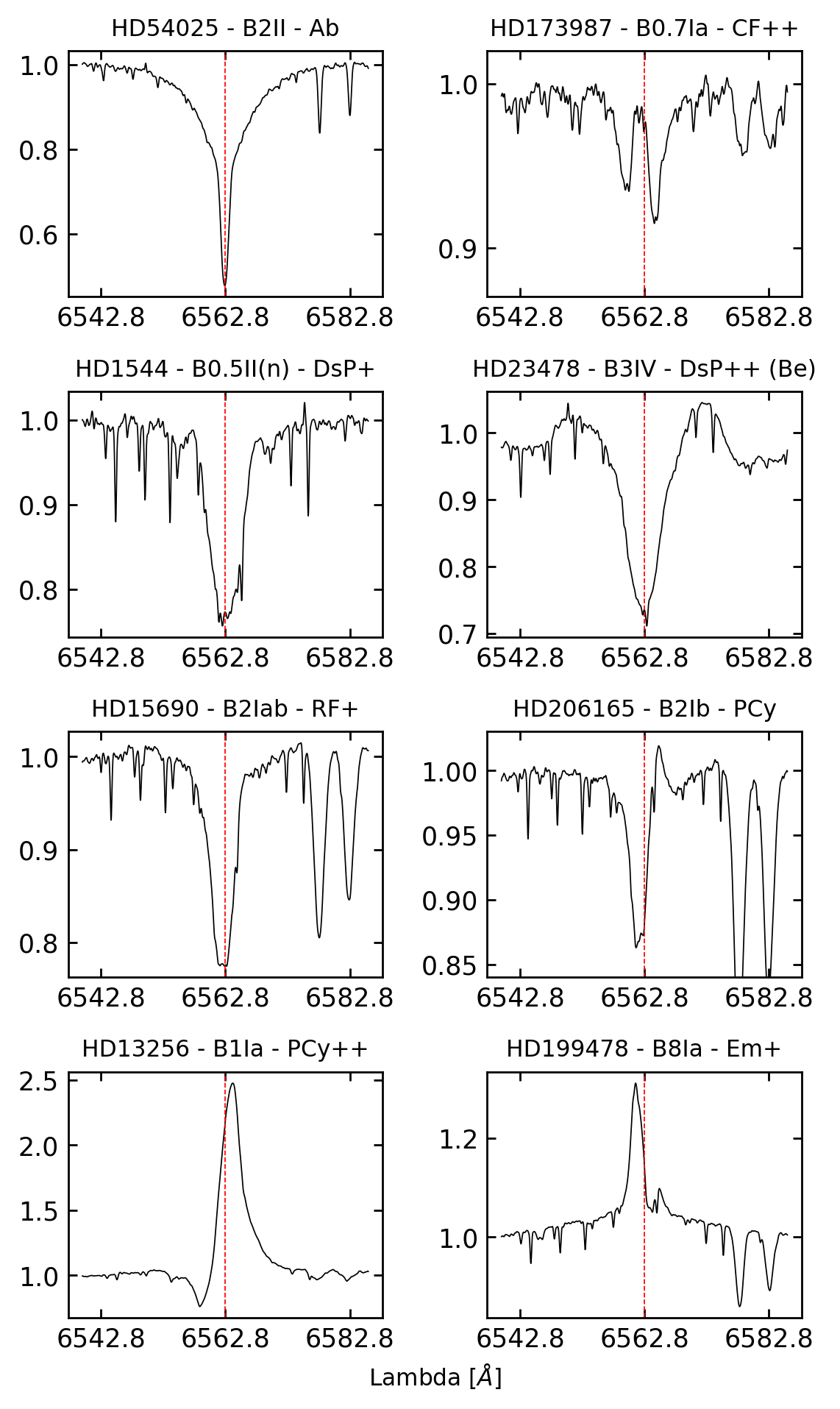}}
    \caption{Examples of different H$\alpha$ profiles found within our sample of study. The name of the stars, spectral types and morphological labels (see Sect.~\ref{subsubsection:321_FFFFFF}) are included. The spectra are corrected from radial velocity by using a set of photospheric metal lines. The vertical red dashed lines indicate the reference position of the line.}
    \label{fig:ha_profiles}
\end{figure}

Fig.~\ref{fig:ha_profiles} shows some examples of the different types of H$\alpha$ profiles found in the spectra of the investigated sample of stars. These go from pure absorption to strong emission (e.g. HD\,54025 and HD\,199478, respectively), cases in which the red wing is refilled (e.g. HD\,15690), a P\,Cygni-type profile is detected (e.g., HD\,206165 and HD\,13256), or the profile appears contaminated by a more or less strong double peak emission (e.g., HD\,1544 and HD\,23478). After a careful inspection of all the H$\alpha$ profiles in our spectroscopic data\,set, we decided to establish a morphological classification which is based on five main different features (with some sub-types) as described below:

\begin{itemize}
    \item Absorption (Ab): When the line appears in pure absorption. \smallskip
    
    \item Core filled (CF): When the core of the line is filled with emission, but this emission does not pass the continuum level of the normalized flux. We set sub-labels with ``CF", ``CF+" and ``CF++" to indicate when the central emission reaches approximately 1/3, 2/3 or up to the continuum level from the expected depth (see top-right panel of Fig.~\ref{fig:ha_profiles} for the later case). \smallskip
    
    \item Double sub-peak (DsP): When both wings of the line are filled or are in emission above the normalized flux. This mainly accounts for fast rotating stars with confined winds \citep[see][Fig 8 for more information]{1996A&A...312..195P}. The labels ``DsP", ``DsP+" and ``DsP++" correspond with the situations in which the filling of the lobes reaches 1/3, 2/3 or above the continuum of the line depth. The later generally includes the case of the Be stars, where both sides of the line are in strong emission due to the presence of a disk. Two cases for ``DsP+" and ``DsP++ (for a confirmed ``Be" star) are included in the left and right second row of panels in Fig.~\ref{fig:ha_profiles}, respectively. \smallskip

    \item Red filling (RF): When only the red wing of the line is filled. The labels ``RF" and ``RF+" separate those situations in which the filling reaches 1/2, or up to the continuum of the line depth, respectively. An example of ``RF+" profile is in the left panel of the third row in Fig.~\ref{fig:ha_profiles}. \smallskip
    
    \item P-Cygni shape (PCy): When the emission is in the red wing of the line profile. In the temperature domain of the stars in the sample, the shape is generally produced by the presence of stellar winds with specific characteristics \citep[see, e.g.,][]{1979ApJS...39..481C}. We use the sub-labels ``PCy", ``PCy+" and ``PCy++" to indicate the observed the emission up to $\sim$1.25, between this and $\sim$1.5, and any value above $\sim$1.5 (traditional P-Cygni profiles) from the continuum, respectively. One example of ``PCy" and ``PCy++" shapes are included in the right panel of the third row and the bottom left panel of Fig.~\ref{fig:ha_profiles}, respectively. \smallskip
    
    \item Pure emission (Em): When there is emission above the normalized flux. Similarly to the P-Cygni shape, we set the labels ``Em", ``Em+" and ``Em++" to indicate when the emission reaches up to $\sim$1.25, between $\sim$1.25 and $\sim$1.5, and any value above $\sim$1.5 from the continuum, respectively. The latter case is found in many ``Be" star. One example of a ``Em+" profile is included in the bottom-right panel of Fig.~\ref{fig:ha_profiles}. \smallskip     
    
\end{itemize}

This classification scheme was also applied to all H$\beta$ profiles. Notably, there are many situations in which the labels assigned to both profiles are not necessarily the same (see Sect.~\ref{subsubsection:425_YYYYY}). For example, in the case of classical Be stars with circumstellar disks, the H$\beta$ profile may be labeled as "DsP+" or "DsP++", while the H$\alpha$ profile has "Em+" or "Em++". A similar situation may occur in the case of stars with strong winds, where the emission in H$\alpha$ is expected to be more significant than in H$\beta$.


\subsubsection{Identification of spectroscopic binaries}
\label{subsubsection:322_YYYYY}

We also carried out a visual identification of SB2+ systems in the sample (See~Sect.~\ref{subsection:32_YYYYY}). To do this, we inspected the temporal behavior of several diagnostic lines (mainly He~{\sc i} $\lambda$5875.62\,{\AA}, but also the Si~{\sc iii} $\lambda\lambda$4552.62,4567.84,4574.76\,{\AA} triplet, or other lines like Mg~{\sc ii} $\lambda$4481, C~{\sc ii} $\lambda$4267, or Si~{\sc ii} $\lambda$6371) using all the available multi-epoch spectra. Some examples of identified SB2+ systems can be found in Fig.~\ref{fig:sb2} of Appendix~\ref{apen.sb2}.

We also refer the reader to a forthcoming paper \citep{simondiaz2023} centered in the detailed identification of the single line spectroscopic binaries (SB1) using spectroscopic and photometric data.


\subsection{Line-broadening analysis}
\label{subsection:33_XXXXXX}

Although the full quantitative spectroscopic analysis of the stars in the sample (see Sect.~\ref{subsection:42_YYYYY}) will be presented in forthcoming papers, here we followed the guidelines in \citet{2014A&A...562A.135S} and \citet{2017A&A...597A..22S}, to obtain estimates of the projected rotational velocities of all star in our working sample.

In brief, we used the {\tt IACOB-BROAD} tool to perform a line-broadening analysis of a set of adequate diagnostic lines. In particular, for those stars with \vsini\,\ls\,200\,\kms\ we used a reduced set of not-blended and strong absorption lines optimized for two ranges of spectral types. For stars earlier than B4, we used Si~{\sc iii} $\lambda$4567.85\,{\AA} except when the line was weak, in which case we used Si~{\sc iii} $\lambda$4552.622\,{\AA}. For B4 and later spectral types we used Si~{\sc ii} $\lambda$6371.37\,{\AA} preferably, but depending on the spectrum we also used Si~{\sc ii} $\lambda$6347.11\,{\AA}. However, the latter is blended with Mg~{\sc ii} lines and therefore we only used in a very few occasions. For stars with \vsini\,\gs\,200\,\kms\ most of these lines are blended by other surrounding lines or became too diluted to provide reliable results. In this case we decided to use He~{\sc i} $\lambda\lambda$4387.93,5015.68\,{\AA} lines, which are less affected by wind than other He lines and are well isolated.


\section{Results and discussion}
\label{section:4_ZZZZZ}


\subsection{Sample selection}
\label{subsection:41_YYYYY}

\begin{figure}[!t]
\centering
\resizebox{\columnwidth}{!}{\includegraphics{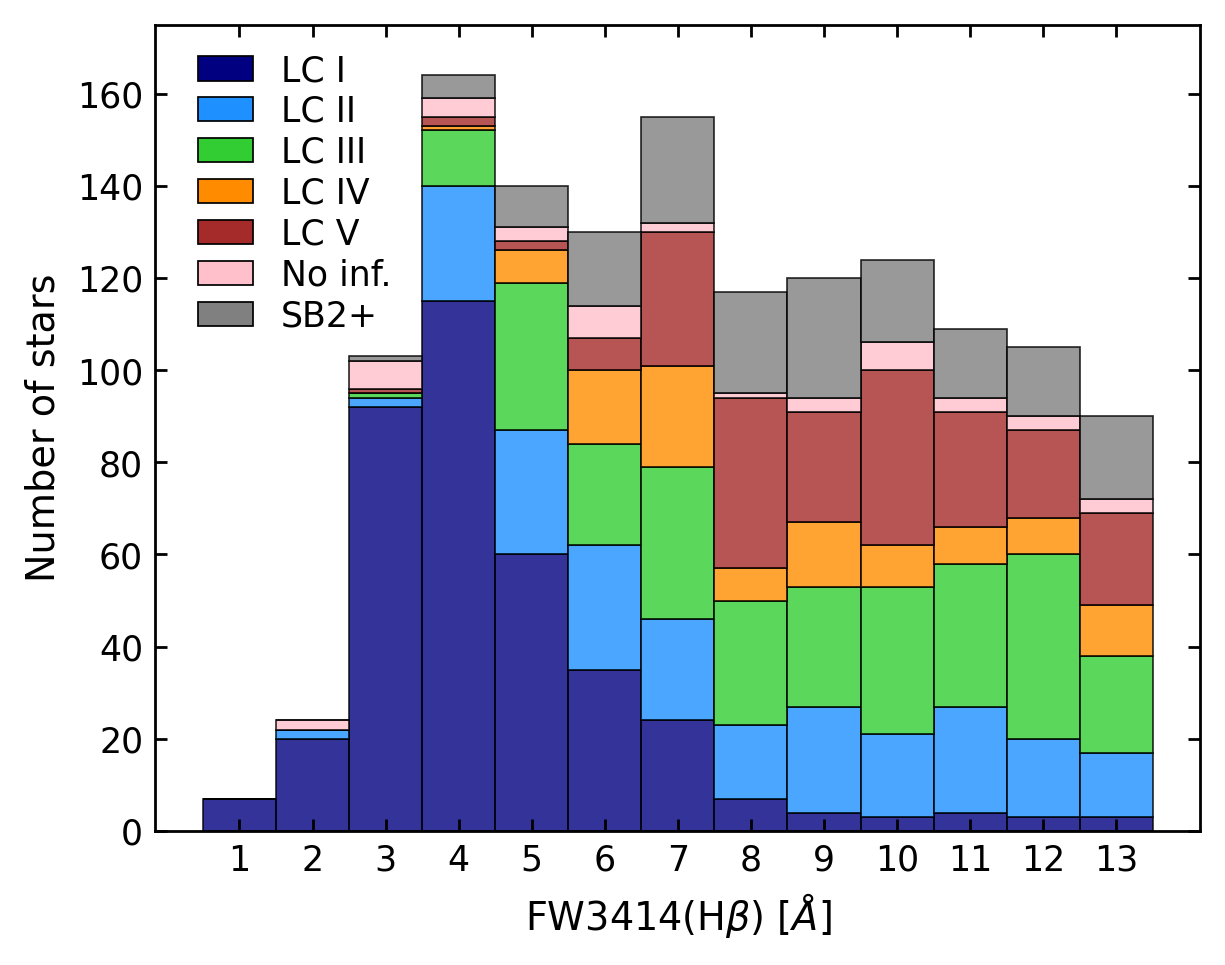}}
\caption{Histogram of the number of stars in bins of \fwhb\ for the initial sample of O9\,--\,B9 type stars. In each bin, the number of stars are stacked by luminosity class with different colors. The spectral classifications have been taken from \textit{Simbad}. The sources without a luminosity class are grouped in pink color under the ``No inf." label. Independently from their LC, all stars classified as SB2+ are grouped in bins with gray color.}
\label{fig:hist_fw}
\end{figure}

Fig.~\ref{fig:hist_fw} shows a histogram with the results of the \fwhb\ measurements for the complete initial sample of stars described in Sect~\ref{subsection:21_YYYYY} (except for $\sim$100 stars for which the fitting of the H$\beta$ failed\footnote{Among these 100 objects, $\sim$75\% correspond to Be stars, $\sim$15\% to SB2+ stars and $\sim$10\% to hypergiants.}). We separate different groups accordingly to the LC quoted by default in \textit{Simbad} and include two additional groups comprising, on the one hand, those stars for which \textit{Simbad} does no provide any LC (pink), and, on the other hand, those stars which we have identified as SB2+ (gray, see Sect.~\ref{subsubsection:322_YYYYY}).

We can see that the measured \fwhb\ for the LC I stars peaks at $\approx$\,4\,\AA\ and, except for a few cases, this LC group extends up to \fwhb\,$\approx$\,7.5\,\AA. Instead, stars with LC II, III and IV are present in a wider range, with \fwhb\,\gs\,4\,\AA, while stars with LC V mostly concentrate at \fwhb\,\gs\,6.5\,\AA.

\begin{figure*}[!t]
\centering
\includegraphics[width=1\textwidth]{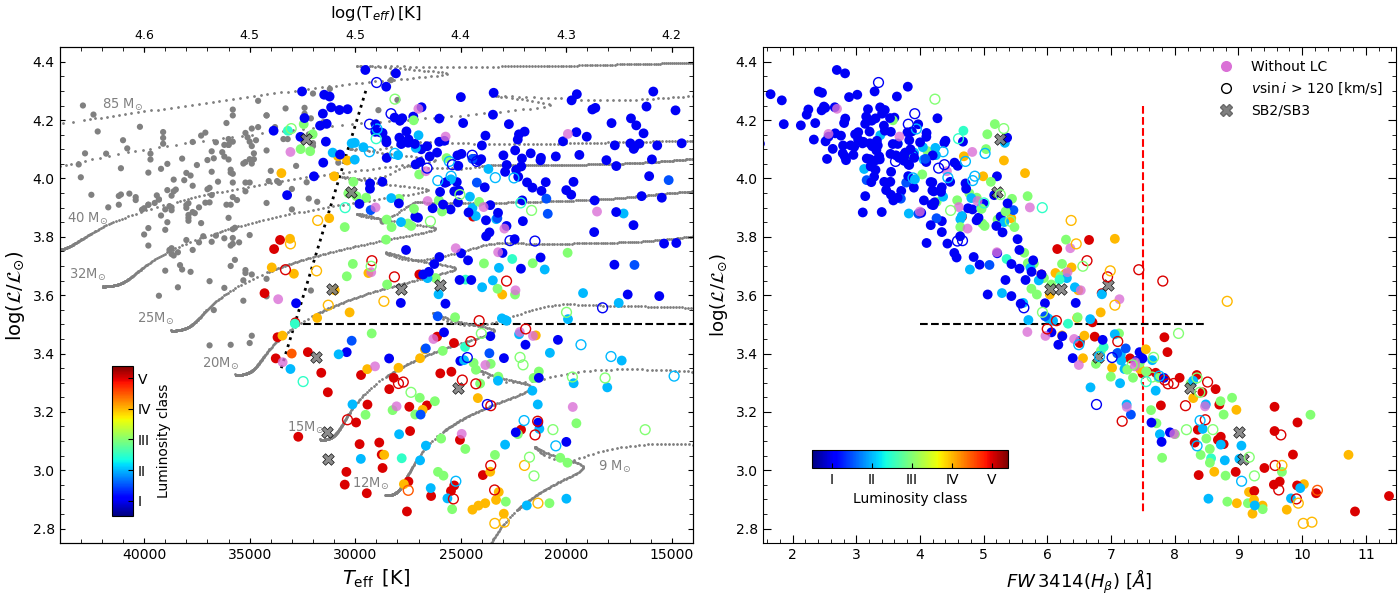}
\caption{Preliminary results from the quantitative spectroscopic analysis of $\sim$500 O9\,--\,B6 type stars presented in this work for which the \fwhb\ has been measured. The color-code in both panels indicates the LC taken from \textit{Simbad}. Left panel: sHRD for these stars and additional results from \citet{2018A&A...613A..65Hol}, indicated with gray dots. The evolutionary tracks are taken from the MESA Isochrones \& Stellar Tracks online tool for solar metallicity and no initial rotation. The approximate separation between the O-type stars and the B-type stars is indicated with a black dotted diagonal line. The black dashed horizontal line at \logLs\ = 3.5\,dex used in the selection of the working sample is indicated (also shown in the right panel). Right panel: \logLs\ against \fwhb\ for the same $\sim$500 stars. The vertical red dashed line shows the adopted value to select the working sample. Open circles indicate stars with \vsini\ > 120\,\kms, and gray crosses are SB2+ systems.}
\label{fig:shrd_fw_adb}
\end{figure*}

Fig.~\ref{fig:shrd_fw_adb} helps to better understand the distribution of the various luminosity class groups in the above-mentioned histogram. Similarly to Fig.~\ref{fig:shrd_fw_ss}, the left panel of this figure shows a sHRD, while the corresponding \logLs\,--\,\fwhb\ diagram is shown on the right panel. This time, both diagrams are populated with a much larger sample, resulting from the preliminary quantitative spectroscopic analysis of a sub-sample of the stars used to build Fig.~\ref{fig:hist_fw}, and comprising $\sim$500 O9\,--\,B6 type stars. For reference purposes, and adding extra information to what we already pointed out when introducing Fig.~\ref{fig:shrd_fw_ss}, open circles in both panels indicate stars with \vsini\ > 120\,\kms, gray crosses indicate 9 SB2+ systems in which the effect of the secondary in the spectrum was such that allowed for determination of some rough spectroscopic parameters. In addition, stars for which \textit{Simbad} does not provide a luminosity class are represented as pink circles. The sHRD includes an additional sample (gray circles) of 280 likely single and SB1 stars investigated by \citet{2020A&A...638A.157H}.

First of all, by comparing the middle panel of Fig.~\ref{fig:Emul_Hb_Lspec} and the right panel of Fig.~\ref{fig:shrd_fw_adb} we can confirm the good agreement existing between model predictions and empirical measurements. Interestingly, we see that fast-rotating stars tend to be located to the right of the rest of stars for each luminosity class, as predicted. Lastly, we note that the location of those stars identified as SB2+ in the \logLs\,--\,\fwhb\ diagram is not particularly different to the rest of the stars with similar LC, hence indicating that the contribution of the two stellar components is not importantly affecting the \fwhb\ estimation, at least in those cases in which one of the two components is dominating the spectrum.

From the location of the stars in the left panel of Fig.~\ref{fig:shrd_fw_adb}, most of the stars with LC I are comprised in the range of \logLs\,$\approx$\,3.8\,--\,4.3\,dex. If we look now at the right panel of the same figure, we can see that this range in \logLs\ implies a range in \fwhb\,$\approx$\,1.8\,--\,6\,\AA, hence explaining the distribution of LC I stars in Fig.~\ref{fig:hist_fw}. For lower values of \logLs, the situation is more mixed. For example the stars in the \logLs\,$\approx$\,3.5\,--\,3.8\,dex range have LCs V\,--\,IV at \Teff\ > 30\,000\,K but also II\,--\,I at \Teff\ < 20\,000\,K. Essentially, all stars except those with LC I span between \logLs\,$\approx$2.8 and 4.2\,dex, which also explains their wider distribution in Fig.~\ref{fig:hist_fw}.

A closer inspection of the two panels of Fig.~\ref{fig:shrd_fw_adb} drives us to also conclude that some of the classifications taken from \textit{Simbad} may not be correct. For instance, it was especially suspicious for those stars with LC I and some II with \logLs\ < 3.5\,dex that are embedded in the sHRD where those with LCs III\,--\,IV\,--\,V are. To confirm this situation, we reviewed the spectral classifications of many of these stars using our own available spectra or alternatively looking at previous classifications that, to our judgment, come from reliable references. The outcome of this exercise is summarized in Table~\ref{tab:newclass_BSGs} of Appendix~\ref{apen.newclass}. As expected, most of them are early B-type stars that should be actually classified as giants, sub-giants, or dwarfs. Also, we have found that, in many cases, we can attribute miss-classifications to the fact that they were made using spectra at low resolution, as it is the case of many classifications from \citet{1999mctd.book.....H} at R\,\ls\,2\,500.

This result highlights once more the risk of using spectral classifications (especially if they come from heterogeneous sources) to select specific groups of stars, making our proposed strategy much more robust and reliable. Moreover, using this strategy has also allowed us to recover a non-negligible number of adequate candidates of interest among those stars quoted as "OB" or "O" in \textit{Simbad}, or for which luminosity classes are not provided in \textit{Simbad} that we would have otherwise missed (all of them indicated with a pink color in Figs.~\ref{fig:hist_fw} and \ref{fig:shrd_fw_adb}). Revised spectral classification for these stars, following the guidelines indicated above are provided in Tables~\ref{tab:newclass_BSGs} and \ref{tab:newclass_noLC} of Appendix~\ref{apen.newclass}. 

As mentioned in Sect.~\ref{subsubsection:312_ZZZZZ}, a good threshold value to select the evolved descendants of O-type stars is at \logLs\ = 3.5\,dex. Based on the empirical correlation between \logLs\ and \fwhb\ shown in the right panel of Fig.~\ref{fig:shrd_fw_adb}, we decided to use a filtering value of \fwhb\ = 7.5\,\AA\ to select our sample (see below), which comprises a total number of 728 stars.

By choosing this value we first see that our selection criteria became very effective for the selection of stars with LCs I\,--\,II, especially now that we have confirmed that almost all stars with \logLs\ < 3.5\,dex quoted in \textit{Simbad} as LC I and II are actually LCs III, IV or V objects. In addition, as already pointed out in Sect~\ref{section:311_XXXXX}, our method is also selecting a significant number of stars with LCs III-IV-V (mostly early B giants and sub-giants) which are of interest for the purposes of our study, aimed to perform an empirical characterization of those B-type stars evolving from the main sequence O-type stars. 

Due to the scatter of the \logLs\,--\,\fwhb\ correlation, we are, however, selecting some stars with \logLs\ in the range 3.1\,--\,3.5\,dex (see right panel of Fig.~\ref{fig:shrd_fw_adb}). This inherent limitation of the proposed methodology does now allow us to exclude them without determining first their \Teff\ and \logLs, or at least using additional information such as photometric data to filter these stars (see Sect.~\ref{subsubsection:422_FFFF}). 

To end this section about the selection process, we define now our final (working) sample as those stars selected with the above-mentioned strategy, plus some additional hypergiant stars identified from the initial sample from their characteristic P-Cygni profiles in the H$\beta$ line \citep{1992A&AS...94..569L}, leading to a final number of 733 stars. Although we could not obtain a reliable \fwhb\ measurement for all the hypergiants, they are, in principle, also natural descendants from the O-stars. All of them (12) are listed in Appendix~\ref{apen.hypergiants}. All the other stars without \fwhb\ are considered for the completeness of the sample (see Sect.~\ref{subsubsection:423_LLLLL}), as also in the forthcoming paper dedicated to spectroscopic binaries in the case of the SB2+ systems, but are all excluded in the following sections.


\subsection{Sample description}
\label{subsection:42_YYYYY}

Table~\ref{tab:observations} gathers the relevant information for the 733 stars resulting from the selection process described in Sect.~\ref{subsection:41_YYYYY}. It includes: an identifier of the star (ID), its apparent $B$ magnitude, and spectral classification (following \textit{Simbad}); the name of the fits-file in the format of the IACOB spectroscopic database corresponding to the best spectrum (see Sect.~\ref{subsection:21_YYYYY}) and its characteristic S/N in the 4\,000-5000\,\AA\ region; the measured value of \fwhb, whether the star has been identified as a spectroscopic binary (see Sect~\ref{subsubsection:424_XXXXX}); the morphological classification of the H$\alpha$ and H$\beta$ lines (see Sect.~\ref{subsubsection:425_YYYYY}) and, lastly, the estimated \vsini\ (see Sect.~\ref{subsection:33_XXXXXX}).

Although in Table~\ref{tab:observations} we quote the spectral classifications provided by default by \textit{Simbad}, we already pointed out that some of these classifications are inaccurate or even wrong in some cases. Those stars for which we have reviewed the spectral classifications (see Sect.~\ref{subsection:41_YYYYY}) have their SpT and LC in parenthesis. For the rest of this section and sub-sections, we have adopted all the new classifications included in Appendix~\ref{apen.newclass} to better interpret the results from the selection method.

In addition, for all the above-mentioned stars, Tables~\ref{tab:phot_kinem_Gaia} and \ref{tab:phot_kinem_Hipp} gather most of the relevant photometric and astrometric information gathered from {\it Gaia} or \textit{Hipparcos}, respectively. Both tables include the same identifier of the star as in Table~\ref{tab:observations}, the Galactic coordinates (l,b), the parallax ($\varpi$) and proper motions ($\mu_{\alpha}\cos{\delta}$, $\mu_{\delta}$), and the distance, which in the first case it has been retrieved from \citet{2021AJ....161..147B} ($Distance {B-J}$), and in the second case it has been derived from the inverse of the parallax as a rough estimate ($Distance$). Table~\ref{tab:phot_kinem_Gaia} additionally includes the {\it Gaia} \textit{G, G$_{BP}$, G$_{RP}$} bands magnitudes, and the RUWE value, while Table~\ref{tab:phot_kinem_Hipp} includes the \textit{B-V} magnitude. In some cases when the available data had very large errors, negative parallaxes or any other issue, we have not include the distance. These cases usually match with bright targets.


\subsubsection{Spectral types and luminosity classes}
\label{subsubsection:421_EEEEE}

\begin{figure}[!t]
\centering
\resizebox{\columnwidth}{!}{\includegraphics{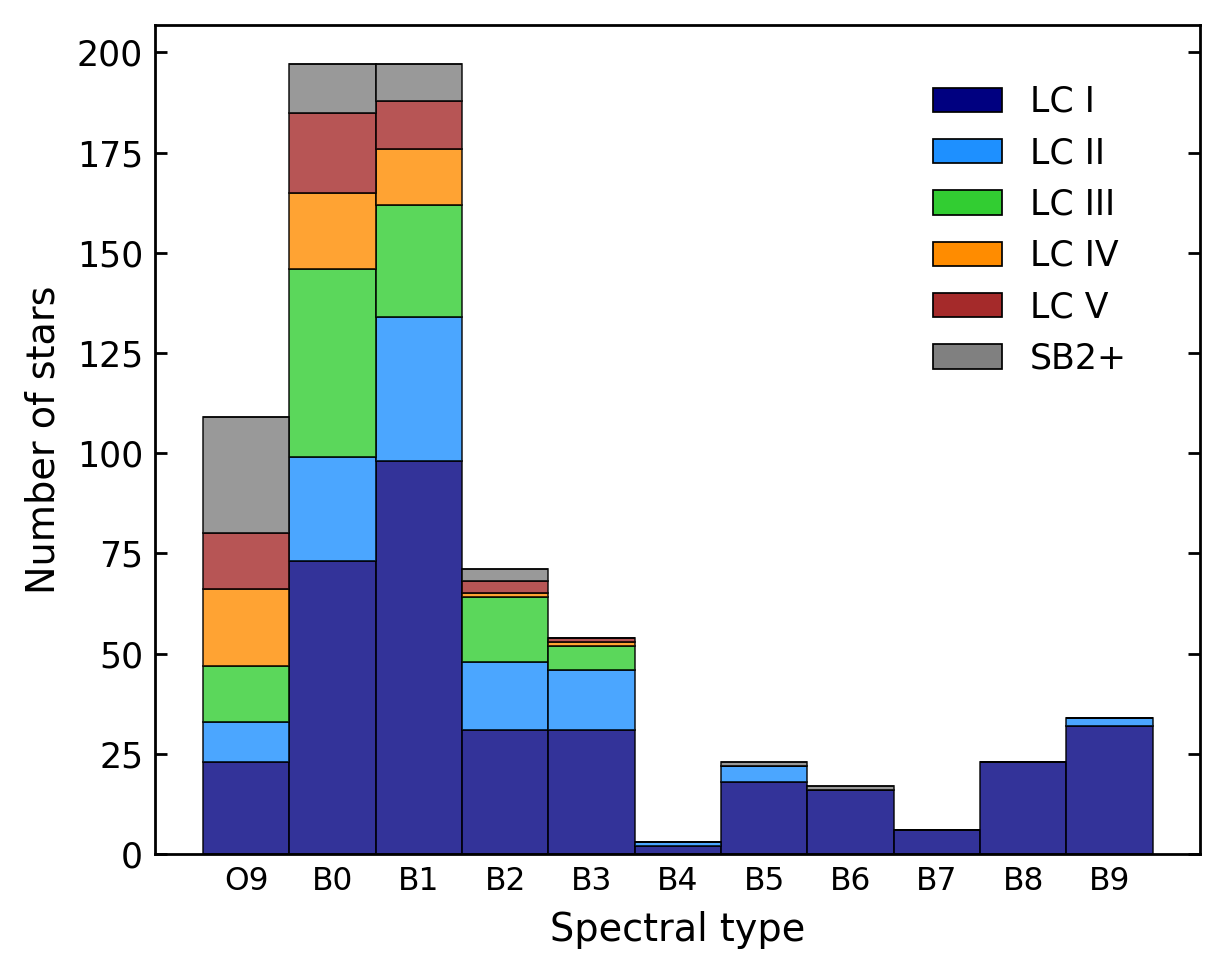}}
\caption{Histogram of the number of stars in bins of spectral types for the final sample of stars. In each bin, the number of stars are stacked by luminosity class with different colors. Independently from their LC, all stars classified as SB2+ are stacked in bins of gray color.} 
\label{fig:hist_spt}
\end{figure}

Fig.~\ref{fig:hist_spt} shows a histogram summarizing the spectral classifications of the stars in the final sample. As expected from the selection criteria (see Sect.~\ref{subsection:41_YYYYY}), the majority of the sample consists of supergiants (364) and bright giants (119) compared to stars of other classes. In particular, it shows that while stars between O9 and B1 cover all luminosity classes, from B2 and especially B3 onwards, the relative number of giants and especially sub-giants and dwarfs noticeably decreases. From B4 and towards later type stars, practically all our sources are classified as supergiants. Interestingly, despite the large number of stars included in our investigated sample, there is a clear lack of stars classified as B4 supergiants and bright giants. 
    
\subsubsection{Astrometry and photometry}
\label{subsubsection:422_FFFF}

\begin{figure}[!t]
\centering
\resizebox{\columnwidth}{!}{\includegraphics{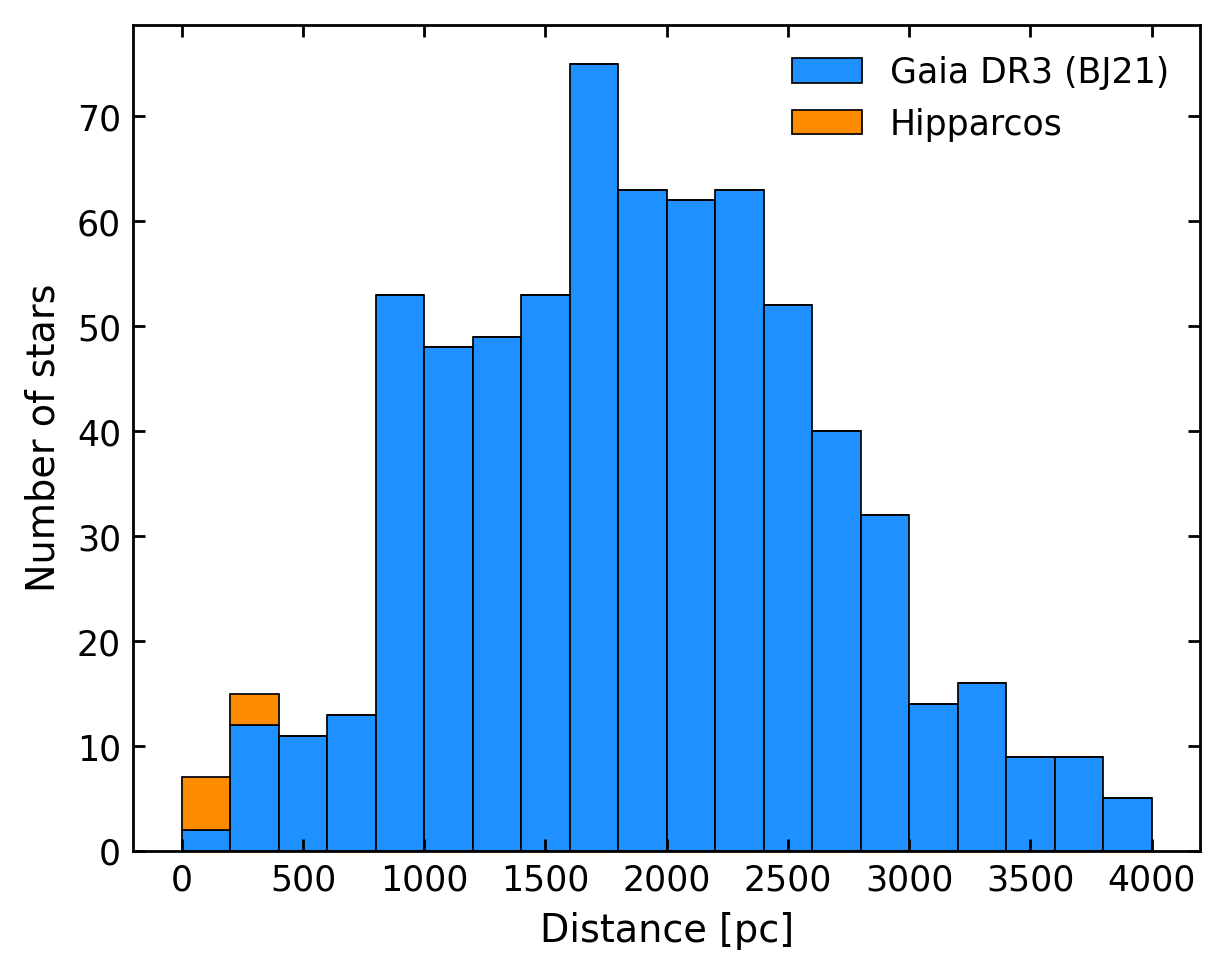}}
\caption{Stacked histogram of the number of stars in bins 200\,pc for the final sample of stars up to 4\,000\,pc. The distances have obtained either from \citet{2021AJ....161..147B} (blue color) or using the inverse of the parallax from \textit{Hipparcos} for stars without {\it Gaia} (E)DR3 data (orange color). The later ones are limited to those with $\sigma_{\varpi}/\varpi \leq$ 0.1.}
\label{fig:hist_dist}
\end{figure}

\begin{figure}[t!]
\centering
\resizebox{\columnwidth}{!}{\includegraphics{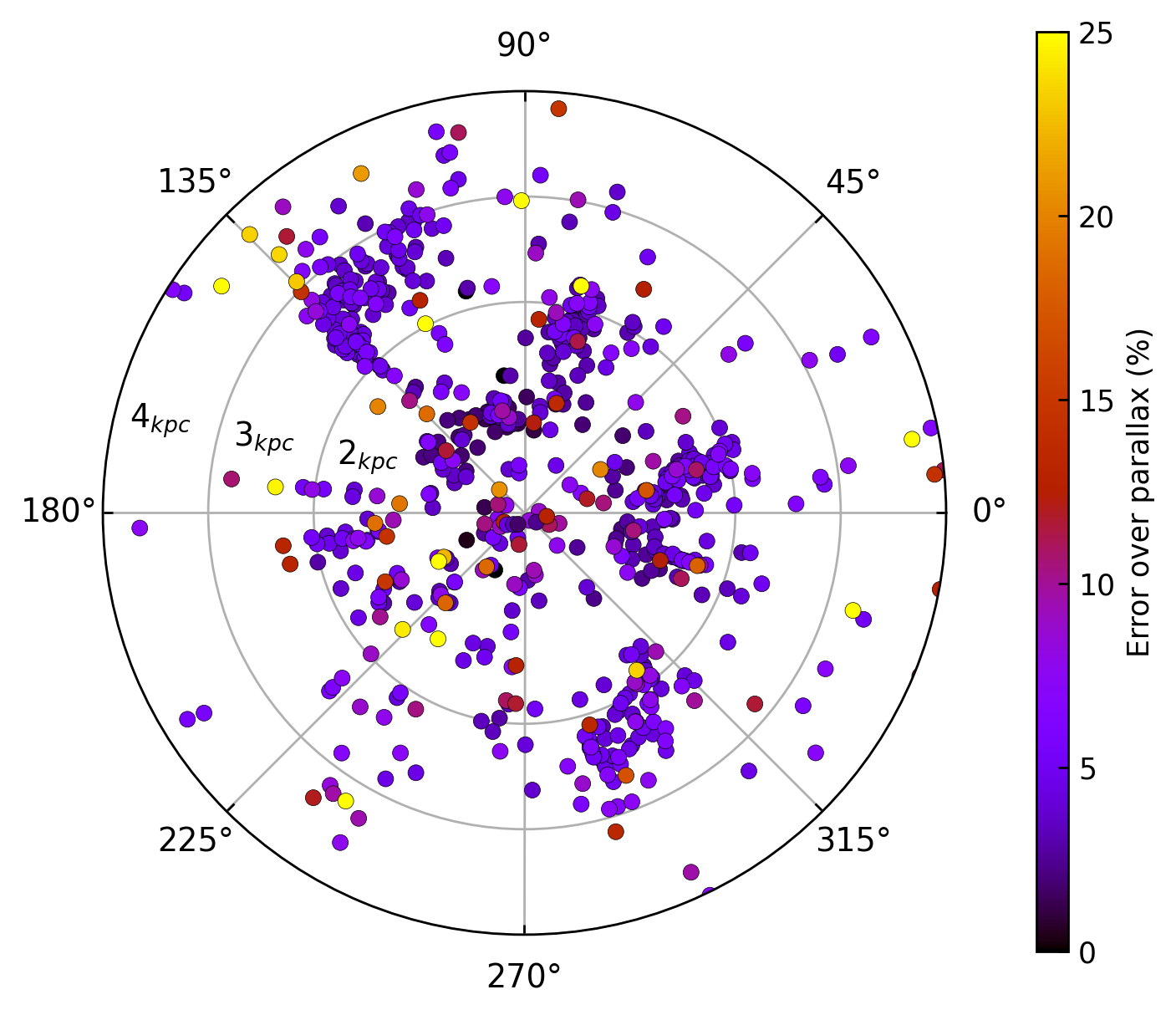}}
\caption{Polar plot in Galactic coordinates of the stars in the final sample, centered at the position of the Sun and up to a distance of 4\,kpc. The stars are colored by the $\sigma_{\varpi}/\varpi$ parameter between 0\% and 25\%. Stars with distances taken from \textit{Hipparcos} are limited to those with $\sigma_{\varpi}/\varpi \leq$ 0.1.}
\label{fig:gaia_lb}
\end{figure}

Fig.~\ref{fig:hist_dist} shows a stacked histogram of the adopted distances for all the stars up to 4\,000\,pc. Those stars that have distances from \citet{2021AJ....161..147B} are shown in blue, while in orange are indicated the remaining stars, whose distances were derived from \textit{Hipparcos} with $\sigma_{\varpi}/\varpi \leq$ 0.1 ($\sim$50\% of the total; see also Table~\ref{tab:phot_kinem_Hipp}). This histogram is complemented with Fig.~\ref{fig:gaia_lb}, which shows the distribution of the stars in a polar plot using Galactic coordinates, and centered at the position of the Sun.

We can see that below 700\,--\,800\,pc the total number of stars is relatively low ($\ls$\,50). These stars are considered to form the Gould Belt System \citep{1997FCPh...18....1P}, however, recent studies suggest that this belt-structure is in fact not real \citep[see, e.g.][]{2015A&A...584A..26B,2018A&A...620A.172Z}. Then, the number of stars suddenly increases with the distance as is expected from the volume increase, but also due to the inclusion of several important clusters and association of massive stars such as, e.g., Cep~OB2 \citep[$l\approx100\degr$, $d\approx$\,900\,pc,][]{2002AJ....124.1585C}, Sgr~OB1 \citep[$l\approx5\degr$, $d$\,\ls\,1\,400\,pc,][]{2020Ap&SS.365..112M}, Sco~OB1 \citep[$l\approx340\degr$, $d\approx$\,1\,600\,pc,][]{2016A&A...596A..82D,2020MNRAS.495.1349Y} or Cyg~OB2 \citep[$l\approx80\degr$, $d$\,\ls\,1\,700\,pc,][]{2019MNRAS.484.1838B,1991AJ....101.1408M}, many of which are embedded in the Sagittarius arm or in the newly discovered Cepheus spur \citep{2021MNRAS.504.2968P}. 

The number of stars at 1\,000\,--\,2\,400\,pc remains more or less constant up to $\sim$2\,500\,pc, where the number of stars begins to decrease significantly. One would expect the number of stars to significantly increase with the increased volume after 1\,000\,pc. However we do not observe this. While one possible and simple explanation could be because of observational biases (discussed in Sect.~\ref{subsubsection:423_LLLLL}) it can also be caused by the lack of structures with young formed regions. The latter is something that has already been observed from distances above 3\,000\,pc where we also observe a drop. However, at those distances stars begin to be too faint to be observed using the facilities used to build this sample (see below).

The color-code used in Fig.~\ref{fig:gaia_lb} indicates the error over the parallax. We can see that, in general, there are no regions with average higher errors than others, but only some dispersed stars in the plane and likely with wrong distances. In addition to the above-mentioned associations and galactic structures, we can see there the location of many others extending up to 3\,kpc such as the Per~OB1 association \citep[$l\approx130\degr$, $d\approx$\,2\,400\,$\pm$\,200\,pc,][]{1992A&AS...94..211G,2020A&A...643A.116D,2020Ap&SS.365..112M} or the Car~OB1 \citep[$l\approx300\degr$, $d\approx$\,2\,300\,pc,][]{2021ApJ...914...18S}. This result tells us about the fact that our sample concentrate in relative close groups in the Galaxy following star forming episodes similarly to what found by \citet[][Figure 5]{2021MNRAS.504.2968P}. Additionally, we also see large empty areas of very low density of stars (many of which are likely runaways from the main clusters). While some of the empty areas can be attributed to gaps between the spiral arms where no star formation occur, others can be attributed to a very high extinction (e.g. towards the direction of $l\approx45\degr$), the later affecting especially towards the direction of the Galactic center \citep[see][]{2018A&A...616A.132L,2019A&A...625A.135L}.

\begin{figure}[t!]
\centering
\resizebox{\columnwidth}{!}{\includegraphics{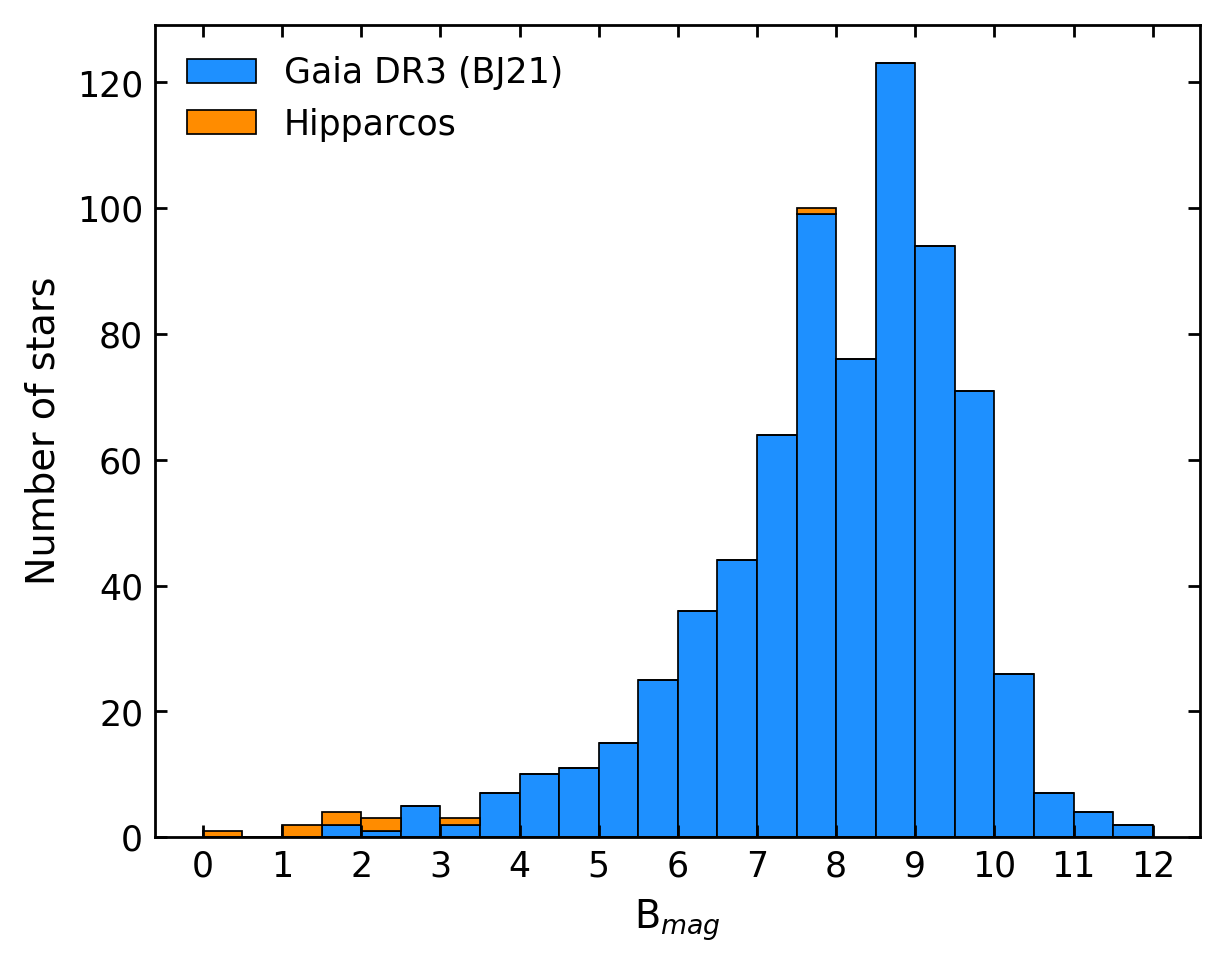}}
\caption{Histogram of the number stars in the sample with the B$_{mag}$ magnitude for the stars in the sample in bins of 0.5 magnitudes. Stars with distances taken from \textit{Hipparcos} with $\sigma_{\varpi}/\varpi \leq$ 0.1 are indicated with orange color.}
\label{fig:hist_Bmag}
\end{figure}

Lastly, Fig.~\ref{fig:hist_Bmag} shows the stars in the sample in a histogram of the apparent B$_{mag}$ magnitude. We can see that the stars with distances taken from \textit{Hipparcos} are all located within the first hundred parsecs in Fig.~\ref{fig:hist_dist}. The fact that most of these stars are among the brightest ones is not casual, as {\it Gaia} is known to have saturation issues for the brightest targets \citep[see][]{2018A&A...616A...2L}, thus missing the astrometric and photometric data. We also see in this histogram that the number of stars exponentially increases up to B$_{mag}\approx$\,9.5 and then rapidly decreases. This is expected as fainter stars become less suitable for an adequate use of the considered observing facilities (see Sect.~\ref{subsection:21_YYYYY}), in order to reach a balance between the exposure times and S/N. 

\begin{figure}[!t]
\centering
\resizebox{\columnwidth}{!}{\includegraphics{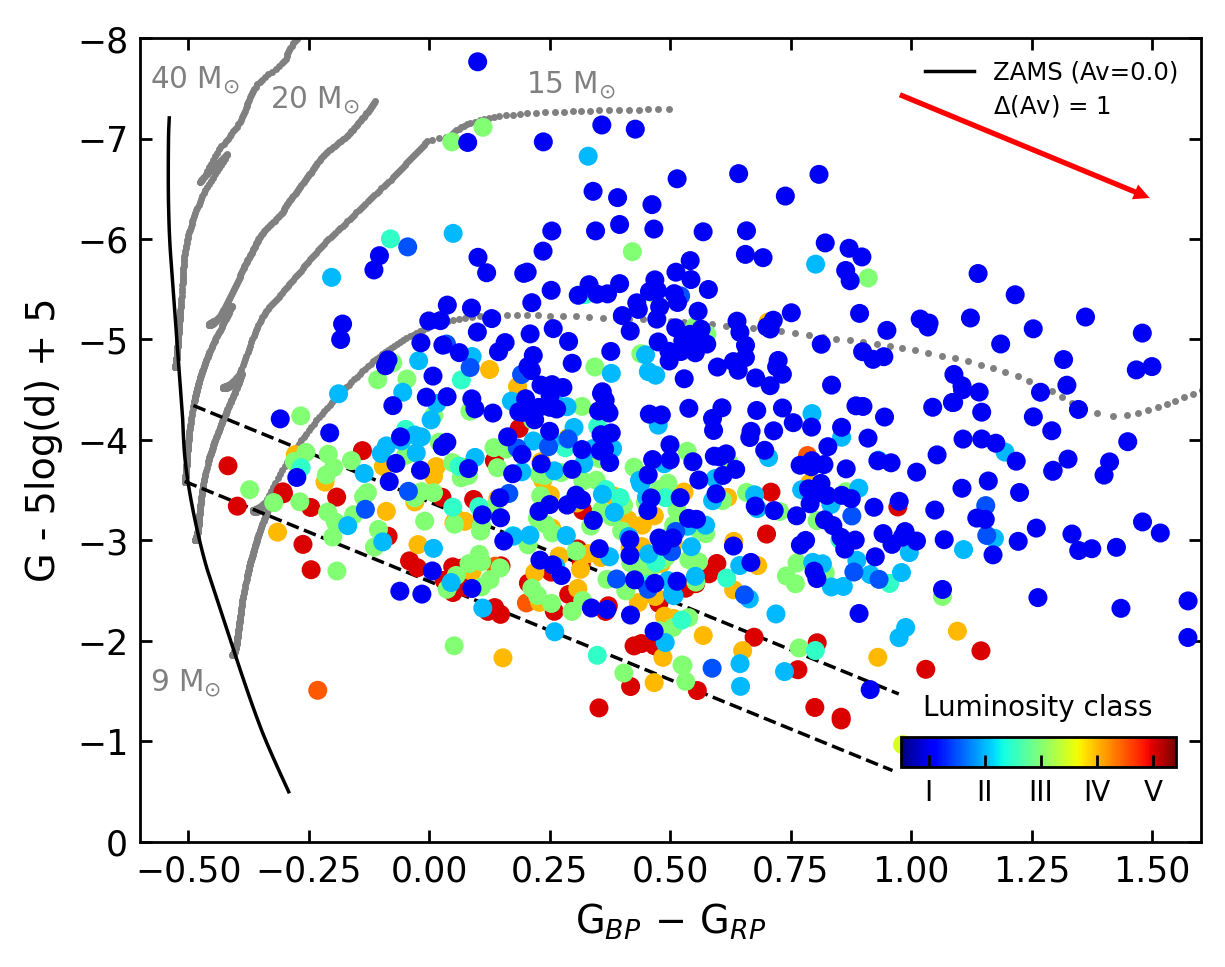}}
\caption{{\em Gaia} DR3 G$_{mag}$ corrected from distance against $G_{BP}$ - $G_{RP}$ for the selected sample, color-coded with the LC. Black solid line is the ZAMS. Evolutionary tracks for 9, 15, 20 and 40\,\MSol\ obtained from MESA with no initial rotation and no extinction are shown in gray. A reddening vector with $\Delta$(Av) = 1 is indicated with a red arrow. Two diagonal black dashed lines show two extinction lines for a 20\,\MSol\ star and for the approximate position in Fig.~\ref{fig:shrd_fw_adb} where the same track crosses the value of \logLs\,$\approx$3.5\,dex, respectively.}
\label{fig:gaia_phot}
\end{figure}

Fig.~\ref{fig:gaia_phot} provides a different and complementary view of the sample, this time in a color-magnitude diagram (CMD) constructed using data from {\em Gaia} DR3 (in particular, the $G$ magnitude corrected from distance against $G_{BP}$ - $G_{RP}$ for the stars in the sample, color-coded by the LC). The figure includes the location of the ZAMS and four non-rotating evolutionary tracks computed with the MESA code for 9, 15, 20 and 40\,\MSol\ (up to the end of the He burning phase), where no reddening was applied. In addition, the inclined red arrow located at the top right corner of the figure indicates the expected effect of extinction in the location of stars in this diagram. 

Similarly to the case of Fig.~\ref{fig:shrd_fw_adb}, we can see a gradient in the position of the stars with LC from I-II down to III-IV-V. More interestingly, we notice that, compared to the expected location of the investigated sample of stars in the CMD (if they were not affected by extinction), namely, between the depicted 9\,\MSol\ and 40\,\MSol\ evolutionary tracks, almost no stars populate this region. This indicates that an important fraction of stars in our sample is expected to be affected by extinction higher than 1 magnitude (and up to 3\,--\,4 mags in some cases).
Indeed, we can see how inefficient this diagram is to classify O- and B-type stars beyond providing a rough estimate of the luminosity class: the range of variation in $G_{BP}$ - $G_{RP}$ of the evolutionary tracks is negligible compared to the scatter of the empirical sample due to the extinction. More specifically, we have accounted for $\sim$50 O9\,--\,B1 LC I stars with $G_{BP}$ - $G_{RP}$ > 0.80, i.e. with a very high reddening compared to those of the same kind present at $G_{BP}$ - $G_{RP}$\,$\sim$\,0.0.

Fig.~\ref{fig:gaia_phot} also includes two diagonal black dashed lines. The bottom one follows the extinction line from the ZAMS of a 20\,\MSol\ star, while the top one has been chosen to start at the approximate position in Fig.~\ref{fig:shrd_fw_adb} where the same track crosses the value of \logLs\,$\approx$3.5\,dex, which also coincides with the approximate separation between O- and B-type stars. This has been done to locate the approximate position of the stars evolving from the O-type and containing all the BSGs. Although this approach could also be used together with \fwhb\ to filter unwanted stars, we have decided not to use it, due to the large uncertainties in the distance of some of the stars in the sample (especially for the bright targets). In fact, while some stars are wrongly classified with LC I (Sect.~\ref{subsection:41_YYYYY}) some others are likely below this top extinction line due to underestimated distances. The latter is compatible with the scenario presented in \cite{2021MNRAS.504.2968P} where the use of a different prior produces longer distances compared to the ones used in this work from \cite{2021AJ....161..147B}. Nevertheless we can see that most of the stars in the sample are located above the top extinction line. 

We recall here that our selection method based on the values of \fwhb\ also selects stars with 3.5\,\gs\,\logLs\,\gs\,3.1\,dex. Despite the possibility of underestimated distances, we locate most of these stars between the two reddening lines, which indicates that by using the correct distances, this method is effective for selecting stars.

\begin{figure}[!t]
\centering
\resizebox{\columnwidth}{!}{\includegraphics{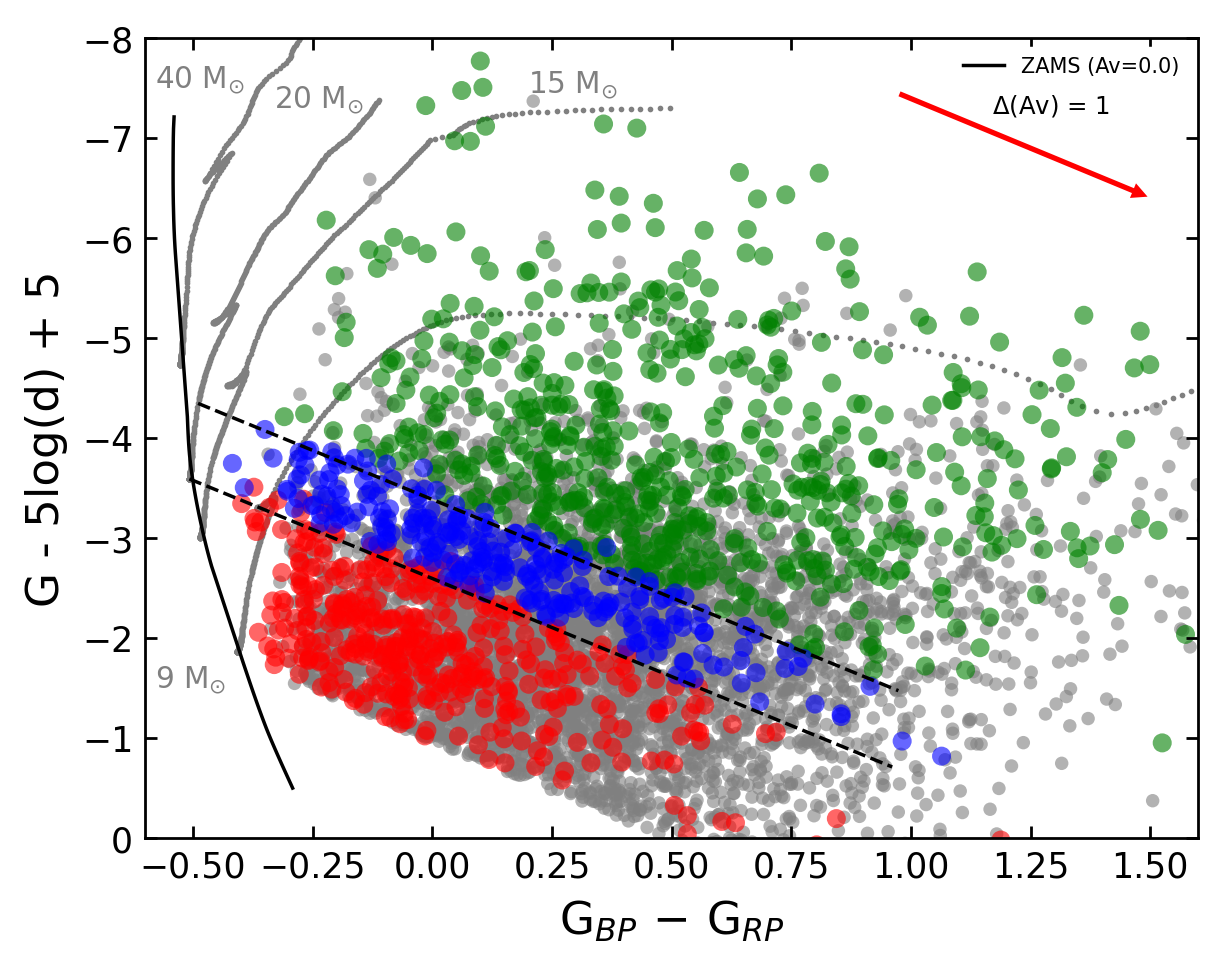}}
\caption{Same figure as in Fig.~\ref{fig:gaia_phot} but now including O9 to B9 stars from the ALS~III catalog (Pantaleoni González, et al., in prep.) limited to stars $B_{mag}$ < 11, and distances of 4\,000\,pc. Two diagonal black-dashed lines mark the reddening line of a 20\,\MSol\ star, one starting from the ZAMS (bottom) and the other at the approximate age where intersects the horizontal black-dashed line in the left panel of Fig.~\ref{fig:shrd_fw_adb} at \logLs\ = 3.5\,dex, (top) both extended up to a $\Delta$(A$_{v}$) = 2. See Sect.~\ref{subsubsection:423_LLLLL} for the different colors. Four tracks from MESA with A$_{V}$ = 0.0 and no initial rotation are included for 9\MSol, 15\MSol, 20\MSol\ and 40\MSol.} 
\label{fig:gaia_alsIII}
\end{figure}


\subsubsection{Completeness and observational biases}
\label{subsubsection:423_LLLLL}

This section is dedicated to provide an overview of the completeness and potential observational biases affecting our sample. As shown in Fig.~\ref{fig:hist_Bmag}, this sample was mostly built to be a magnitude limited sample. In this regard, it is important to remark that any use of this sample to empirically constraint e.g. evolutionary models must take into account that, for a given limiting magnitude, there will be more mid-late B-type stars compared to the early-B or O-type as the first are intrinsically brighter in this band compared to the latter group, thus being observable at larger distances (ignoring extinction). However, as shown in Sect.~\ref{subsubsection:422_FFFF}, many stars within the sample have very different reddening, dimming their brightness up to several magnitudes. As a consequence, one has to take the extinction into account to ``unbias" the sample from potential fainter stars that are in fact at closer distances. To overcome these biases, one possibility is to investigate to what extent we can build a volume-limited sample (see below).

In order to assess the completeness of our sample within a specific volume or distance, we have cross-matched our database with the third update of the Alma Luminous Star catalog (Pantaleoni González, et al., in prep.), which not only includes all the stars in our sample but also provides an increased number of stars compared to earlier releases by including fainter objects (down to $B_{\text{mag}} \approx 16$) and covering a larger distance (up to 10\,000\,pc). However, despite the authors' claim of high completeness within the first 5\,000\,pc, the fraction of stars with $B_{\text{mag}} > 16$ within this distance is not negligible due to extinction along the line of sight. For instance, if we consider stars up to 2\,000\,pc and use the absolute V-band magnitudes from \citet{1968ApJS...17..371L}, the intrinsic faintest stars within our sample could have a maximum extinction of A$_{v}$\,=\,8\,--\,9 in order to have $B_{\text{mag}} \leq 16$ (see further notes in Appendix~\ref{apen.completeness}). While some star-forming regions within this distance exhibit average extinction values close to this limit \citep[e.g. Cyg-OB2, A$_{v}$\,=\,5\,--\,7,][]{1991AJ....101.1408M}, highly obscured massive stars have been found to have extinctions exceeding this limit \citep[see, e.g.,][]{2020MNRAS.495.3323C}, thus indicating that a complete sample cannot be assumed. Nevertheless, we consider the ALS~III to be a very good reference to find missing stars within the observing magnitude limitations.

Fig.~\ref{fig:gaia_alsIII} presents a similar CMD diagram to the one shown in Fig.~\ref{fig:gaia_phot}, but this time including all stars quoted in the ALS~III catalog which fulfill the following criteria: (1) being classified as O9\,--\,B9 in \textit{Simbad} (2) being located above the extinction line corresponding to a ZAMS star of 9\,\MSol, (3) having a $B_{mag}$ < 11, and (4) having a estimated distance\footnote{Derived as described in Sect.~\ref{subsection:22_YYYYY} for the case of our working sample} below 4\,000\,pc.

We use the two diagonal black-dashed lines to establish three regions in the diagram which, as described in Sect.~\ref{subsubsection:422_FFFF}, are expected to be mostly correlated with \logLs\ (and hence also with the luminosity) of the stars. We then highlight using green, blue and red circles those stars for which we have spectra available in each of these regions, respectively. In particular, with different colors: stars in the full sample of O9\,--\,B9 type 
stars for which the upper and lower distance based on the errors both lie above the top reddening line are marked in green, those for which one of the distances lie between both reddening lines in blue, and those for which one of the distances lie below the bottom reddening line in red. The remaining stars populating the CMD in gray color correspond to those not yet observed.

Since we are mostly interested in those stars located above \logLs\ = 3.5\,dex, which correspond to the sample located above the top reddening line in Fig.~\ref{fig:gaia_phot} (green stars), hereafter we only discuss about the completeness and potential observational biases affecting this region. But before entering into that discussion, we consider of interest to remark that, even considering the full sample of O9\,--\,B9 stars in ALS~III, the lack of stars in the region of the diagram in between the non-reddened 15\,--\,40\,\MSol\ evolutionary tracks still persists. 

\begin{figure}[!t]
\centering
\resizebox{\columnwidth}{!}{\includegraphics{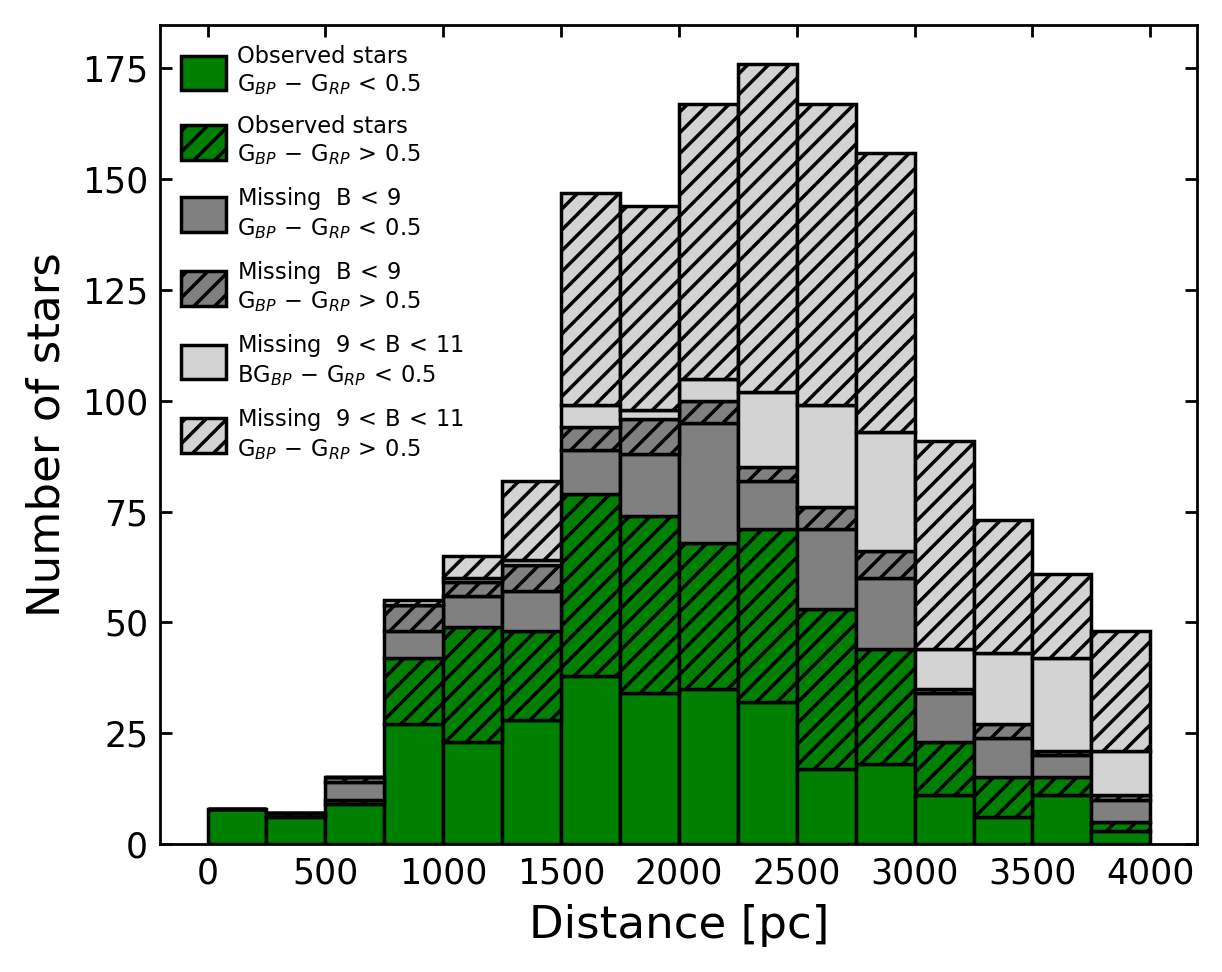}}
\caption{Histogram summarizing the completeness of the sample respect to the ALS~III catalog for stars above the top reddening line of Fig.~\ref{fig:gaia_alsIII} (see Sect.~\ref{subsubsection:423_LLLLL} for more details), in bins of 250\,pc of distance. Observed O9\,--\,B9 stars are in green, while missing stars from the ALS~III are colored in two gray colors: in dark gray, missing stars with $B_{mag}$ < 9, and in light gray, stars with 9 < $B_{mag}$ < 11. For each group, hatched areas indicate those stars with $G_{BP}$ - $G_{RP}$ > 0.5.} 
\label{fig:hist_als}
\end{figure}

To visually evaluate the completeness, Fig.~\ref{fig:hist_als} shows a stacked histogram of the selected sample of stars from ALS~III which are located above the top reddening line of Fig.~\ref{fig:gaia_alsIII}, indicating in green those stars for which we have spectra, but separating the missing stars in two gray colors: dark gray if $B_{mag}$ < 9 and light gray for those stars with 9 < $B_{mag}$ < 11. In addition, we highlight with hatched areas those stars with a $G_{BP}$ - $G_{RP}$ > 0.5, i.e., indicating those targets which are more importantly affected by extinction.

\begin{table}[!t]
    \centering
    \caption{Summary of the completeness of the sample for four different ranges of distances.}
    \label{tab:completeness}
    \begin{tabular}{lcccc}
        &  Total \#       & \% obs.         & Total \#          & \% obs. \\
        &  $B_{mag}\leq$9 & $B_{mag}\leq$9  & 9<$B_{mag}\leq$11 & 9<$B_{mag}\leq$11 \\
        \noalign{\smallskip}\hline\noalign{\smallskip}
        \multicolumn{5}{c}{Distance 0\,--\,1\,000\,pc}\\
        \hline\noalign{\smallskip}
          Any $\delta$   &  83 & 80\% &  2 & 50\% \\
          $\delta$ > -20 &  59 & 80\% &  1 &  0\% \\
        \noalign{\smallskip}\hline\noalign{\smallskip}
        \multicolumn{5}{c}{Distance 1\,000\,--\,2\,000\,pc}\\
        \hline\noalign{\smallskip}
          Any $\delta$   &  265 & 77\% & 173 & 27\% \\
          $\delta$ > -20 &  150 & 90\% & 104 & 38\% \\
        \noalign{\smallskip}\hline\noalign{\smallskip}
        \multicolumn{5}{c}{Distance 2\,000\,--\,3\,000\,pc}\\
        \hline\noalign{\smallskip}
          Any $\delta$   & 235 & 61\% & 431 & 21\% \\
          $\delta$ > -20 & 105 & 91\% & 234 & 34\% \\
        \noalign{\smallskip}\hline\noalign{\smallskip}
        \multicolumn{5}{c}{Distance 3\,000\,--\,4\,000\,pc}\\
        \hline\noalign{\smallskip}
          Any $\delta$   & 68 & 47\% & 205 & 13\% \\
          $\delta$ > -20 & 26 & 73\% & 123 & 15\% \\
        \noalign{\smallskip}\hline
    \end{tabular}
    \tablefoot{For each range of distance, we include the statistics for stars in the ``green" and ``blue" regions where stars of each kind are described as in Sect.~\ref{subsubsection:423_LLLLL} (see also Fig.~\ref{fig:gaia_alsIII}). In particular, the first two columns indicate the total number of stars with $B_{mag}$ < 9, and the corresponding percentage of observed ones with respect to the total. Last two columns indicate the same but for stars with 9 < $B_{mag}$ < 11.
    }
\end{table}

\begin{figure}[!t]
\centering
\resizebox{\columnwidth}{!}{\includegraphics{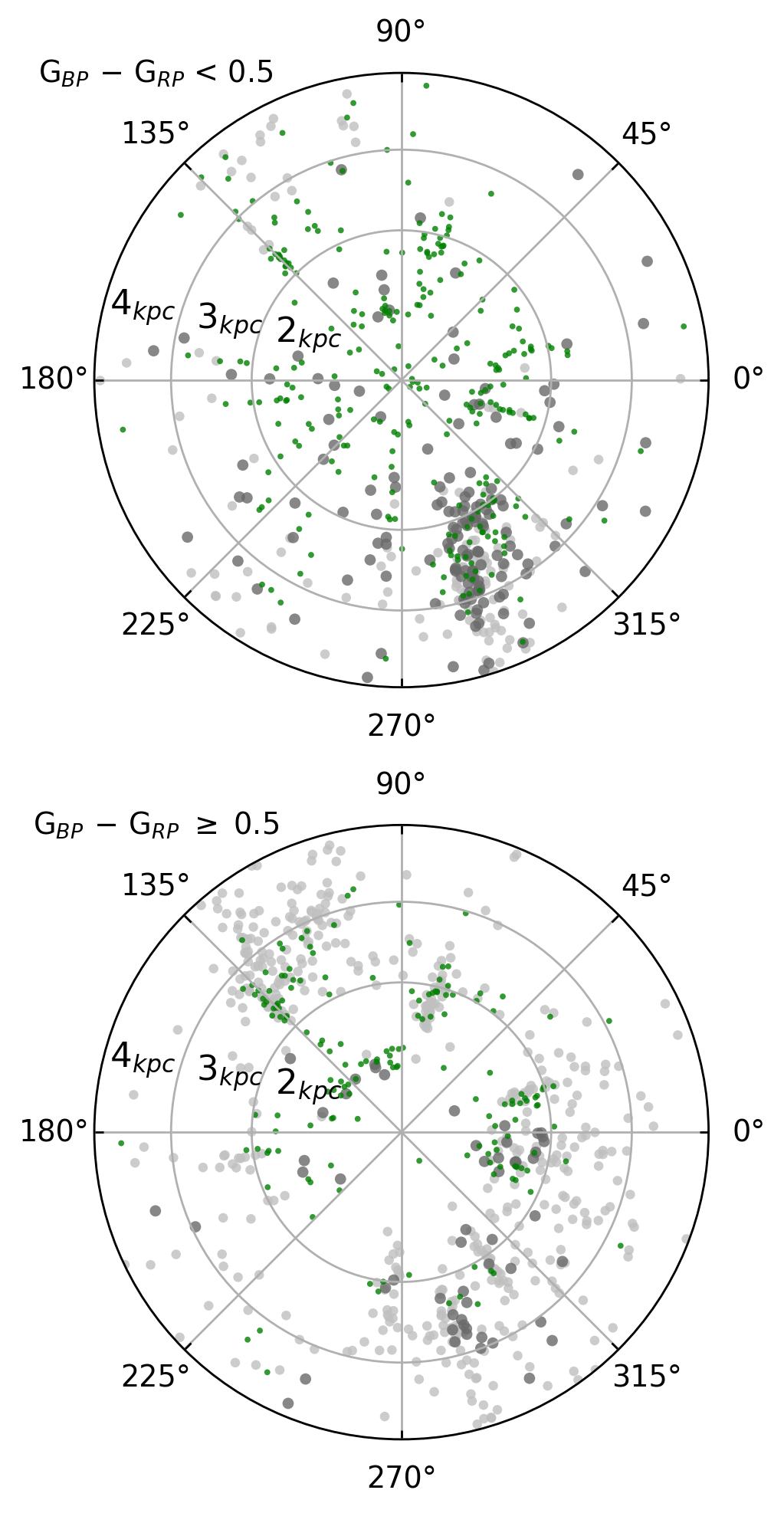}}
\caption{Two polar plot in Galactic coordinates of all the stars above the top reddening line of Fig.~\ref{fig:gaia_alsIII}, separating in the top and bottom panels those with $G_{BP}$ - $G_{RP}\mathop{\lessgtr}$\,0.5. In both panels light-gray or gray circles indicate missing stars separating those with 9 < $B_{mag}$ < 11 or $B_{mag}$ < 9, respectively, and with green circles the stars for which spectroscopic data is available.} 
\label{fig:gaia_lb_als}
\end{figure}

This figure is complemented with the information provided in Table~\ref{tab:completeness}, were we present a summary of the total number of stars and relative completeness of stars in the above-mentioned region of the $Gaia$-CMD for four different bins of distance, and separating stars below or above the $B_{mag}$ = 9 threshold. The corresponding statistics are given for the complete sample as well as for those stars observable from the Northern hemisphere (i.e. with $\delta$ > -20\,$\degr$ using either the NOT or Mercator telescopes).

We can see that up to 2\,000\,pc, we are complete up to 77\% of the total stars with $B_{mag}$ < 9, this number being reduced to 58\% for the stars located further away. It is important to point out that the IACOB project observes only with the NOT and Mercator telescopes from the Northern hemisphere (see Sect.~\ref{subsection:21_YYYYY}). In the case of the Southern hemisphere, we are currently limited to FEROS objects retrieved from the ESO public archive. In fact, only 20\% of the spectra comes from FEROS. To put this in numbers, if we only consider the stars with declination above -20\,$\degr$, from those up to 2\,000\,pc we are currently missing 30\% (93 stars) with $B_{mag}$ < 11, and only 13\% (44 stars) with $B_{mag}$ < 9 up to 4\,000\,pc. These missing (Northern) stars are currently in observing queues for upcoming IACOB observing campaigns. Regarding the missing sources with declination below -20\,$\degr$ up to 4\,000\,pc (162 stars with $B_{mag}$ < 9, and 321 with 9 < $B_{mag}$ < 11), we are presently in the process of improving the situation thanks to a recently approved proposal\footnote{The awarded time, in collaboration with Drs. Ram\'irez-Tanus and E. Zari, from the MPIA, Heidelberg, is $\approx$250h distributed in two semesters} with FEROS. Lastly, we have evaluated the contamination of classical Be stars in the completeness, as these stars are not of our interest. We concluded that while they only represent $\sim$11\% of all the stars in Fig.~\ref{fig:hist_als}, they are homogeneously distributed in Fig.~\ref{fig:gaia_alsIII} above the top reddening line (but close to the it), and therefore not varying the results in Table~\ref{tab:completeness}.

Figure~\ref{fig:gaia_lb_als} provides a complementary view of the available and missing stars, this time showing their location in the Galaxy and following the same color code as in previous figures, but separating the stars by their $G_{BP}$ - $G_{RP}$ value in two groups. Globally, it can be seen that for the region between $l$ = 0\,$\degr$ and 230\,$\degr$, we have a very high degree of completeness, while the missing Southern stars are located between 230\,$\degr$ and 0\,$\degr$. The top panel for stars with $G_{BP}$ - $G_{RP}$ < 0.5 shows those stars less reddened where it can be clearly seen the reason for separating stars with -20\,$\degr$ in Table~\ref{tab:completeness}, as the vast majority of missing stars concentrate in the Carina region of the Southern hemisphere ($l\approx270\degr$--\,$315\degr$, $d\approx$\,2\,500\,pc). Additionally we can see that most missing stars in this panel have $B_{mag}$ < 9 (dark gray circles), while almost no stars with 9 < $B_{mag}$ < 11 (light gray circles) are located within 2\,000\,pc. The bottom panel instead, shows the more reddened stars among which most missing have $B_{mag}$ > 9. They are located not only in the Carina region, but also in some other specific regions such as the Sagittarius region ($l\approx340\degr$--\,$25\degr$, $d\approx$\,1\,000\,--\,2\,500\,pc), the Cygnus region ($l\approx80\degr$, $d\approx$\,2\,000\,pc) as well as the Cassiopeia region ($l\approx115\degr$--\,$135\degr$, $d\approx$\,2\,500\,--\,3\,500\,pc). The fact that we have many stars observed in some of these regions tells us about the very different degrees of extinction present in them (see e.g. the Cygnus region). From the Northern hemisphere the Cassiopeia region got our attention as it becomes the region with more missing stars spreading in a wide space. 

Regarding the observational biases within the spectral classifications, we have carried out homogeneous observations in the Northern hemisphere trying to ensure the completeness within the B0\,--\,B3 I-III and B3\,--\,B9 I-II type stars up to $B_{mag}$ = 9. In fact, the percentages given in Table~\ref{tab:completeness} increase 5\% to 10\% if only the stars of LCs I and II are considered. Regarding the stars with LCs III-IV-V, currently $\sim$80\% are already observed, and we expect this percentage to reach $\sim$95\% in the next upcoming campaigns, taking into account that, in this case, a non negligible number will be Be stars not of our interest.

To sum up, we can confidently say that we have achieved an homogeneous and unbiased sample with a high degree of completeness respect to the ALS~III catalog except for those stars with $\delta$\,\ls\,-20\,$\deg$, and also shows the success of the IACOB project within the last 10 years observing in the Northern hemisphere.


\subsubsection{Identification of double-line spectroscopic binaries}
\label{subsubsection:424_XXXXX}

As explained in Sect.~\ref{subsubsection:322_YYYYY}, we also performed a visual inspection of all the multi-epoch spectra looking for potential SB2+ systems. In Fig.~\ref{fig:hist_fw} we can see that the fraction of SB2 systems is less towards higher \fwhb\ values, being almost none below 4.5\,\AA. Within the stars fitting the selection criteria (see Fig.~\ref{fig:hist_spt}), we account for 56 SB2+ systems, and additional 7 for which it is not clear whether they are SB2 systems or the features present in the diagnostic lines are caused by long profile variability. The latter are indicated as ``LPV/SB2?" in the ``SB" column of Table~\ref{tab:observations} and will require from additional spectroscopic observations and possibly from photometric analyses to unveil whether or not they are binary systems or the features observed in the spectra correspond to variability such as pulsations. Of the total 56 SB2+ systems, 5 of them have a third component (formally SB3 systems). Of all of them, 21 correspond to B-type stars while the rest are O9-type stars. We have identified 8 new systems not identified in the literature. New identifications are marked with an ``*'' in the ``SB" column. In some of the cases (HD\,7252, HD\,12150 and HD\,46484) they were already proposed as multiple systems from the astrometric anomalies \citep{2019A&A...623A..72K} or photometrically observed as eclipsing binaries (HD\,142634, HD\,153140 and HD\,170159). 

From Fig.~\ref{fig:hist_spt}, we can also see that, compared to the number of binary systems with a O9-type primary star, the number of systems with a B-type primary within the sample is much lower. Although several studies have estimated multiplicity for O-type stars \citep[e.g.][among some others]{2010RMxAC..38...30B,2013A&A...550A.107S}, not so many have explored the B-type domain, and most of them have only covered the dwarf stars regime with many classical Be stars within their samples \citep[e.g.][]{2006A&A...456..623E,2021A&A...652A..70B,2022A&A...658A..69B}. The number of studies dedicated to cover the B-type supergiants is even smaller \citep[][]{2015A&A...580A..93D,2015A&A...575A..70M}. In this work, we have found that the fraction of O9-type binaries in the sample is 23\% while for the B0\,--\,B9 in the sample (i.e. mostly including I-II type stars) is 3.6\%, this is $\approx$\,6 times less. This observed decrease is expected from the natural evolution of these objects as some of them may have either become much more luminous than the companion \citep[by evolution or mass transfer,][]{2012ARA&A..50..107L,2013ApJ...764..166D}, merged into a single more massive object \citep[e.g.][]{1992ApJ...391..246P,2013A&A...552A.105V}, or even separated after supernova \citep[e.g.][]{2011BSRSL..80..543D,2014ApJ...782....7D}, remaining only the less evolved companion. 


\begin{figure*}[!t]
\centering
\includegraphics[width=0.9\textwidth]{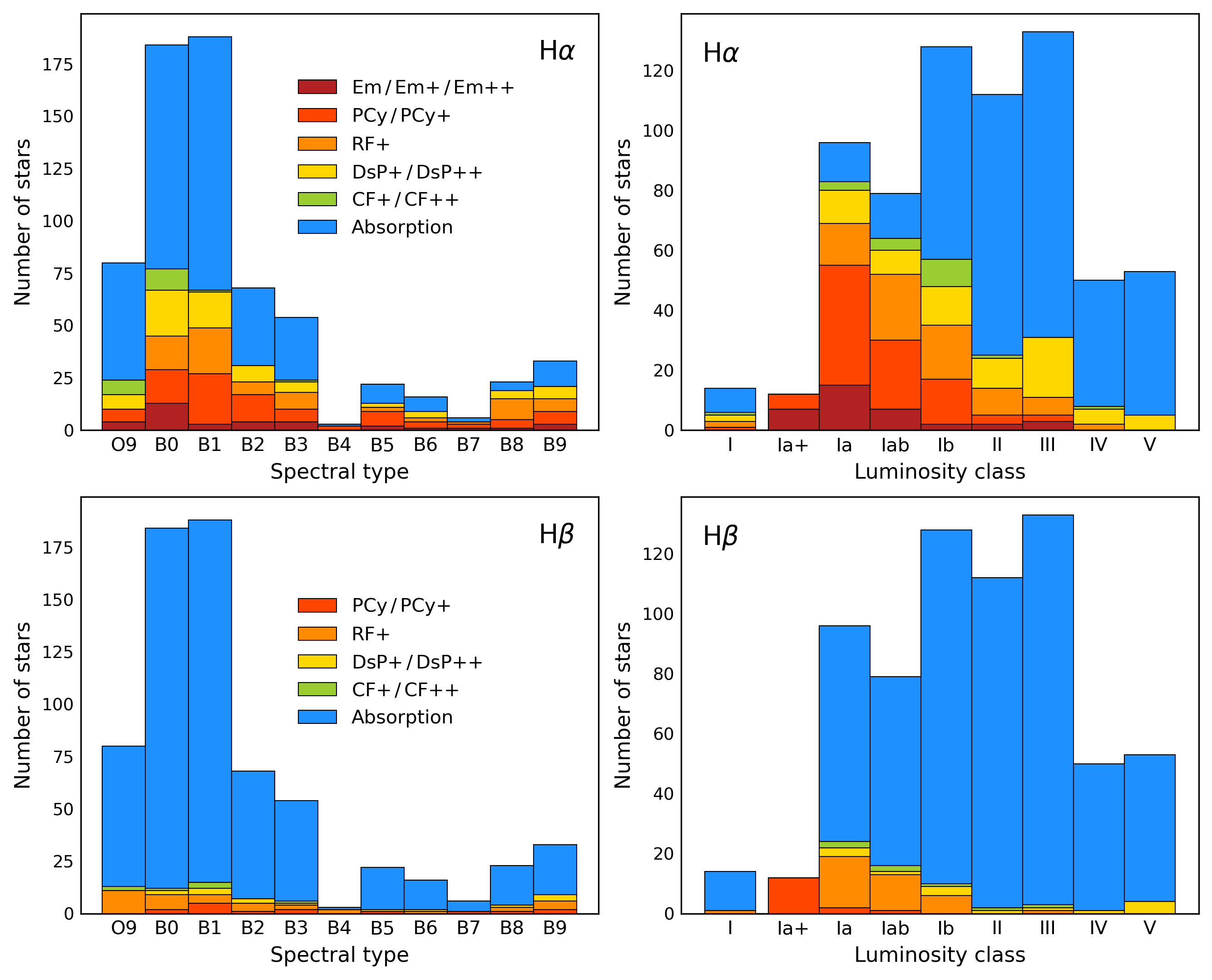}
\caption{Results of the number of stars against their spectral type (two leftmost panels) and luminosity class (two rightmost panels), represented in histograms. Each bin stacks the number of stars with different morphologies found for the H$\beta$ and H$\alpha$ profiles following the classification in Sect.\ref{subsubsection:321_FFFFFF} and grouped as in the legend. Stars labeled within the ``Absorption" label include profiles ``CF, DsP, PCy, RF". The sources with LC I without sub-type in the literature are grouped in a separate bin. The sources classified as SB2+ are removed from the bins.}
\label{fig:hab_morph}
\end{figure*}

\subsubsection{\texorpdfstring{Morphology of the H$\beta$ and H$\alpha$ profiles}{Morphology of the Hb and Ha profiles}}
\label{subsubsection:425_YYYYY}

Fig.~\ref{fig:hab_morph} shows four histograms indicating the number of stars as a function of SpT and LC for which the various types of H$\beta$ and H$\alpha$ profiles presented in Sect.\ref{subsubsection:321_FFFFFF} are identified. We note that, since those stars labeled as ``CF, DsP, PCy, RF" still have a strong absorption shape, they have been joined together with the stars labeled as ``Ab" (Absorption) for the purposes of this figure. Also in this case, we removed all the SB2+ systems.

As illustrated by the bottom two panels of Fig.~\ref{fig:hab_morph}, the majority of the stars have H$\beta$ profiles in absorption, with the exception of a group of stars labeled as ``RF+" (red wing filled) and ``PCy/PCy+" (P-Cygni shape), both evidencing the presence of winds \citep{1970ApJ...159..879L,1975ApJ...195..157C}. This group is concentrated within SpT O9\,--\,B2 and LC I (and sub-types). Of those labeled with ``PCy+", the majority correspond to the hypergiants (Ia+) as it can be seen in the bottom-right panel (see also Appendix~\ref{apen.hypergiants}).

The two panels on the top show a more mixed situation for the H$\alpha$ line. This is just a consequence of the more important effect of the stellar winds on the H$\alpha$ profile than the other hydrogen lines of the Balmer series. Interestingly, the relative fraction of stars with H$\alpha$ in absorption respect to the total for each bin decreases towards later spectral types, being $\approx$50\% on average. Regarding the luminosity classes, a similar decrease occur from dwarf stars towards the supergiants. In particular, we can see that most of the stars with LCs Ia+, Ia, Iab are labeled with ``RF+" and ``PCy/PCy+", which is also the case for the B8\,--\,B9 type stars. The extended presence of P-Cygni shapes is expected as it is well known that B-type supergiants have stronger stellar winds than other less luminous stars \citep{1999A&A...350..970K,2008A&A...478..823M}. We can also see that stars with H$\alpha$ in emission (7\% of the total) span across all spectral types, but mostly concentrate at LC I. In fact, nearly 90\% of the stars with LCs IV and V have absorption profiles in H$\alpha$. This indicates that stars affected by winds are located at higher luminosities. Lastly, profiles with sub-double peak emission gather 10\% of the total while those with the core of the lines filled are only 2\%.

\begin{figure}[!t]
\centering
\includegraphics[width=0.48\textwidth]{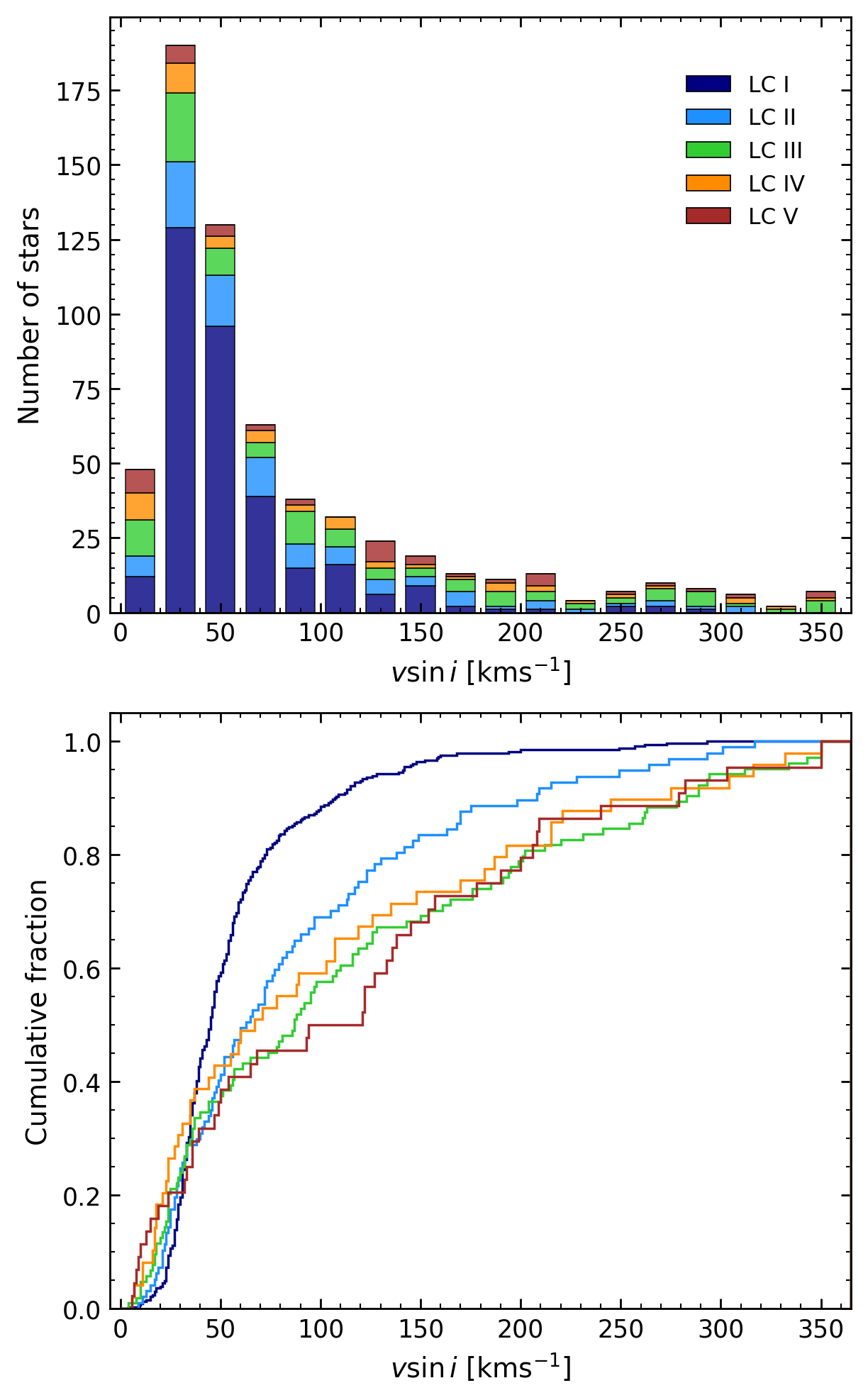}
\caption{Histogram and cumulative distribution of the \vsini\ in the sample stars, excluding the identified SB2+ stars. In the top panel, each bin of the histogram is 20\kms\ wide and stacks the stars separated and color-coded by their LC. In the bottom panel, cumulative distributions separated by LCs.} 
\label{fig:hist_vsini}
\end{figure}


\subsection{First hints about the rotational properties of the sample}
\label{subsection:43_TTTTTTT}

Top panel of Fig.~\ref{fig:hist_vsini} shows a histogram of the projected rotational velocities of our working sample (excluding those stars labeled as SB2+, and some problematic cases), highlighting in different colors the different LCs of the stars, and including in the last bin 7 stars with \vsini\,>\,340\,\kms. This distribution has a mean and median values of 80\,\kms\ and 51\,\kms, respectively (64\,\kms\ and 47\,\kms\ for stars with LCs I and II), and a standard deviation of 74\,\kms (50\,\kms), with $\approx$85\% of the sample having values below 150\,\kms, which is followed by a tail of fast-rotating stars. In addition, the bottom panel of the same figure shows the corresponding cumulative distribution. There, we see that stars with brighter luminosity classes reach higher fractions earlier than those less luminous. In particular for the supergiant stars they reach 80\% within only the first 70\,\kms. The rest have a more or less gradual increase.

\begin{figure}[!t]
\centering
\resizebox{\columnwidth}{!}{\includegraphics{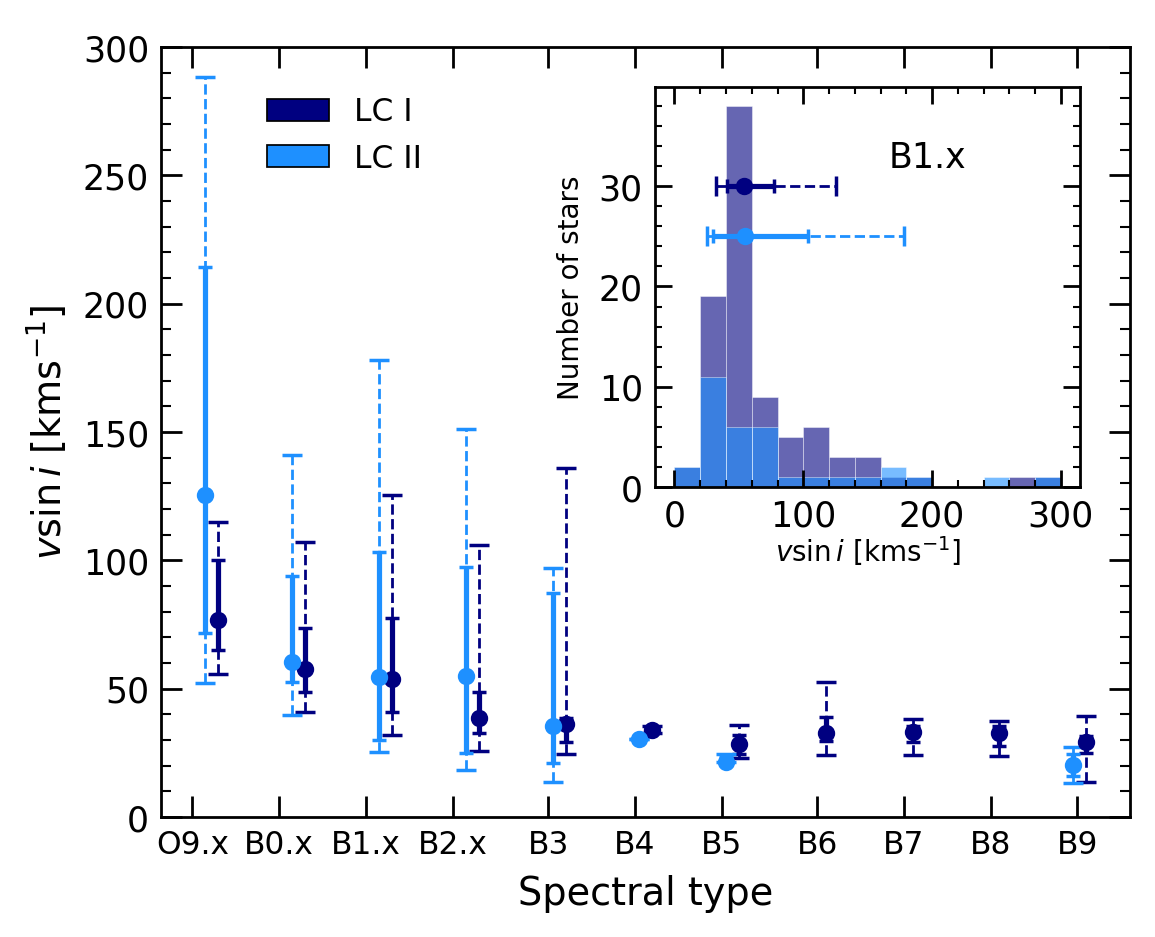}}
\caption{Main figure: Distribution of the median of \vsini\ obtained through the GoF method of {\tt IACOB-BROAD} against the SpT, separating in stars with LCs I and II. Solid error-bars indicate the upper and lower limits corresponding to the percentiles 75\% and 25\% while the dashed error-bars correspond to percentiles 90\% and 10\%. Stars without LC or SB2+ systems are not included. Sub-figure: histogram of the sub-sample of B1 I-II stars against the \vsini\ for better interpretation of the main figure and Sect.~\ref{subsection:43_TTTTTTT}. The error-bars corresponding to the B1 I-II stars are plotted horizontally.} 
\label{fig:vsini_spt_bars}
\end{figure}

Focusing our attention in the stars with LCs I and II, which are the main drivers of our study, the main plot of Fig.~\ref{fig:vsini_spt_bars} shows, for each SpT in the O9\,--\,B9 range, the median of the \vsini\ distribution and upper and lower limits corresponding to percentiles 75\% and 25\% (with solid error-bars), and percentiles 90\% and 10\% (with dashed error-bars). The main characteristic of the figure is that the mean values for both LCs gradually decrease down to 30\,\kms\ for the B3-type stars, and remain in that value up to the B9-type stars. This behavior is also present for the lower limits of the error-bars (slow rotating stars). The second main characteristic is that the upper limits of the percentiles (mainly corresponding to those more fast rotating stars) also follow the same trend up to the B3-type, where their number suddenly drops. In more detail, the sub-figure inside includes histograms of the \vsini\ distribution for the B1-type stars separating LCs I from II, illustrating the above-mentioned characteristic. There, we observe a main distribution centered at $\sim$50\,\kms\ followed by a relatively smooth tail of these fast-rotating objects with \vsini\ extending up to 300\,\kms.

To compare with some other results of \vsini\ from the literature we mainly considered those studies also using large samples of O-stars and BSGs. For instance in \citet{1997MNRAS.284..265H}, they obtained a median value of 91\,\kms\ for 373 stars, with an average difference between O- and B-type of 25\,\kms, being higher for the former ones. Compared to our results, the slightly higher values obtained there can be explained if one takes into account the effect of macroturbulence broadening, which can shift \vsini\ by $\sim$25\,\kms \citep{2014A&A...562A.135S}. This fact is actually shown in \citet{2010MNRAS.404.1306F}, where they found a $\sim$30\,\kms\ offset with those stars in common with \citet{1997MNRAS.284..265H} from their sample of 57 Galactic BSGs. These authors obtained mean values of $\approx$60\,\kms with a standard deviation of 50\,\kms, very similar to our results \citep[see also][for more examples including BSGs]{2008A&A...479..541H,2014A&A...562A.135S}.

Similarly to Fig.~\ref{fig:vsini_spt_bars}, \citet{2010MNRAS.404.1306F} and \citet{2014A&A...562A.135S} included distributions of \vsini\ with respect to the SpT for stars with LCs I and II, although the later only reached B2-type stars. In both cases their results are very similar to ours. On the one hand, in all the cases we can see a decrease in \vsini\ values towards mid-B-type stars. However, our significantly larger sample (comprising 7 to 8 times more stars) clearly extends the presence of fast-rotating stars with \vsini\,$\geq$\,100\,\kms\ up to the B3-type stars. On the one hand, the slow decrease given by the mean values could be explained by the natural evolution of the stars, decreasing their \vsini\ as they increase their radii. However, this decrease does not continue after the B4-type stars, suggesting that either they loose enough angular momentum along their evolution \citep[e.g. via stellar winds,][]{2010A&A...512L...7V} or that our methodology does now allow to measure the actual \vsini\ below a certain threshold limit \citep[see][]{2017A&A...597A..22S}.

We have also compared our results for the \vsini\ with those from \citet{2022A&A...665A.150H} for the O-type stars (see gray dots in Fig~\ref{fig:shrd_fw_adb}) to see the differences with the BSGs as the first are the progenitors of the second. A simple comparison with Fig.~\ref{fig:hist_vsini} indicates a similar distribution, having both a main group of stars at lower \vsini\ values followed by a tail of fast-rotators. Moreover, we can see that the stars with LC I concentrate at \vsini\,\ls\,150\,\kms, similarly to our sample of stars of the same LC. Moreover, the cumulative distributions for this LC also look very similar. Lastly, the results from \citet{2022A&A...665A.150H} for those stars located between the 20\,--\,32\,\MSol\ evolutionary tracks (i.e., those evolving into the early B-type III-IV-V stars) also show higher \vsini\ values than those directly above, which is the same situation as what shown in Fig.~\ref{fig:hist_vsini}, even though our sample is biased by the relative number of stars with LCs III-IV-V with respect to I and II.

These results clearly indicate that there is not a significant decrease in the \vsini\ values along the transition from the O-type stars towards the BSGs (and those additional B-type III-IV-V stars), thus favoring those evolutionary models where the braking effect is not so strong and/or the angular momentum transfer from the cores to the outer layers is very efficient and counteracts the effect \citep[see][]{2011A&A...530A.115B}.


\section{Summary, conclusions and future work}
\label{section:5_XXXXXX}

In this work we have built an homogeneous spectroscopic sample of 733 O9\,--\,B9 stars of which the majority (483) have luminosity classes I and II, while the rest (250) are late O- and early B-type stars with classes III-IV-V. We have benefited from over two decades of observations using the same three high-resolution spectrographs of high stability, resolution and precision.

To create the sample, we have explored the possibility to use one simple measurement on the H$\beta$ line to select stars above a certain \logLs\ in the sHRD, filtering out other unwanted stars from initial sample of stars with mixed luminosities. With the only exception of some stars strongly affected by disks or stellar winds where the H$\beta$ line could not be measured (e.g. hypergiants, classical Be stars), this new method has been proved to be of much utility using either low or high-resolution spectra. Furthermore, despite the fact that we have limited the applicability of this method to O9\,--\,B9 stars, its usage can also be extended to also include the A-type supergiants, or include all the O-type stars.

We have also performed a morphological classification of H$\beta$ and H$\alpha$ lines, identified 56 SB2+, and studied the distribution of \vsini\ among the stars in the sample, and in particular those with LC I-II, finding that the presence of fast-rotating stars extends up to B3-type stars.

Compared to other works using large samples of stars observed at high-resolution, our sample of BSGs is $\sim$3.5 times larger (see \citet{2017A&A...597A..22S}). For other studies in which spectroscopic parameters for Galactic BSGs are also derived \citep[e.g.][]{2006A&A...446..279C,2008A&A...478..823M,2022A&A...668A..92W}, our sample comprises between $\sim$8-25 times more stars. 

With the only limitation of stars too faint to be observed with the facilities used, we have evaluated the completeness of the sample (Sect.~\ref{subsubsection:423_LLLLL}) and found that, for stars below $B_{mag}\,\leq$\,9, we have $\approx$90\% completeness of those in the Northern hemisphere ($\delta$ > -20) up to 4\,000\,pc, or $\approx$80\% within the first 2\,000\,pc in both hemispheres. Regarding the missing stars, we are carrying out large observing campaigns in both hemispheres to be complete up to $B_{mag}$ = 9. 

Some main conclusions can be extracted from this work. Firstly, that the method based on the \fwhb\ will not only (but certainly) help to filter new observations, but also will be of extreme use in next generation spectroscopic surveys such WEAVE-SCIP \citep{2020SPIE11447E..14D,2023MNRAS.tmp..715J} or 4MIDABLE-LR \citep{2019Msngr.175...30C} as they will increase in some orders of magnitude the number of available spectra of observed BSGs in the Milky Way, which have been never classified before. Secondly, by using this method we have also found about the risk of using spectral classifications from the literature. In particular, we have found many stars classified as supergiants in the regions of the sHRD where sub-giants and dwarfs are located. Lastly and very important, the sample of $\approx$750 O9\,--\,B9 stars introduced in this work is large enough to set (with enough confidence) new empirical constraints to be used as anchor points in our understanding of the nature of massive stars, and more specifically in the regime of the Galactic BSGs.
In particular, spectroscopic parameters for these stars derived from the quantitative spectroscopic analysis will contribute to improve the state-of-the-art evolutionary models \citep[e.g. those from][]{2011A&A...530A.115B,2012A&A...537A.146E}, constraining some input parameters of their convective cores \citep[such the core-overshooting,][]{2019A&A...625A.132S,2021A&A...648A.126M}, and also helping to constrain the location end of the main sequence \citep{2010A&A...512L...7V,2014A&A...570L..13C}. The analysis of the multi-epoch spectroscopic data and the identification of the SB1 and SB2 systems together with additional empirical information from {\em Gaia} (such the proper motions) can also help to constraint the fraction of binaries that can interact or even become mergers at some point during their evolution \citep{2012Sci...337..444S,2014ApJ...782....7D}. Moreover, spectroscopic variability in combination with the pulsational properties derived from photometric lightcurves can also be used to continue the work by \citet{2020A&A...639A..81B} to study the stellar interiors \citep{2013MNRAS.433.1246S,2020svos.conf...53B,2021RvMP...93a5001A}. The inclusion of derived abundances can as well help to disentangle the different populations of stars heading to, or coming from the RSg phase \citep{2021A&A...650A.128G}.

In this first paper, we have given a glimpse of the potential of the sample by obtaining first results of their rotational properties, and in particular the \vsini. Interestingly, we found the presence of fast-rotating stars of LCs I and II up to B3-type. Considering results from \citet{2006A&A...446..279C,2022A&A...668A..92W}, B3-type stars can have temperatures down to $\approx$16\,000\,K for LC I. This means that our results extends the presence of fast-rotating stars some thousand kelvins below previously observed \citep[e.g.][]{2010MNRAS.404.1306F,2010A&A...512L...7V}, probably as a consequence of the importantly increased statistics. Although further analyses using derived \Teff\ and \logg\ will provide additional information, as pointed out in \citet{2010A&A...512L...7V}, the presence of fast-rotating stars could have a very important role for locating the Terminal Age Main Sequence, which is yet an open question regarding the BSGs \citep[see also][]{2014A&A...570L..13C}. Nevertheless, as also pointed out in \citet{2010A&A...512L...7V}, additional mechanisms \citep[e.g. enhanced mass losses,][]{2010A&A...512L...7V} could also slow down the rotation of these stars, not forgetting that some of them could actually be the result of binary interaction \citep{2013ApJ...764..166D}.


\begin{acknowledgements}

This research acknowledges the support by the Spanish Government Ministerio de Ciencia e Innovaci\'on through grants PGC-2018-091,3741-B-C22, PID2021-122397NB-C21 and SEV 2015-0548, and from the Canarian Agency for Research, Innovation and Information Society (ACIISI), of the Canary Islands Government, and the European Regional Development Fund (ERDF), under grant with reference ProID2017010115.

We give special thanks to all the observers who contributed to the acquisition of the spectra used here in this work. Among them, especially to C.~Konstantopoulou and G.~Holgado.  

Regarding the observing facilities, this research is based on observations made with the Mercator Telescope, operated by the Flemish Community at the Observatorio del Roque de los Muchachos (La Palma, Spain), of the Instituto de Astrof\'isica de Canarias. In particular, obtained with the HERMES spectrograph, which is supported by the Research Foundation - Flanders (FWO), Belgium, the Research Council of KU Leuven, Belgium, the Fonds National de la Recherche Scientifique (F.R.S.-FNRS), Belgium, the Royal Observatory of Belgium, the Observatoire de Genève, Switzerland and the Thüringer Landessternwarte Tautenburg, Germany.

This research also based on observations with the Nordic Optical Telescope, owned in collaboration by the University of Turku and Aarhus University, and operated jointly by Aarhus University, the University of Turku and the University of Oslo, representing Denmark, Finland and Norway, the University of Iceland and Stockholm University, at the Observatorio del Roque de los Muchachos, of the Instituto de Astrof\'isica de Canarias.

Regarding the data retrieved, this work has made use of data from the European Space Agency (ESA) mission {\it Gaia}\footnote{\url{https://www.cosmos.esa.int/gaia/}}, processed by the {\it Gaia} Data Processing and Analysis Consortium (DPAC\footnote{\url{https://www.cosmos.esa.int/web/gaia/dpac/consortium/}}). Funding for the DPAC has been provided by national institutions, in particular the institutions participating in the {\it Gaia} Multilateral Agreement.

Additionally, this work has made use of observations collected from the ESO Science Archive Facility under ESO programs: 60.A-9700(A), 72.D-0235(B), 73.C-0337(A), 73.D-0234(A), 73.D-0609(A), 74.D-0008(B), 74.D-0300(A), 75.D-0103(A), 75.D-0369(A), 76.C-0431(A), 77.D-0025(A), 77.D-0635(A), 79.A-9008(A), 79.B-0856(A), 81.A-9005(A), 81.A-9006(A), 81.C-2003(A), 81.D-2008(A), 81.D-2008(B), 82.D-0933(A), 83.D-0589(A), 83.D-0589(B), 85.D-0262(A), 86.D-0997(B), 87.D-0946(A), 88.A-9003(A), 89.D-0975(A), 90.D-0358(A), 91.C-0713(A), 91.D-0061(A), 91.D-0221(A), 92.A-9020(A), 95.A-9029(D), 97.A-9039(C), 102.A-9010(A) and 179.C-0197(C).

\end{acknowledgements}


\typeout{}
\bibliographystyle{aa} 
\bibliography{biblio} 

\begin{thebibliography}{132}
\expandafter\ifx\csname natexlab\endcsname\relax\def\natexlab#1{#1}\fi

\bibitem[{{Abbott} \& {Conti}(1987)}]{1987ARA&A..25..113A}
{Abbott}, D.~C. \& {Conti}, P.~S. 1987, \araa, 25, 113

\bibitem[{{Adams} {et~al.}(2017){Adams}, {Kochanek}, {Gerke}, {Stanek}, \&
  {Dai}}]{2017MNRAS.468.4968A}
{Adams}, S.~M., {Kochanek}, C.~S., {Gerke}, J.~R., {Stanek}, K.~Z., \& {Dai},
  X. 2017, \mnras, 468, 4968

\bibitem[{{Aerts}(2021)}]{2021RvMP...93a5001A}
{Aerts}, C. 2021, Reviews of Modern Physics, 93, 015001

\bibitem[{{Aerts} {et~al.}(2009){Aerts}, {Puls}, {Godart}, \&
  {Dupret}}]{2009A&A...508..409A}
{Aerts}, C., {Puls}, J., {Godart}, M., \& {Dupret}, M.~A. 2009, \aap, 508, 409

\bibitem[{{Babusiaux} {et~al.}(2022){Babusiaux}, {Fabricius}, {Khanna},
  {Muraveva}, {Reyl{\'e}}, {Spoto}, {Vallenari}, {Luri}, {Arenou}, {Alvarez},
  {Anders}, {Antoja}, {Balbinot}, {Barache}, {Bauchet}, {Bossini}, {Busonero},
  {Cantat-Gaudin}, {Carrasco}, {Dafonte}, {Diakite}, {Figueras},
  {Garcia-Gutierrez}, {Garofalo}, {Helmi}, {Jimenez-Arranz}, {Jordi},
  {Kervella}, {Kostrzewa-Rutkowska}, {Leclerc}, {Licata}, {Manteiga}, {Masip},
  {Monguio}, {Ramos}, {Robichon}, {Robin}, {Romero-Gomez}, {Saez}, {Santovena},
  {Spina}, {Torralba Elipe}, \& {Weiler}}]{2022arXiv220605989B}
{Babusiaux}, C., {Fabricius}, C., {Khanna}, S., {et~al.} 2022, arXiv e-prints,
  arXiv:2206.05989

\bibitem[{{Bailer-Jones} {et~al.}(2021){Bailer-Jones}, {Rybizki}, {Fouesneau},
  {Demleitner}, \& {Andrae}}]{2021AJ....161..147B}
{Bailer-Jones}, C.~A.~L., {Rybizki}, J., {Fouesneau}, M., {Demleitner}, M., \&
  {Andrae}, R. 2021, \aj, 161, 147

\bibitem[{{Banyard} {et~al.}(2022){Banyard}, {Sana}, {Mahy}, {Bodensteiner},
  {Villase{\~n}or}, \& {Evans}}]{2022A&A...658A..69B}
{Banyard}, G., {Sana}, H., {Mahy}, L., {et~al.} 2022, \aap, 658, A69

\bibitem[{{Barb{\'a}} {et~al.}(2010){Barb{\'a}}, {Gamen}, {Arias}, {Morrell},
  {Ma{\'\i}z Apell{\'a}niz}, {Alfaro}, {Walborn}, \&
  {Sota}}]{2010RMxAC..38...30B}
{Barb{\'a}}, R.~H., {Gamen}, R., {Arias}, J.~I., {et~al.} 2010, in Revista
  Mexicana de Astronomia y Astrofisica Conference Series, Vol.~38, Revista
  Mexicana de Astronomia y Astrofisica Conference Series, 30--32

\bibitem[{{Belczynski} {et~al.}(2016){Belczynski}, {Holz}, {Bulik}, \&
  {O'Shaughnessy}}]{2016Natur.534..512B}
{Belczynski}, K., {Holz}, D.~E., {Bulik}, T., \& {O'Shaughnessy}, R. 2016,
  \nat, 534, 512

\bibitem[{{Berlanas} {et~al.}(2019){Berlanas}, {Wright}, {Herrero}, {Drew}, \&
  {Lennon}}]{2019MNRAS.484.1838B}
{Berlanas}, S.~R., {Wright}, N.~J., {Herrero}, A., {Drew}, J.~E., \& {Lennon},
  D.~J. 2019, \mnras, 484, 1838

\bibitem[{{Blomme} {et~al.}(2022){Blomme}, {Fremat}, {Sartoretti}, {Guerrier},
  {Panuzzo}, {Katz}, {Seabroke}, {Thevenin}, {Cropper}, {Benson}, {Damerdji},
  {Haigron}, {Marchal}, {Smith}, {Baker}, {Chemin}, {David}, {Dolding},
  {Gosset}, {Janssen}, {Jasniewicz}, {Lobel}, {Plum}, {Samaras}, {Snaith},
  {Soubiran}, {Vanel}, {Zwitter}, {Brouillet}, {Caffau}, {Crifo}, {Fabre},
  {Frakgoudi}, {Huckle}, {Jean-Antoine Piccolo}, {Lasne}, {Leclerc},
  {Mastrobuono-Battisti}, {Royer}, {Viala}, \& {Zorec}}]{2022arXiv220605486B}
{Blomme}, R., {Fremat}, Y., {Sartoretti}, P., {et~al.} 2022, arXiv e-prints,
  arXiv:2206.05486

\bibitem[{{Bodensteiner} {et~al.}(2021){Bodensteiner}, {Sana}, {Wang},
  {Langer}, {Mahy}, {Banyard}, {de Koter}, {de Mink}, {Evans}, {G{\"o}tberg},
  {Patrick}, {Schneider}, \& {Tramper}}]{2021A&A...652A..70B}
{Bodensteiner}, J., {Sana}, H., {Wang}, C., {et~al.} 2021, \aap, 652, A70

\bibitem[{{Bouy} \& {Alves}(2015)}]{2015A&A...584A..26B}
{Bouy}, H. \& {Alves}, J. 2015, \aap, 584, A26

\bibitem[{{Bowman}(2020)}]{2020svos.conf...53B}
{Bowman}, D.~M. 2020, in Stars and their Variability Observed from Space, ed.
  C.~{Neiner}, W.~W. {Weiss}, D.~{Baade}, R.~E. {Griffin}, C.~C. {Lovekin}, \&
  A.~F.~J. {Moffat}, 53--59

\bibitem[{{Brott} {et~al.}(2011){Brott}, {de Mink}, {Cantiello}, {Langer}, {de
  Koter}, {Evans}, {Hunter}, {Trundle}, \& {Vink}}]{2011A&A...530A.115B}
{Brott}, I., {de Mink}, S.~E., {Cantiello}, M., {et~al.} 2011, \aap, 530, A115

\bibitem[{{Burssens} {et~al.}(2020){Burssens}, {Sim{\'o}n-D{\'\i}az}, {Bowman},
  {Holgado}, {Michielsen}, {de Burgos}, {Castro}, {Barb{\'a}}, \&
  {Aerts}}]{2020A&A...639A..81B}
{Burssens}, S., {Sim{\'o}n-D{\'\i}az}, S., {Bowman}, D.~M., {et~al.} 2020,
  \aap, 639, A81

\bibitem[{{Callingham} {et~al.}(2020){Callingham}, {Crowther}, {Williams},
  {Tuthill}, {Han}, {Pope}, \& {Marcote}}]{2020MNRAS.495.3323C}
{Callingham}, J.~R., {Crowther}, P.~A., {Williams}, P.~M., {et~al.} 2020,
  \mnras, 495, 3323

\bibitem[{{Cantiello} {et~al.}(2009){Cantiello}, {Langer}, {Brott}, {de Koter},
  {Shore}, {Vink}, {Voegler}, {Lennon}, \& {Yoon}}]{2009A&A...499..279C}
{Cantiello}, M., {Langer}, N., {Brott}, I., {et~al.} 2009, \aap, 499, 279

\bibitem[{{Castor} {et~al.}(1975){Castor}, {Abbott}, \&
  {Klein}}]{1975ApJ...195..157C}
{Castor}, J.~I., {Abbott}, D.~C., \& {Klein}, R.~I. 1975, \apj, 195, 157

\bibitem[{{Castor} \& {Lamers}(1979)}]{1979ApJS...39..481C}
{Castor}, J.~I. \& {Lamers}, H.~J.~G.~L.~M. 1979, \apjs, 39, 481

\bibitem[{{Castro} {et~al.}(2014){Castro}, {Fossati}, {Langer},
  {Sim{\'o}n-D{\'\i}az}, {Schneider}, \& {Izzard}}]{2014A&A...570L..13C}
{Castro}, N., {Fossati}, L., {Langer}, N., {et~al.} 2014, \aap, 570, L13

\bibitem[{{Chiappini} {et~al.}(2019){Chiappini}, {Minchev}, {Starkenburg},
  {Anders}, {Gentile Fusillo}, {Gerhard}, {Guiglion}, {Khalatyan},
  {Kordopatis}, {Lemasle}, {Matijevic}, {Queiroz}, {Schwope}, {Steinmetz},
  {Storm}, {Traven}, {Tremblay}, {Valentini}, {Andrae}, {Arentsen}, {Asplund},
  {Bensby}, {Bergemann}, {Casagrande}, {Church}, {Cescutti}, {Feltzing},
  {Fouesneau}, {Grebel}, {Kovalev}, {McMillan}, {Monari}, {Rybizki}, {Ryde},
  {Rix}, {Walton}, {Xiang}, {Zucker}, \& {4MIDABLE-Lr
  Team}}]{2019Msngr.175...30C}
{Chiappini}, C., {Minchev}, I., {Starkenburg}, E., {et~al.} 2019, The
  Messenger, 175, 30

\bibitem[{{Choi} {et~al.}(2016){Choi}, {Dotter}, {Conroy}, {Cantiello},
  {Paxton}, \& {Johnson}}]{2016ApJ...823..102C}
{Choi}, J., {Dotter}, A., {Conroy}, C., {et~al.} 2016, \apj, 823, 102

\bibitem[{{Clark} {et~al.}(2012){Clark}, {Najarro}, {Negueruela}, {Ritchie},
  {Urbaneja}, \& {Howarth}}]{2012A&A...541A.145C}
{Clark}, J.~S., {Najarro}, F., {Negueruela}, I., {et~al.} 2012, \aap, 541, A145

\bibitem[{{Contreras} {et~al.}(2002){Contreras}, {Sicilia-Aguilar},
  {Muzerolle}, {Calvet}, {Berlind}, \& {Hartmann}}]{2002AJ....124.1585C}
{Contreras}, M.~E., {Sicilia-Aguilar}, A., {Muzerolle}, J., {et~al.} 2002, \aj,
  124, 1585

\bibitem[{{Crowther} {et~al.}(2006){Crowther}, {Lennon}, \&
  {Walborn}}]{2006A&A...446..279C}
{Crowther}, P.~A., {Lennon}, D.~J., \& {Walborn}, N.~R. 2006, \aap, 446, 279

\bibitem[{{Dalton} {et~al.}(2020){Dalton}, {Trager}, {Abrams}, {Bonifacio},
  {Aguerri}, {Vallenari}, {Bishop}, {Middleton}, {Benn}, {Dee}, {Mignot},
  {Lewis}, {Pragt}, {Pico}, {Walton}, {Rey}, {Allende Prieto}, {Lhom{\'e}},
  {Balcells}, {Terrett}, {Brock}, {Ridings}, {Skvar{\v{c}}}, {Verheijen},
  {Steele}, {Stuik}, {Kroes}, {Tromp}, {Kragt}, {Lesman}, {Mottram}, {Bates},
  {Gribbin}, {Burgal}, {Herreros}, {Delgado}, {Martin}, {Cano}, {Navarro},
  {Irwin}, {Peralta de Arriba}, {O'Mahoney}, {Bianco}, {Moleinezhad}, {ter
  Horst}, {Molinari}, {Lodi}, {Guerra}, {Baruffalo}, {Carrasco}, {Farcas},
  {Schallig}, {Hughes}, {Hill}, {Smith}, {Drew}, {Poggianti}, {Iovino},
  {Pieri}, {Jin}, {Dominguez Palmero}, {Fari{\~n}a}, {Mart{\'\i}n}, {Worley},
  {Murphy}, {Guest}, {Morris}, {Elswijk}, {de Haan}, {Hanenburg}, {Salasnich},
  {Mayya}, {Izazaga-P{\'e}rez}, {Gafton}, {Caffau}, {Horville}, {Paz
  Chinch{\'o}n}, {Falcon-Barosso}, {G{\"a}nsicke}, {San Juan}, \&
  {Hernandez}}]{2020SPIE11447E..14D}
{Dalton}, G., {Trager}, S., {Abrams}, D.~C., {et~al.} 2020, in Society of
  Photo-Optical Instrumentation Engineers (SPIE) Conference Series, Vol. 11447,
  Society of Photo-Optical Instrumentation Engineers (SPIE) Conference Series,
  1144714

\bibitem[{{Damiani} {et~al.}(2016){Damiani}, {Micela}, \&
  {Sciortino}}]{2016A&A...596A..82D}
{Damiani}, F., {Micela}, G., \& {Sciortino}, S. 2016, \aap, 596, A82

\bibitem[{{de Burgos} {et~al.}(2020){de Burgos}, {Simon-D{\'\i}az}, {Lennon},
  {Dorda}, {Negueruela}, {Urbaneja}, {Patrick}, \&
  {Herrero}}]{2020A&A...643A.116D}
{de Burgos}, A., {Simon-D{\'\i}az}, S., {Lennon}, D.~J., {et~al.} 2020, \aap,
  643, A116

\bibitem[{{de Mink} {et~al.}(2011){de Mink}, {Langer}, \&
  {Izzard}}]{2011BSRSL..80..543D}
{de Mink}, S.~E., {Langer}, N., \& {Izzard}, R.~G. 2011, Bulletin de la Societe
  Royale des Sciences de Liege, 80, 543

\bibitem[{{de Mink} {et~al.}(2013){de Mink}, {Langer}, {Izzard}, {Sana}, \& {de
  Koter}}]{2013ApJ...764..166D}
{de Mink}, S.~E., {Langer}, N., {Izzard}, R.~G., {Sana}, H., \& {de Koter}, A.
  2013, \apj, 764, 166

\bibitem[{{de Mink} {et~al.}(2014){de Mink}, {Sana}, {Langer}, {Izzard}, \&
  {Schneider}}]{2014ApJ...782....7D}
{de Mink}, S.~E., {Sana}, H., {Langer}, N., {Izzard}, R.~G., \& {Schneider},
  F.~R.~N. 2014, \apj, 782, 7

\bibitem[{{Dotter}(2016)}]{2016ApJS..222....8D}
{Dotter}, A. 2016, \apjs, 222, 8

\bibitem[{{Dunstall} {et~al.}(2015){Dunstall}, {Dufton}, {Sana}, {Evans},
  {Howarth}, {Sim{\'o}n-D{\'\i}az}, {de Mink}, {Langer}, {Ma{\'\i}z
  Apell{\'a}niz}, \& {Taylor}}]{2015A&A...580A..93D}
{Dunstall}, P.~R., {Dufton}, P.~L., {Sana}, H., {et~al.} 2015, \aap, 580, A93

\bibitem[{{Ekstr{\"o}m} {et~al.}(2012){Ekstr{\"o}m}, {Georgy}, {Eggenberger},
  {Meynet}, {Mowlavi}, {Wyttenbach}, {Granada}, {Decressin}, {Hirschi},
  {Frischknecht}, {Charbonnel}, \& {Maeder}}]{2012A&A...537A.146E}
{Ekstr{\"o}m}, S., {Georgy}, C., {Eggenberger}, P., {et~al.} 2012, \aap, 537,
  A146

\bibitem[{{Evans} {et~al.}(2006){Evans}, {Lennon}, {Smartt}, \&
  {Trundle}}]{2006A&A...456..623E}
{Evans}, C.~J., {Lennon}, D.~J., {Smartt}, S.~J., \& {Trundle}, C. 2006, \aap,
  456, 623

\bibitem[{{Fitzpatrick} \& {Garmany}(1990)}]{1990ApJ...363..119F}
{Fitzpatrick}, E.~L. \& {Garmany}, C.~D. 1990, \apj, 363, 119

\bibitem[{{Fraser} {et~al.}(2010){Fraser}, {Dufton}, {Hunter}, \&
  {Ryans}}]{2010MNRAS.404.1306F}
{Fraser}, M., {Dufton}, P.~L., {Hunter}, I., \& {Ryans}, R.~S.~I. 2010, \mnras,
  404, 1306

\bibitem[{{Fr{\'e}mat} {et~al.}(2022){Fr{\'e}mat}, {Royer}, {Marchal},
  {Blomme}, {Sartoretti}, {Guerrier}, {Panuzzo}, {Katz}, {Seabroke},
  {Th{\'e}venin}, {Cropper}, {Benson}, {Damerdji}, {Haigron}, {Lobel}, {Smith},
  {Baker}, {Chemin}, {David}, {Dolding}, {Gosset}, {Jan{\ss}en}, {Jasniewicz},
  {Plum}, {Samaras}, {Snaith}, {Soubiran}, {Vanel}, {Zorec}, {Zwitter},
  {Brouillet}, {Caffau}, {Crifo}, {Fabre}, {Fragkoudi}, {Huckle}, {Lasne},
  {Leclerc}, {Mastrobuono-Battisti}, {Jean-Antoine Piccolo}, \&
  {Viala}}]{2022arXiv220610986F}
{Fr{\'e}mat}, Y., {Royer}, F., {Marchal}, O., {et~al.} 2022, arXiv e-prints,
  arXiv:2206.10986

\bibitem[{{Gaia Collaboration} {et~al.}(2021){Gaia Collaboration}, {Brown},
  {Vallenari}, {Prusti}, {de Bruijne}, {Babusiaux}, {Biermann}, {Creevey},
  {Evans}, {Eyer}, {Hutton}, {Jansen}, {Jordi}, {Klioner}, {Lammers},
  {Lindegren}, {Luri}, {Mignard}, {Panem}, {Pourbaix}, {Randich}, {Sartoretti},
  {Soubiran}, {Walton}, {Arenou}, {Bailer-Jones}, {Bastian}, {Cropper},
  {Drimmel}, {Katz}, {Lattanzi}, {van Leeuwen}, {Bakker}, {Cacciari},
  {Casta{\~n}eda}, {De Angeli}, {Ducourant}, {Fabricius}, {Fouesneau},
  {Fr{\'e}mat}, {Guerra}, {Guerrier}, {Guiraud}, {Jean-Antoine Piccolo},
  {Masana}, {Messineo}, {Mowlavi}, {Nicolas}, {Nienartowicz}, {Pailler},
  {Panuzzo}, {Riclet}, {Roux}, {Seabroke}, {Sordo}, {Tanga}, {Th{\'e}venin},
  {Gracia-Abril}, {Portell}, {Teyssier}, {Altmann}, {Andrae}, {Bellas-Velidis},
  {Benson}, {Berthier}, {Blomme}, {Brugaletta}, {Burgess}, {Busso}, {Carry},
  {Cellino}, {Cheek}, {Clementini}, {Damerdji}, {Davidson}, {Delchambre},
  {Dell'Oro}, {Fern{\'a}ndez-Hern{\'a}ndez}, {Galluccio}, {Garc{\'\i}a-Lario},
  {Garcia-Reinaldos}, {Gonz{\'a}lez-N{\'u}{\~n}ez}, {Gosset}, {Haigron},
  {Halbwachs}, {Hambly}, {Harrison}, {Hatzidimitriou}, {Heiter},
  {Hern{\'a}ndez}, {Hestroffer}, {Hodgkin}, {Holl}, {Jan{\ss}en}, {Jevardat de
  Fombelle}, {Jordan}, {Krone-Martins}, {Lanzafame}, {L{\"o}ffler}, {Lorca},
  {Manteiga}, {Marchal}, {Marrese}, {Moitinho}, {Mora}, {Muinonen}, {Osborne},
  {Pancino}, {Pauwels}, {Petit}, {Recio-Blanco}, {Richards}, {Riello},
  {Rimoldini}, {Robin}, {Roegiers}, {Rybizki}, {Sarro}, {Siopis}, {Smith},
  {Sozzetti}, {Ulla}, {Utrilla}, {van Leeuwen}, {van Reeven}, {Abbas}, {Abreu
  Aramburu}, {Accart}, {Aerts}, {Aguado}, {Ajaj}, {Altavilla}, {{\'A}lvarez},
  {{\'A}lvarez Cid-Fuentes}, {Alves}, {Anderson}, {Anglada Varela}, {Antoja},
  {Audard}, {Baines}, {Baker}, {Balaguer-N{\'u}{\~n}ez}, {Balbinot}, {Balog},
  {Barache}, {Barbato}, {Barros}, {Barstow}, {Bartolom{\'e}}, {Bassilana},
  {Bauchet}, {Baudesson-Stella}, {Becciani}, {Bellazzini}, {Bernet}, {Bertone},
  {Bianchi}, {Blanco-Cuaresma}, {Boch}, {Bombrun}, {Bossini}, {Bouquillon},
  {Bragaglia}, {Bramante}, {Breedt}, {Bressan}, {Brouillet}, {Bucciarelli},
  {Burlacu}, {Busonero}, {Butkevich}, {Buzzi}, {Caffau}, {Cancelliere},
  {C{\'a}novas}, {Cantat-Gaudin}, {Carballo}, {Carlucci}, {Carnerero},
  {Carrasco}, {Casamiquela}, {Castellani}, {Castro-Ginard}, {Castro Sampol},
  {Chaoul}, {Charlot}, {Chemin}, {Chiavassa}, {Cioni}, {Comoretto}, {Cooper},
  {Cornez}, {Cowell}, {Crifo}, {Crosta}, {Crowley}, {Dafonte}, {Dapergolas},
  {David}, {David}, {de Laverny}, {De Luise}, {De March}, {De Ridder}, {de
  Souza}, {de Teodoro}, {de Torres}, {del Peloso}, {del Pozo}, {Delbo},
  {Delgado}, {Delgado}, {Delisle}, {Di Matteo}, {Diakite}, {Diener},
  {Distefano}, {Dolding}, {Eappachen}, {Edvardsson}, {Enke}, {Esquej}, {Fabre},
  {Fabrizio}, {Faigler}, {Fedorets}, {Fernique}, {Fienga}, {Figueras},
  {Fouron}, {Fragkoudi}, {Fraile}, {Franke}, {Gai}, {Garabato},
  {Garcia-Gutierrez}, {Garc{\'\i}a-Torres}, {Garofalo}, {Gavras}, {Gerlach},
  {Geyer}, {Giacobbe}, {Gilmore}, {Girona}, {Giuffrida}, {Gomel}, {Gomez},
  {Gonzalez-Santamaria}, {Gonz{\'a}lez-Vidal}, {Granvik},
  {Guti{\'e}rrez-S{\'a}nchez}, {Guy}, {Hauser}, {Haywood}, {Helmi}, {Hidalgo},
  {Hilger}, {H{\l}adczuk}, {Hobbs}, {Holland}, {Huckle}, {Jasniewicz},
  {Jonker}, {Juaristi Campillo}, {Julbe}, {Karbevska}, {Kervella}, {Khanna},
  {Kochoska}, {Kontizas}, {Kordopatis}, {Korn}, {Kostrzewa-Rutkowska},
  {Kruszy{\'n}ska}, {Lambert}, {Lanza}, {Lasne}, {Le Campion}, {Le Fustec},
  {Lebreton}, {Lebzelter}, {Leccia}, {Leclerc}, {Lecoeur-Taibi}, {Liao},
  {Licata}, {Lindstr{\o}m}, {Lister}, {Livanou}, {Lobel}, {Madrero Pardo},
  {Managau}, {Mann}, {Marchant}, {Marconi}, {Marcos Santos}, {Marinoni},
  {Marocco}, {Marshall}, {Martin Polo}, {Mart{\'\i}n-Fleitas}, {Masip},
  {Massari}, {Mastrobuono-Battisti}, {Mazeh}, {McMillan}, {Messina},
  {Michalik}, {Millar}, {Mints}, {Molina}, {Molinaro}, {Moln{\'a}r},
  {Montegriffo}, {Mor}, {Morbidelli}, {Morel}, {Morris}, {Mulone}, {Munoz},
  {Muraveva}, {Murphy}, {Musella}, {Noval}, {Ord{\'e}novic}, {Orr{\`u}},
  {Osinde}, {Pagani}, {Pagano}, {Palaversa}, {Palicio}, {Panahi}, {Pawlak},
  {Pe{\~n}alosa Esteller}, {Penttil{\"a}}, {Piersimoni}, {Pineau}, {Plachy},
  {Plum}, {Poggio}, {Poretti}, {Poujoulet}, {Pr{\v{s}}a}, {Pulone}, {Racero},
  {Ragaini}, {Rainer}, {Raiteri}, {Rambaux}, {Ramos}, {Ramos-Lerate}, {Re
  Fiorentin}, {Regibo}, {Reyl{\'e}}, {Ripepi}, {Riva}, {Rixon}, {Robichon},
  {Robin}, {Roelens}, {Rohrbasser}, {Romero-G{\'o}mez}, {Rowell}, {Royer},
  {Rybicki}, {Sadowski}, {Sagrist{\`a} Sell{\'e}s}, {Sahlmann}, {Salgado},
  {Salguero}, {Samaras}, {Sanchez Gimenez}, {Sanna}, {Santove{\~n}a},
  {Sarasso}, {Schultheis}, {Sciacca}, {Segol}, {Segovia}, {S{\'e}gransan},
  {Semeux}, {Shahaf}, {Siddiqui}, {Siebert}, {Siltala}, {Slezak}, {Smart},
  {Solano}, {Solitro}, {Souami}, {Souchay}, {Spagna}, {Spoto}, {Steele},
  {Steidelm{\"u}ller}, {Stephenson}, {S{\"u}veges}, {Szabados}, {Szegedi-Elek},
  {Taris}, {Tauran}, {Taylor}, {Teixeira}, {Thuillot}, {Tonello}, {Torra},
  {Torra}, {Turon}, {Unger}, {Vaillant}, {van Dillen}, {Vanel}, {Vecchiato},
  {Viala}, {Vicente}, {Voutsinas}, {Weiler}, {Wevers}, {Wyrzykowski}, {Yoldas},
  {Yvard}, {Zhao}, {Zorec}, {Zucker}, {Zurbach}, \&
  {Zwitter}}]{2021A&A...649A...1G}
{Gaia Collaboration}, {Brown}, A.~G.~A., {Vallenari}, A., {et~al.} 2021, \aap,
  649, A1

\bibitem[{{Gaia Collaboration} {et~al.}(2016){Gaia Collaboration}, {Prusti},
  {de Bruijne}, {Brown}, {Vallenari}, {Babusiaux}, {Bailer-Jones}, {Bastian},
  {Biermann}, {Evans}, {Eyer}, {Jansen}, {Jordi}, {Klioner}, {Lammers},
  {Lindegren}, {Luri}, {Mignard}, {Milligan}, {Panem}, {Poinsignon},
  {Pourbaix}, {Randich}, {Sarri}, {Sartoretti}, {Siddiqui}, {Soubiran},
  {Valette}, {van Leeuwen}, {Walton}, {Aerts}, {Arenou}, {Cropper}, {Drimmel},
  {H{\o}g}, {Katz}, {Lattanzi}, {O'Mullane}, {Grebel}, {Holland}, {Huc},
  {Passot}, {Bramante}, {Cacciari}, {Casta{\~n}eda}, {Chaoul}, {Cheek}, {De
  Angeli}, {Fabricius}, {Guerra}, {Hern{\'a}ndez}, {Jean-Antoine-Piccolo},
  {Masana}, {Messineo}, {Mowlavi}, {Nienartowicz}, {Ord{\'o}{\~n}ez-Blanco},
  {Panuzzo}, {Portell}, {Richards}, {Riello}, {Seabroke}, {Tanga},
  {Th{\'e}venin}, {Torra}, {Els}, {Gracia-Abril}, {Comoretto},
  {Garcia-Reinaldos}, {Lock}, {Mercier}, {Altmann}, {Andrae}, {Astraatmadja},
  {Bellas-Velidis}, {Benson}, {Berthier}, {Blomme}, {Busso}, {Carry},
  {Cellino}, {Clementini}, {Cowell}, {Creevey}, {Cuypers}, {Davidson}, {De
  Ridder}, {de Torres}, {Delchambre}, {Dell'Oro}, {Ducourant}, {Fr{\'e}mat},
  {Garc{\'\i}a-Torres}, {Gosset}, {Halbwachs}, {Hambly}, {Harrison}, {Hauser},
  {Hestroffer}, {Hodgkin}, {Huckle}, {Hutton}, {Jasniewicz}, {Jordan},
  {Kontizas}, {Korn}, {Lanzafame}, {Manteiga}, {Moitinho}, {Muinonen},
  {Osinde}, {Pancino}, {Pauwels}, {Petit}, {Recio-Blanco}, {Robin}, {Sarro},
  {Siopis}, {Smith}, {Smith}, {Sozzetti}, {Thuillot}, {van Reeven}, {Viala},
  {Abbas}, {Abreu Aramburu}, {Accart}, {Aguado}, {Allan}, {Allasia},
  {Altavilla}, {{\'A}lvarez}, {Alves}, {Anderson}, {Andrei}, {Anglada Varela},
  {Antiche}, {Antoja}, {Ant{\'o}n}, {Arcay}, {Atzei}, {Ayache}, {Bach},
  {Baker}, {Balaguer-N{\'u}{\~n}ez}, {Barache}, {Barata}, {Barbier}, {Barblan},
  {Baroni}, {Barrado y Navascu{\'e}s}, {Barros}, {Barstow}, {Becciani},
  {Bellazzini}, {Bellei}, {Bello Garc{\'\i}a}, {Belokurov}, {Bendjoya},
  {Berihuete}, {Bianchi}, {Bienaym{\'e}}, {Billebaud}, {Blagorodnova},
  {Blanco-Cuaresma}, {Boch}, {Bombrun}, {Borrachero}, {Bouquillon}, {Bourda},
  {Bouy}, {Bragaglia}, {Breddels}, {Brouillet}, {Br{\"u}semeister},
  {Bucciarelli}, {Budnik}, {Burgess}, {Burgon}, {Burlacu}, {Busonero}, {Buzzi},
  {Caffau}, {Cambras}, {Campbell}, {Cancelliere}, {Cantat-Gaudin}, {Carlucci},
  {Carrasco}, {Castellani}, {Charlot}, {Charnas}, {Charvet}, {Chassat},
  {Chiavassa}, {Clotet}, {Cocozza}, {Collins}, {Collins}, {Costigan}, {Crifo},
  {Cross}, {Crosta}, {Crowley}, {Dafonte}, {Damerdji}, {Dapergolas}, {David},
  {David}, {De Cat}, {de Felice}, {de Laverny}, {De Luise}, {De March}, {de
  Martino}, {de Souza}, {Debosscher}, {del Pozo}, {Delbo}, {Delgado},
  {Delgado}, {di Marco}, {Di Matteo}, {Diakite}, {Distefano}, {Dolding}, {Dos
  Anjos}, {Drazinos}, {Dur{\'a}n}, {Dzigan}, {Ecale}, {Edvardsson}, {Enke},
  {Erdmann}, {Escolar}, {Espina}, {Evans}, {Eynard Bontemps}, {Fabre},
  {Fabrizio}, {Faigler}, {Falc{\~a}o}, {Farr{\`a}s Casas}, {Faye}, {Federici},
  {Fedorets}, {Fern{\'a}ndez-Hern{\'a}ndez}, {Fernique}, {Fienga}, {Figueras},
  {Filippi}, {Findeisen}, {Fonti}, {Fouesneau}, {Fraile}, {Fraser}, {Fuchs},
  {Furnell}, {Gai}, {Galleti}, {Galluccio}, {Garabato}, {Garc{\'\i}a-Sedano},
  {Gar{\'e}}, {Garofalo}, {Garralda}, {Gavras}, {Gerssen}, {Geyer}, {Gilmore},
  {Girona}, {Giuffrida}, {Gomes}, {Gonz{\'a}lez-Marcos},
  {Gonz{\'a}lez-N{\'u}{\~n}ez}, {Gonz{\'a}lez-Vidal}, {Granvik}, {Guerrier},
  {Guillout}, {Guiraud}, {G{\'u}rpide}, {Guti{\'e}rrez-S{\'a}nchez}, {Guy},
  {Haigron}, {Hatzidimitriou}, {Haywood}, {Heiter}, {Helmi}, {Hobbs},
  {Hofmann}, {Holl}, {Holland}, {Hunt}, {Hypki}, {Icardi}, {Irwin}, {Jevardat
  de Fombelle}, {Jofr{\'e}}, {Jonker}, {Jorissen}, {Julbe}, {Karampelas},
  {Kochoska}, {Kohley}, {Kolenberg}, {Kontizas}, {Koposov}, {Kordopatis},
  {Koubsky}, {Kowalczyk}, {Krone-Martins}, {Kudryashova}, {Kull}, {Bachchan},
  {Lacoste-Seris}, {Lanza}, {Lavigne}, {Le Poncin-Lafitte}, {Lebreton},
  {Lebzelter}, {Leccia}, {Leclerc}, {Lecoeur-Taibi}, {Lemaitre}, {Lenhardt},
  {Leroux}, {Liao}, {Licata}, {Lindstr{\o}m}, {Lister}, {Livanou}, {Lobel},
  {L{\"o}ffler}, {L{\'o}pez}, {Lopez-Lozano}, {Lorenz}, {Loureiro},
  {MacDonald}, {Magalh{\~a}es Fernandes}, {Managau}, {Mann}, {Mantelet},
  {Marchal}, {Marchant}, {Marconi}, {Marie}, {Marinoni}, {Marrese},
  {Marschalk{\'o}}, {Marshall}, {Mart{\'\i}n-Fleitas}, {Martino}, {Mary},
  {Matijevi{\v{c}}}, {Mazeh}, {McMillan}, {Messina}, {Mestre}, {Michalik},
  {Millar}, {Miranda}, {Molina}, {Molinaro}, {Molinaro}, {Moln{\'a}r},
  {Moniez}, {Montegriffo}, {Monteiro}, {Mor}, {Mora}, {Morbidelli}, {Morel},
  {Morgenthaler}, {Morley}, {Morris}, {Mulone}, {Muraveva}, {Musella},
  {Narbonne}, {Nelemans}, {Nicastro}, {Noval}, {Ord{\'e}novic},
  {Ordieres-Mer{\'e}}, {Osborne}, {Pagani}, {Pagano}, {Pailler}, {Palacin},
  {Palaversa}, {Parsons}, {Paulsen}, {Pecoraro}, {Pedrosa}, {Pentik{\"a}inen},
  {Pereira}, {Pichon}, {Piersimoni}, {Pineau}, {Plachy}, {Plum}, {Poujoulet},
  {Pr{\v{s}}a}, {Pulone}, {Ragaini}, {Rago}, {Rambaux}, {Ramos-Lerate},
  {Ranalli}, {Rauw}, {Read}, {Regibo}, {Renk}, {Reyl{\'e}}, {Ribeiro},
  {Rimoldini}, {Ripepi}, {Riva}, {Rixon}, {Roelens}, {Romero-G{\'o}mez},
  {Rowell}, {Royer}, {Rudolph}, {Ruiz-Dern}, {Sadowski}, {Sagrist{\`a}
  Sell{\'e}s}, {Sahlmann}, {Salgado}, {Salguero}, {Sarasso}, {Savietto},
  {Schnorhk}, {Schultheis}, {Sciacca}, {Segol}, {Segovia}, {Segransan},
  {Serpell}, {Shih}, {Smareglia}, {Smart}, {Smith}, {Solano}, {Solitro},
  {Sordo}, {Soria Nieto}, {Souchay}, {Spagna}, {Spoto}, {Stampa}, {Steele},
  {Steidelm{\"u}ller}, {Stephenson}, {Stoev}, {Suess}, {S{\"u}veges}, {Surdej},
  {Szabados}, {Szegedi-Elek}, {Tapiador}, {Taris}, {Tauran}, {Taylor},
  {Teixeira}, {Terrett}, {Tingley}, {Trager}, {Turon}, {Ulla}, {Utrilla},
  {Valentini}, {van Elteren}, {Van Hemelryck}, {van Leeuwen}, {Varadi},
  {Vecchiato}, {Veljanoski}, {Via}, {Vicente}, {Vogt}, {Voss}, {Votruba},
  {Voutsinas}, {Walmsley}, {Weiler}, {Weingrill}, {Werner}, {Wevers},
  {Whitehead}, {Wyrzykowski}, {Yoldas}, {{\v{Z}}erjal}, {Zucker}, {Zurbach},
  {Zwitter}, {Alecu}, {Allen}, {Allende Prieto}, {Amorim},
  {Anglada-Escud{\'e}}, {Arsenijevic}, {Azaz}, {Balm}, {Beck}, {Bernstein},
  {Bigot}, {Bijaoui}, {Blasco}, {Bonfigli}, {Bono}, {Boudreault}, {Bressan},
  {Brown}, {Brunet}, {Bunclark}, {Buonanno}, {Butkevich}, {Carret}, {Carrion},
  {Chemin}, {Ch{\'e}reau}, {Corcione}, {Darmigny}, {de Boer}, {de Teodoro}, {de
  Zeeuw}, {Delle Luche}, {Domingues}, {Dubath}, {Fodor}, {Fr{\'e}zouls},
  {Fries}, {Fustes}, {Fyfe}, {Gallardo}, {Gallegos}, {Gardiol}, {Gebran},
  {Gomboc}, {G{\'o}mez}, {Grux}, {Gueguen}, {Heyrovsky}, {Hoar}, {Iannicola},
  {Isasi Parache}, {Janotto}, {Joliet}, {Jonckheere}, {Keil}, {Kim},
  {Klagyivik}, {Klar}, {Knude}, {Kochukhov}, {Kolka}, {Kos}, {Kutka}, {Lainey},
  {LeBouquin}, {Liu}, {Loreggia}, {Makarov}, {Marseille}, {Martayan},
  {Martinez-Rubi}, {Massart}, {Meynadier}, {Mignot}, {Munari}, {Nguyen},
  {Nordlander}, {Ocvirk}, {O'Flaherty}, {Olias Sanz}, {Ortiz}, {Osorio},
  {Oszkiewicz}, {Ouzounis}, {Palmer}, {Park}, {Pasquato}, {Peltzer}, {Peralta},
  {P{\'e}turaud}, {Pieniluoma}, {Pigozzi}, {Poels}, {Prat}, {Prod'homme},
  {Raison}, {Rebordao}, {Risquez}, {Rocca-Volmerange}, {Rosen}, {Ruiz-Fuertes},
  {Russo}, {Sembay}, {Serraller Vizcaino}, {Short}, {Siebert}, {Silva},
  {Sinachopoulos}, {Slezak}, {Soffel}, {Sosnowska}, {Strai{\v{z}}ys}, {ter
  Linden}, {Terrell}, {Theil}, {Tiede}, {Troisi}, {Tsalmantza}, {Tur},
  {Vaccari}, {Vachier}, {Valles}, {Van Hamme}, {Veltz}, {Virtanen}, {Wallut},
  {Wichmann}, {Wilkinson}, {Ziaeepour}, \& {Zschocke}}]{2016A&A...595A...1G}
{Gaia Collaboration}, {Prusti}, T., {de Bruijne}, J.~H.~J., {et~al.} 2016,
  \aap, 595, A1

\bibitem[{{Gaia Collaboration} {et~al.}(2022){Gaia Collaboration}, {Vallenari},
  {Brown}, {Prusti}, {de Bruijne}, {Arenou}, {Babusiaux}, {Biermann},
  {Creevey}, {Ducourant}, {Evans}, {Eyer}, {Guerra}, {Hutton}, {Jordi},
  {Klioner}, {Lammers}, {Lindegren}, {Luri}, {Mignard}, {Panem}, {Pourbaix},
  {Randich}, {Sartoretti}, {Soubiran}, {Tanga}, {Walton}, {Bailer-Jones},
  {Bastian}, {Drimmel}, {Jansen}, {Katz}, {Lattanzi}, {van Leeuwen}, {Bakker},
  {Cacciari}, {Casta{\~n}eda}, {De Angeli}, {Fabricius}, {Fouesneau},
  {Fr{\'e}mat}, {Galluccio}, {Guerrier}, {Heiter}, {Masana}, {Messineo},
  {Mowlavi}, {Nicolas}, {Nienartowicz}, {Pailler}, {Panuzzo}, {Riclet}, {Roux},
  {Seabroke}, {Sordo{\o}rcit}, {Th{\'e}venin}, {Gracia-Abril}, {Portell},
  {Teyssier}, {Altmann}, {Andrae}, {Audard}, {Bellas-Velidis}, {Benson},
  {Berthier}, {Blomme}, {Burgess}, {Busonero}, {Busso}, {C{\'a}novas}, {Carry},
  {Cellino}, {Cheek}, {Clementini}, {Damerdji}, {Davidson}, {de Teodoro},
  {Nu{\~n}ez Campos}, {Delchambre}, {Dell'Oro}, {Esquej},
  {Fern{\'a}ndez-Hern{\'a}ndez}, {Fraile}, {Garabato}, {Garc{\'\i}a-Lario},
  {Gosset}, {Haigron}, {Halbwachs}, {Hambly}, {Harrison}, {Hern{\'a}ndez},
  {Hestroffer}, {Hodgkin}, {Holl}, {Jan{\ss}en}, {Jevardat de Fombelle},
  {Jordan}, {Krone-Martins}, {Lanzafame}, {L{\"o}ffler}, {Marchal}, {Marrese},
  {Moitinho}, {Muinonen}, {Osborne}, {Pancino}, {Pauwels}, {Recio-Blanco},
  {Reyl{\'e}}, {Riello}, {Rimoldini}, {Roegiers}, {Rybizki}, {Sarro}, {Siopis},
  {Smith}, {Sozzetti}, {Utrilla}, {van Leeuwen}, {Abbas}, {{\'A}brah{\'a}m},
  {Abreu Aramburu}, {Aerts}, {Aguado}, {Ajaj}, {Aldea-Montero}, {Altavilla},
  {{\'A}lvarez}, {Alves}, {Anders}, {Anderson}, {Anglada Varela}, {Antoja},
  {Baines}, {Baker}, {Balaguer-N{\'u}{\~n}ez}, {Balbinot}, {Balog}, {Barache},
  {Barbato}, {Barros}, {Barstow}, {Bartolom{\'e}}, {Bassilana}, {Bauchet},
  {Becciani}, {Bellazzini}, {Berihuete}, {Bernet}, {Bertone}, {Bianchi},
  {Binnenfeld}, {Blanco-Cuaresma}, {Blazere}, {Boch}, {Bombrun}, {Bossini},
  {Bouquillon}, {Bragaglia}, {Bramante}, {Breedt}, {Bressan}, {Brouillet},
  {Brugaletta}, {Bucciarelli}, {Burlacu}, {Butkevich}, {Buzzi}, {Caffau},
  {Cancelliere}, {Cantat-Gaudin}, {Carballo}, {Carlucci}, {Carnerero},
  {Carrasco}, {Casamiquela}, {Castellani}, {Castro-Ginard}, {Chaoul},
  {Charlot}, {Chemin}, {Chiaramida}, {Chiavassa}, {Chornay}, {Comoretto},
  {Contursi}, {Cooper}, {Cornez}, {Cowell}, {Crifo}, {Cropper}, {Crosta},
  {Crowley}, {Dafonte}, {Dapergolas}, {David}, {David}, {de Laverny}, {De
  Luise}, {De March}, {De Ridder}, {de Souza}, {de Torres}, {del Peloso}, {del
  Pozo}, {Delbo}, {Delgado}, {Delisle}, {Demouchy}, {Dharmawardena}, {Di
  Matteo}, {Diakite}, {Diener}, {Distefano}, {Dolding}, {Edvardsson}, {Enke},
  {Fabre}, {Fabrizio}, {Faigler}, {Fedorets}, {Fernique}, {Fienga}, {Figueras},
  {Fournier}, {Fouron}, {Fragkoudi}, {Gai}, {Garcia-Gutierrez},
  {Garcia-Reinaldos}, {Garc{\'\i}a-Torres}, {Garofalo}, {Gavel}, {Gavras},
  {Gerlach}, {Geyer}, {Giacobbe}, {Gilmore}, {Girona}, {Giuffrida}, {Gomel},
  {Gomez}, {Gonz{\'a}lez-N{\'u}{\~n}ez}, {Gonz{\'a}lez-Santamar{\'\i}a},
  {Gonz{\'a}lez-Vidal}, {Granvik}, {Guillout}, {Guiraud},
  {Guti{\'e}rrez-S{\'a}nchez}, {Guy}, {Hatzidimitriou}, {Hauser}, {Haywood},
  {Helmer}, {Helmi}, {Sarmiento}, {Hidalgo}, {Hilger}, {H{\l}adczuk}, {Hobbs},
  {Holland}, {Huckle}, {Jardine}, {Jasniewicz}, {Jean-Antoine Piccolo},
  {Jim{\'e}nez-Arranz}, {Jorissen}, {Juaristi Campillo}, {Julbe}, {Karbevska},
  {Kervella}, {Khanna}, {Kontizas}, {Kordopatis}, {Korn}, {K{\'o}sp{\'a}l},
  {Kostrzewa-Rutkowska}, {Kruszy{\'n}ska}, {Kun}, {Laizeau}, {Lambert},
  {Lanza}, {Lasne}, {Le Campion}, {Lebreton}, {Lebzelter}, {Leccia}, {Leclerc},
  {Lecoeur-Taibi}, {Liao}, {Licata}, {Lindstr{\o}m}, {Lister}, {Livanou},
  {Lobel}, {Lorca}, {Loup}, {Madrero Pardo}, {Magdaleno Romeo}, {Managau},
  {Mann}, {Manteiga}, {Marchant}, {Marconi}, {Marcos}, {Marcos Santos},
  {Mar{\'\i}n Pina}, {Marinoni}, {Marocco}, {Marshall}, {Polo},
  {Mart{\'\i}n-Fleitas}, {Marton}, {Mary}, {Masip}, {Massari},
  {Mastrobuono-Battisti}, {Mazeh}, {McMillan}, {Messina}, {Michalik}, {Millar},
  {Mints}, {Molina}, {Molinaro}, {Moln{\'a}r}, {Monari}, {Mongui{\'o}},
  {Montegriffo}, {Montero}, {Mor}, {Mora}, {Morbidelli}, {Morel}, {Morris},
  {Muraveva}, {Murphy}, {Musella}, {Nagy}, {Noval}, {Oca{\~n}a}, {Ogden},
  {Ordenovic}, {Osinde}, {Pagani}, {Pagano}, {Palaversa}, {Palicio},
  {Pallas-Quintela}, {Panahi}, {Payne-Wardenaar}, {Pe{\~n}alosa Esteller},
  {Penttil{\"a}}, {Pichon}, {Piersimoni}, {Pineau}, {Plachy}, {Plum}, {Poggio},
  {Pr{\v{s}}a}, {Pulone}, {Racero}, {Ragaini}, {Rainer}, {Raiteri}, {Rambaux},
  {Ramos}, {Ramos-Lerate}, {Re Fiorentin}, {Regibo}, {Richards}, {Rios Diaz},
  {Ripepi}, {Riva}, {Rix}, {Rixon}, {Robichon}, {Robin}, {Robin}, {Roelens},
  {Rogues}, {Rohrbasser}, {Romero-G{\'o}mez}, {Rowell}, {Royer}, {Ruz Mieres},
  {Rybicki}, {Sadowski}, {S{\'a}ez N{\'u}{\~n}ez}, {Sagrist{\`a} Sell{\'e}s},
  {Sahlmann}, {Salguero}, {Samaras}, {Sanchez Gimenez}, {Sanna},
  {Santove{\~n}a}, {Sarasso}, {Schultheis}, {Sciacca}, {Segol}, {Segovia},
  {S{\'e}gransan}, {Semeux}, {Shahaf}, {Siddiqui}, {Siebert}, {Siltala},
  {Silvelo}, {Slezak}, {Slezak}, {Smart}, {Snaith}, {Solano}, {Solitro},
  {Souami}, {Souchay}, {Spagna}, {Spina}, {Spoto}, {Steele},
  {Steidelm{\"u}ller}, {Stephenson}, {S{\"u}veges}, {Surdej}, {Szabados},
  {Szegedi-Elek}, {Taris}, {Taylo}, {Teixeira}, {Tolomei}, {Tonello}, {Torra},
  {Torra}, {Torralba Elipe}, {Trabucchi}, {Tsounis}, {Turon}, {Ulla}, {Unger},
  {Vaillant}, {van Dillen}, {van Reeven}, {Vanel}, {Vecchiato}, {Viala},
  {Vicente}, {Voutsinas}, {Weiler}, {Wevers}, {Wyrzykowski}, {Yoldas}, {Yvard},
  {Zhao}, {Zorec}, {Zucker}, \& {Zwitter}}]{2022arXiv220800211G}
{Gaia Collaboration}, {Vallenari}, A., {Brown}, A.~G.~A., {et~al.} 2022, arXiv
  e-prints, arXiv:2208.00211

\bibitem[{{Garcia-Segura} {et~al.}(1996){Garcia-Segura}, {Mac Low}, \&
  {Langer}}]{1996A&A...305..229G}
{Garcia-Segura}, G., {Mac Low}, M.~M., \& {Langer}, N. 1996, \aap, 305, 229

\bibitem[{{Garmany} \& {Stencel}(1992)}]{1992A&AS...94..211G}
{Garmany}, C.~D. \& {Stencel}, R.~E. 1992, \aaps, 94, 211

\bibitem[{{Georgy} {et~al.}(2021){Georgy}, {Saio}, \&
  {Meynet}}]{2021A&A...650A.128G}
{Georgy}, C., {Saio}, H., \& {Meynet}, G. 2021, \aap, 650, A128

\bibitem[{{Grassitelli} {et~al.}(2015){Grassitelli}, {Fossati},
  {Sim{\'o}n-Di{\'a}z}, {Langer}, {Castro}, \& {Sanyal}}]{2015ApJ...808L..31G}
{Grassitelli}, L., {Fossati}, L., {Sim{\'o}n-Di{\'a}z}, S., {et~al.} 2015,
  \apjl, 808, L31

\bibitem[{{Heger} \& {Langer}(2000)}]{2000ApJ...544.1016H}
{Heger}, A. \& {Langer}, N. 2000, \apj, 544, 1016

\bibitem[{{Holgado} {et~al.}(2018){Holgado}, {Sim{\'o}n-D{\'\i}az},
  {Barb{\'a}}, {Puls}, {Herrero}, {Castro}, {Garcia}, {Ma{\'\i}z
  Apell{\'a}niz}, {Negueruela}, \&
  {Sab{\'\i}n-Sanjuli{\'a}n}}]{2018A&A...613A..65Hol}
{Holgado}, G., {Sim{\'o}n-D{\'\i}az}, S., {Barb{\'a}}, R.~H., {et~al.} 2018,
  \aap, 613, A65

\bibitem[{{Holgado} {et~al.}(2020){Holgado}, {Sim{\'o}n-D{\'\i}az},
  {Haemmerl{\'e}}, {Lennon}, {Barb{\'a}}, {Cervi{\~n}o}, {Castro}, {Herrero},
  {Meynet}, \& {Arias}}]{2020A&A...638A.157H}
{Holgado}, G., {Sim{\'o}n-D{\'\i}az}, S., {Haemmerl{\'e}}, L., {et~al.} 2020,
  \aap, 638, A157

\bibitem[{{Holgado} {et~al.}(2022){Holgado}, {Sim{\'o}n-D{\'\i}az}, {Herrero},
  \& {Barb{\'a}}}]{2022A&A...665A.150H}
{Holgado}, G., {Sim{\'o}n-D{\'\i}az}, S., {Herrero}, A., \& {Barb{\'a}}, R.~H.
  2022, \aap, 665, A150

\bibitem[{{Houk} \& {Swift}(1999)}]{1999mctd.book.....H}
{Houk}, N. \& {Swift}, C. 1999, {Michigan catalogue of two-dimensional spectral
  types for the HD Stars ; vol. 5}, Vol.~5

\bibitem[{{Howarth} {et~al.}(1997){Howarth}, {Siebert}, {Hussain}, \&
  {Prinja}}]{1997MNRAS.284..265H}
{Howarth}, I.~D., {Siebert}, K.~W., {Hussain}, G. A.~J., \& {Prinja}, R.~K.
  1997, \mnras, 284, 265

\bibitem[{{Humphreys}(1991)}]{1991IAUS..143..485H}
{Humphreys}, R.~M. 1991, in Wolf-Rayet Stars and Interrelations with Other
  Massive Stars in Galaxies, ed. K.~A. {van der Hucht} \& B.~{Hidayat}, Vol.
  143, 485

\bibitem[{{Humphreys} \& {Davidson}(1994)}]{1994PASP..106.1025H}
{Humphreys}, R.~M. \& {Davidson}, K. 1994, \pasp, 106, 1025

\bibitem[{{Hunter} {et~al.}(2009){Hunter}, {Brott}, {Langer}, {Lennon},
  {Dufton}, {Howarth}, {Ryans}, {Trundle}, {Evans}, {de Koter}, \&
  {Smartt}}]{2009A&A...496..841H}
{Hunter}, I., {Brott}, I., {Langer}, N., {et~al.} 2009, \aap, 496, 841

\bibitem[{{Hunter} {et~al.}(2008){Hunter}, {Lennon}, {Dufton}, {Trundle},
  {Sim{\'o}n-D{\'\i}az}, {Smartt}, {Ryans}, \& {Evans}}]{2008A&A...479..541H}
{Hunter}, I., {Lennon}, D.~J., {Dufton}, P.~L., {et~al.} 2008, \aap, 479, 541

\bibitem[{{Jin} {et~al.}(2023){Jin}, {Trager}, {Dalton}, {Aguerri}, {Drew},
  {Falc{\'o}n-Barroso}, {G{\"a}nsicke}, {Hill}, {Iovino}, {Pieri}, {Poggianti},
  {Smith}, {Vallenari}, {Abrams}, {Aguado}, {Antoja}, {Arag{\'o}n-Salamanca},
  {Ascasibar}, {Babusiaux}, {Balcells}, {Barrena}, {Battaglia}, {Belokurov},
  {Bensby}, {Bonifacio}, {Bragaglia}, {Carrasco}, {Carrera}, {Cornwell},
  {Dom{\'\i}nguez-Palmero}, {Duncan}, {Famaey}, {Fari{\~n}a}, {Gonzalez},
  {Guest}, {Hatch}, {Hess}, {Hoskin}, {Irwin}, {Knapen}, {Koposov}, {Kuchner},
  {Laigle}, {Lewis}, {Longhetti}, {Lucatello}, {M{\'e}ndez-Abreu}, {Mercurio},
  {Molaeinezhad}, {Mongui{\'o}}, {Morrison}, {Murphy}, {Peralta de Arriba},
  {P{\'e}rez}, {P{\'e}rez-R{\`a}fols}, {Pic{\'o}}, {Raddi}, {Romero-G{\'o}mez},
  {Royer}, {Siebert}, {Seabroke}, {Som}, {Terrett}, {Thomas}, {Wesson},
  {Worley}, {Alfaro}, {Prieto}, {Alonso-Santiago}, {Amos}, {Ashley},
  {Balaguer-N{\'u} nez}, {Balbinot}, {Bellazzini}, {Benn}, {Berlanas},
  {Bernard}, {Best}, {Bettoni}, {Bianco}, {Bishop}, {Blomqvist}, {Boeche},
  {Bolzonella}, {Bonoli}, {Bosma}, {Britavskiy}, {Busarello}, {Caffau},
  {Cantat-Gaudin}, {Castro-Ginard}, {Couto}, {Carbajo-Hijarrubia}, {Carter},
  {Casamiquela}, {Conrado}, {Corcho-Caballero}, {Costantin}, {Deason}, {de
  Burgos}, {De Grandi}, {Di Matteo}, {Dom{\'\i}nguez-G{\'o}mez}, {Dorda},
  {Drake}, {Dutta}, {Erkal}, {Feltzing}, {Ferr{\'e}-Mateu}, {Feuillet},
  {Figueras}, {Fossati}, {Franciosini}, {Frasca}, {Fumagalli}, {Gallazzi},
  {Garc{\'\i}a-Benito}, {Fusillo}, {Gebran}, {Gilbert}, {Gledhill},
  {Gonz{\'a}lez Delgado}, {Greimel}, {Guarcello}, {Guerra}, {Gullieuszik},
  {Haines}, {Hardcastle}, {Harris}, {Haywood}, {Helmi}, {Hernandez}, {Herrero},
  {Hughes}, {Irsic}, {Jablonka}, {Jarvis}, {Jordi}, {Kondapally}, {Kordopatis},
  {Krogager}, {La Barbera}, {Lam}, {Larsen}, {Lemasle}, {Lewis}, {Lhom{\'e}},
  {Lind}, {Lodi}, {Longobardi}, {Lonoce}, {Magrini}, {Ma{\'\i}z Apell{\'a}niz},
  {Marchal}, {Marco}, {Martin}, {Matsuno}, {Maurogordato}, {Merluzzi},
  {Miralda-Escud{\'e}}, {Molinari}, {Monari}, {Morelli}, {Mottram}, {Naylor},
  {Negueruela}, {Onorbe}, {Pancino}, {Peirani}, {Peletier}, {Pozzetti},
  {Rainer}, {Ramos}, {Read}, {Rossi}, {R{\"o}ttgering},
  {Rubi{\~n}o-Mart{\'\i}n}, {Sabater Montes}, {San Juan}, {Sanna}, {Schallig},
  {Schiavon}, {Schultheis}, {Serra}, {Shimwell}, {Sim{\'o}n-D{\'\i}az},
  {Smith}, {Sordo}, {Sorini}, {Soubiran}, {Starkenburg}, {Steele}, {Stott},
  {Stuik}, {Tolstoy}, {Tortora}, {Tsantaki}, {Van der Swaelmen}, {van Weeren},
  {Vergani}, {Verheijen}, {Verro}, {Vink}, {Vioque}, {Walcher}, {Walton},
  {Wegg}, {Weijmans}, {Williams}, {Wilson}, {Wright}, {Xylakis-Dornbusch},
  {Youakim}, {Zibetti}, \& {Zurita}}]{2023MNRAS.tmp..715J}
{Jin}, S., {Trager}, S.~C., {Dalton}, G.~B., {et~al.} 2023, \mnras
  [\eprint[arXiv]{2212.03981}]

\bibitem[{{Kaufer} {et~al.}(1997){Kaufer}, {Wolf}, {Andersen}, \&
  {Pasquini}}]{1997Msngr..89....1K}
{Kaufer}, A., {Wolf}, B., {Andersen}, J., \& {Pasquini}, L. 1997, The
  Messenger, 89, 1

\bibitem[{{Kervella} {et~al.}(2019){Kervella}, {Arenou}, {Mignard}, \&
  {Th{\'e}venin}}]{2019A&A...623A..72K}
{Kervella}, P., {Arenou}, F., {Mignard}, F., \& {Th{\'e}venin}, F. 2019, \aap,
  623, A72

\bibitem[{{Krause} {et~al.}(2013){Krause}, {Fierlinger}, {Diehl}, {Burkert},
  {Voss}, \& {Ziegler}}]{2013A&A...550A..49K}
{Krause}, M., {Fierlinger}, K., {Diehl}, R., {et~al.} 2013, \aap, 550, A49

\bibitem[{{Kudritzki} {et~al.}(2003){Kudritzki}, {Bresolin}, \&
  {Przybilla}}]{2003ApJ...582L..83K}
{Kudritzki}, R.~P., {Bresolin}, F., \& {Przybilla}, N. 2003, \apjl, 582, L83

\bibitem[{{Kudritzki} \& {Przybilla}(2003)}]{2003LNP...635..123K}
{Kudritzki}, R.-P. \& {Przybilla}, N. 2003, in Stellar Candles for the
  Extragalactic Distance Scale, ed. D.~{Alloin} \& W.~{Gieren}, Vol. 635,
  123--148

\bibitem[{{Kudritzki} \& {Puls}(2000)}]{2000ARA&A..38..613K}
{Kudritzki}, R.-P. \& {Puls}, J. 2000, \araa, 38, 613

\bibitem[{{Kudritzki} {et~al.}(1999){Kudritzki}, {Puls}, {Lennon}, {Venn},
  {Reetz}, {Najarro}, {McCarthy}, \& {Herrero}}]{1999A&A...350..970K}
{Kudritzki}, R.~P., {Puls}, J., {Lennon}, D.~J., {et~al.} 1999, \aap, 350, 970

\bibitem[{{Kudritzki} {et~al.}(2008){Kudritzki}, {Urbaneja}, {Bresolin},
  {Przybilla}, {Gieren}, \& {Pietrzy{\'n}ski}}]{2008ApJ...681..269K}
{Kudritzki}, R.-P., {Urbaneja}, M.~A., {Bresolin}, F., {et~al.} 2008, \apj,
  681, 269

\bibitem[{{Lallement} {et~al.}(2019){Lallement}, {Babusiaux}, {Vergely},
  {Katz}, {Arenou}, {Valette}, {Hottier}, \& {Capitanio}}]{2019A&A...625A.135L}
{Lallement}, R., {Babusiaux}, C., {Vergely}, J.~L., {et~al.} 2019, \aap, 625,
  A135

\bibitem[{{Lallement} {et~al.}(2018){Lallement}, {Capitanio}, {Ruiz-Dern},
  {Danielski}, {Babusiaux}, {Vergely}, {Elyajouri}, {Arenou}, \&
  {Leclerc}}]{2018A&A...616A.132L}
{Lallement}, R., {Capitanio}, L., {Ruiz-Dern}, L., {et~al.} 2018, \aap, 616,
  A132

\bibitem[{{Langer}(2012)}]{2012ARA&A..50..107L}
{Langer}, N. 2012, \araa, 50, 107

\bibitem[{{Langer} \& {Kudritzki}(2014)}]{2014A&A...564A..52L}
{Langer}, N. \& {Kudritzki}, R.~P. 2014, \aap, 564, A52

\bibitem[{{Lennon} {et~al.}(1992){Lennon}, {Dufton}, \&
  {Fitzsimmons}}]{1992A&AS...94..569L}
{Lennon}, D.~J., {Dufton}, P.~L., \& {Fitzsimmons}, A. 1992, \aaps, 94, 569

\bibitem[{{Lennon} {et~al.}(1993){Lennon}, {Dufton}, \&
  {Fitzsimmons}}]{1993A&AS...97..559L}
{Lennon}, D.~J., {Dufton}, P.~L., \& {Fitzsimmons}, A. 1993, \aaps, 97, 559

\bibitem[{{Lesh}(1968)}]{1968ApJS...17..371L}
{Lesh}, J.~R. 1968, \apjs, 17, 371

\bibitem[{{Levesque} {et~al.}(2005){Levesque}, {Massey}, {Olsen}, {Plez},
  {Josselin}, {Maeder}, \& {Meynet}}]{2005ApJ...628..973L}
{Levesque}, E.~M., {Massey}, P., {Olsen}, K.~A.~G., {et~al.} 2005, \apj, 628,
  973

\bibitem[{{Lindegren} {et~al.}(2021{\natexlab{a}}){Lindegren}, {Bastian},
  {Biermann}, {Bombrun}, {de Torres}, {Gerlach}, {Geyer}, {Hern{\'a}ndez},
  {Hilger}, {Hobbs}, {Klioner}, {Lammers}, {McMillan}, {Ramos-Lerate},
  {Steidelm{\"u}ller}, {Stephenson}, \& {van Leeuwen}}]{2021A&A...649A...4L}
{Lindegren}, L., {Bastian}, U., {Biermann}, M., {et~al.} 2021{\natexlab{a}},
  \aap, 649, A4

\bibitem[{{Lindegren} {et~al.}(2018){Lindegren}, {Hern{\'a}ndez}, {Bombrun},
  {Klioner}, {Bastian}, {Ramos-Lerate}, {de Torres}, {Steidelm{\"u}ller},
  {Stephenson}, {Hobbs}, {Lammers}, {Biermann}, {Geyer}, {Hilger}, {Michalik},
  {Stampa}, {McMillan}, {Casta{\~n}eda}, {Clotet}, {Comoretto}, {Davidson},
  {Fabricius}, {Gracia}, {Hambly}, {Hutton}, {Mora}, {Portell}, {van Leeuwen},
  {Abbas}, {Abreu}, {Altmann}, {Andrei}, {Anglada}, {Balaguer-N{\'u}{\~n}ez},
  {Barache}, {Becciani}, {Bertone}, {Bianchi}, {Bouquillon}, {Bourda},
  {Br{\"u}semeister}, {Bucciarelli}, {Busonero}, {Buzzi}, {Cancelliere},
  {Carlucci}, {Charlot}, {Cheek}, {Crosta}, {Crowley}, {de Bruijne}, {de
  Felice}, {Drimmel}, {Esquej}, {Fienga}, {Fraile}, {Gai}, {Garralda},
  {Gonz{\'a}lez-Vidal}, {Guerra}, {Hauser}, {Hofmann}, {Holl}, {Jordan},
  {Lattanzi}, {Lenhardt}, {Liao}, {Licata}, {Lister}, {L{\"o}ffler},
  {Marchant}, {Martin-Fleitas}, {Messineo}, {Mignard}, {Morbidelli}, {Poggio},
  {Riva}, {Rowell}, {Salguero}, {Sarasso}, {Sciacca}, {Siddiqui}, {Smart},
  {Spagna}, {Steele}, {Taris}, {Torra}, {van Elteren}, {van Reeven}, \&
  {Vecchiato}}]{2018A&A...616A...2L}
{Lindegren}, L., {Hern{\'a}ndez}, J., {Bombrun}, A., {et~al.} 2018, \aap, 616,
  A2

\bibitem[{{Lindegren} {et~al.}(2021{\natexlab{b}}){Lindegren}, {Klioner},
  {Hern{\'a}ndez}, {Bombrun}, {Ramos-Lerate}, {Steidelm{\"u}ller}, {Bastian},
  {Biermann}, {de Torres}, {Gerlach}, {Geyer}, {Hilger}, {Hobbs}, {Lammers},
  {McMillan}, {Stephenson}, {Casta{\~n}eda}, {Davidson}, {Fabricius},
  {Gracia-Abril}, {Portell}, {Rowell}, {Teyssier}, {Torra}, {Bartolom{\'e}},
  {Clotet}, {Garralda}, {Gonz{\'a}lez-Vidal}, {Torra}, {Abbas}, {Altmann},
  {Anglada Varela}, {Balaguer-N{\'u}{\~n}ez}, {Balog}, {Barache}, {Becciani},
  {Bernet}, {Bertone}, {Bianchi}, {Bouquillon}, {Brown}, {Bucciarelli},
  {Busonero}, {Butkevich}, {Buzzi}, {Cancelliere}, {Carlucci}, {Charlot},
  {Cioni}, {Crosta}, {Crowley}, {del Peloso}, {del Pozo}, {Drimmel}, {Esquej},
  {Fienga}, {Fraile}, {Gai}, {Garcia-Reinaldos}, {Guerra}, {Hambly}, {Hauser},
  {Jan{\ss}en}, {Jordan}, {Kostrzewa-Rutkowska}, {Lattanzi}, {Liao}, {Licata},
  {Lister}, {L{\"o}ffler}, {Marchant}, {Masip}, {Mignard}, {Mints}, {Molina},
  {Mora}, {Morbidelli}, {Murphy}, {Pagani}, {Panuzzo}, {Pe{\~n}alosa Esteller},
  {Poggio}, {Re Fiorentin}, {Riva}, {Sagrist{\`a} Sell{\'e}s}, {Sanchez
  Gimenez}, {Sarasso}, {Sciacca}, {Siddiqui}, {Smart}, {Souami}, {Spagna},
  {Steele}, {Taris}, {Utrilla}, {van Reeven}, \&
  {Vecchiato}}]{2021A&A...649A...2L}
{Lindegren}, L., {Klioner}, S.~A., {Hern{\'a}ndez}, J., {et~al.}
  2021{\natexlab{b}}, \aap, 649, A2

\bibitem[{{Lucy} \& {Solomon}(1970)}]{1970ApJ...159..879L}
{Lucy}, L.~B. \& {Solomon}, P.~M. 1970, \apj, 159, 879

\bibitem[{{Maeder} \& {Meynet}(2000)}]{2000ARA&A..38..143M}
{Maeder}, A. \& {Meynet}, G. 2000, \araa, 38, 143

\bibitem[{{Maeder} \& {Meynet}(2005)}]{2005A&A...440.1041M}
{Maeder}, A. \& {Meynet}, G. 2005, \aap, 440, 1041

\bibitem[{{Ma{\'\i}z Apell{\'a}niz}(2022)}]{2022A&A...657A.130M}
{Ma{\'\i}z Apell{\'a}niz}, J. 2022, \aap, 657, A130

\bibitem[{{Ma{\'\i}z Apell{\'a}niz} {et~al.}(2021){Ma{\'\i}z Apell{\'a}niz},
  {Pantaleoni Gonz{\'a}lez}, \& {Barb{\'a}}}]{2021A&A...649A..13M}
{Ma{\'\i}z Apell{\'a}niz}, J., {Pantaleoni Gonz{\'a}lez}, M., \& {Barb{\'a}},
  R.~H. 2021, \aap, 649, A13

\bibitem[{{Marchant} {et~al.}(2016){Marchant}, {Langer}, {Podsiadlowski},
  {Tauris}, \& {Moriya}}]{2016A&A...588A..50M}
{Marchant}, P., {Langer}, N., {Podsiadlowski}, P., {Tauris}, T.~M., \&
  {Moriya}, T.~J. 2016, \aap, 588, A50

\bibitem[{{Markova} \& {Puls}(2008)}]{2008A&A...478..823M}
{Markova}, N. \& {Puls}, J. 2008, \aap, 478, 823

\bibitem[{{Martinet} {et~al.}(2021){Martinet}, {Meynet}, {Ekstr{\"o}m},
  {Sim{\'o}n-D{\'\i}az}, {Holgado}, {Castro}, {Georgy}, {Eggenberger},
  {Buldgen}, {Salmon}, {Hirschi}, {Groh}, {Farrell}, \&
  {Murphy}}]{2021A&A...648A.126M}
{Martinet}, S., {Meynet}, G., {Ekstr{\"o}m}, S., {et~al.} 2021, \aap, 648, A126

\bibitem[{{Massey} \& {Thompson}(1991)}]{1991AJ....101.1408M}
{Massey}, P. \& {Thompson}, A.~B. 1991, \aj, 101, 1408

\bibitem[{{Matteucci}(2008)}]{2008IAUS..250..391M}
{Matteucci}, F. 2008, in Massive Stars as Cosmic Engines, ed. F.~{Bresolin},
  P.~A. {Crowther}, \& J.~{Puls}, Vol. 250, 391--400

\bibitem[{{Matzner}(2002)}]{2002ApJ...566..302M}
{Matzner}, C.~D. 2002, \apj, 566, 302

\bibitem[{{McEvoy} {et~al.}(2015){McEvoy}, {Dufton}, {Evans}, {Kalari},
  {Markova}, {Sim{\'o}n-D{\'\i}az}, {Vink}, {Walborn}, {Crowther}, {de Koter},
  {de Mink}, {Dunstall}, {H{\'e}nault-Brunet}, {Herrero}, {Langer}, {Lennon},
  {Ma{\'\i}z Apell{\'a}niz}, {Najarro}, {Puls}, {Sana}, {Schneider}, \&
  {Taylor}}]{2015A&A...575A..70M}
{McEvoy}, C.~M., {Dufton}, P.~L., {Evans}, C.~J., {et~al.} 2015, \aap, 575, A70

\bibitem[{{Melnik} \& {Dambis}(2020)}]{2020Ap&SS.365..112M}
{Melnik}, A.~M. \& {Dambis}, A.~K. 2020, \apss, 365, 112

\bibitem[{{Meynet} \& {Maeder}(2000)}]{2000A&A...361..101M}
{Meynet}, G. \& {Maeder}, A. 2000, \aap, 361, 101

\bibitem[{{Meynet} \& {Maeder}(2005)}]{2005A&A...429..581M}
{Meynet}, G. \& {Maeder}, A. 2005, \aap, 429, 581

\bibitem[{{Morgan} {et~al.}(1955){Morgan}, {Code}, \&
  {Whitford}}]{1955ApJS....2...41M}
{Morgan}, W.~W., {Code}, A.~D., \& {Whitford}, A.~E. 1955, \apjs, 2, 41

\bibitem[{{Morgan} {et~al.}(1953){Morgan}, {Whitford}, \&
  {Code}}]{1953ApJ...118..318M}
{Morgan}, W.~W., {Whitford}, A.~E., \& {Code}, A.~D. 1953, \apj, 118, 318

\bibitem[{{Nugis} \& {Lamers}(2000)}]{2000A&A...360..227N}
{Nugis}, T. \& {Lamers}, H.~J.~G.~L.~M. 2000, \aap, 360, 227

\bibitem[{{Pantaleoni Gonz{\'a}lez} {et~al.}(2021){Pantaleoni Gonz{\'a}lez},
  {Ma{\'\i}z Apell{\'a}niz}, {Barb{\'a}}, \& {Reed}}]{2021MNRAS.504.2968P}
{Pantaleoni Gonz{\'a}lez}, M., {Ma{\'\i}z Apell{\'a}niz}, J., {Barb{\'a}},
  R.~H., \& {Reed}, B.~C. 2021, \mnras, 504, 2968

\bibitem[{{Paxton} {et~al.}(2011){Paxton}, {Bildsten}, {Dotter}, {Herwig},
  {Lesaffre}, \& {Timmes}}]{2011ApJS..192....3P}
{Paxton}, B., {Bildsten}, L., {Dotter}, A., {et~al.} 2011, \apjs, 192, 3

\bibitem[{{Paxton} {et~al.}(2013){Paxton}, {Cantiello}, {Arras}, {Bildsten},
  {Brown}, {Dotter}, {Mankovich}, {Montgomery}, {Stello}, {Timmes}, \&
  {Townsend}}]{2013ApJS..208....4P}
{Paxton}, B., {Cantiello}, M., {Arras}, P., {et~al.} 2013, \apjs, 208, 4

\bibitem[{{Paxton} {et~al.}(2015){Paxton}, {Marchant}, {Schwab}, {Bauer},
  {Bildsten}, {Cantiello}, {Dessart}, {Farmer}, {Hu}, {Langer}, {Townsend},
  {Townsley}, \& {Timmes}}]{2015ApJS..220...15P}
{Paxton}, B., {Marchant}, P., {Schwab}, J., {et~al.} 2015, \apjs, 220, 15

\bibitem[{{Perryman} {et~al.}(1997){Perryman}, {Lindegren}, {Kovalevsky},
  {Hoeg}, {Bastian}, {Bernacca}, {Cr{\'e}z{\'e}}, {Donati}, {Grenon},
  {Grewing}, {van Leeuwen}, {van der Marel}, {Mignard}, {Murray}, {Le Poole},
  {Schrijver}, {Turon}, {Arenou}, {Froeschl{\'e}}, \&
  {Petersen}}]{1997A&A...323L..49P}
{Perryman}, M.~A.~C., {Lindegren}, L., {Kovalevsky}, J., {et~al.} 1997, \aap,
  323, L49

\bibitem[{{Petrenz} \& {Puls}(1996)}]{1996A&A...312..195P}
{Petrenz}, P. \& {Puls}, J. 1996, \aap, 312, 195

\bibitem[{{Podsiadlowski} {et~al.}(1992){Podsiadlowski}, {Joss}, \&
  {Hsu}}]{1992ApJ...391..246P}
{Podsiadlowski}, P., {Joss}, P.~C., \& {Hsu}, J.~J.~L. 1992, \apj, 391, 246

\bibitem[{{Poppel}(1997)}]{1997FCPh...18....1P}
{Poppel}, W. 1997, \fcp, 18, 1

\bibitem[{{Puls} {et~al.}(1996){Puls}, {Kudritzki}, {Herrero}, {Pauldrach},
  {Haser}, {Lennon}, {Gabler}, {Voels}, {Vilchez}, {Wachter}, \&
  {Feldmeier}}]{1996A&A...305..171P}
{Puls}, J., {Kudritzki}, R.~P., {Herrero}, A., {et~al.} 1996, \aap, 305, 171

\bibitem[{{Puls} {et~al.}(2005){Puls}, {Urbaneja}, {Venero}, {Repolust},
  {Springmann}, {Jokuthy}, \& {Mokiem}}]{2005A&A...435..669P}
{Puls}, J., {Urbaneja}, M.~A., {Venero}, R., {et~al.} 2005, \aap, 435, 669

\bibitem[{{Puls} {et~al.}(2008){Puls}, {Vink}, \&
  {Najarro}}]{2008A&ARv..16..209P}
{Puls}, J., {Vink}, J.~S., \& {Najarro}, F. 2008, \aapr, 16, 209

\bibitem[{{Raskin} {et~al.}(2011){Raskin}, {van Winckel}, {Hensberge},
  {Jorissen}, {Lehmann}, {Waelkens}, {Avila}, {de Cuyper}, {Degroote},
  {Dubosson}, {Dumortier}, {Fr{\'e}mat}, {Laux}, {Michaud}, {Morren}, {Perez
  Padilla}, {Pessemier}, {Prins}, {Smolders}, {van Eck}, \&
  {Winkler}}]{2011A&A...526A..69R}
{Raskin}, G., {van Winckel}, H., {Hensberge}, H., {et~al.} 2011, A\&A, 526, A69

\bibitem[{{Rogers} \& {Pittard}(2013)}]{2013MNRAS.431.1337R}
{Rogers}, H. \& {Pittard}, J.~M. 2013, \mnras, 431, 1337

\bibitem[{{Saio} {et~al.}(2013){Saio}, {Georgy}, \&
  {Meynet}}]{2013MNRAS.433.1246S}
{Saio}, H., {Georgy}, C., \& {Meynet}, G. 2013, \mnras, 433, 1246

\bibitem[{{Sana} {et~al.}(2013){Sana}, {de Koter}, {de Mink}, {Dunstall},
  {Evans}, {H{\'e}nault-Brunet}, {Ma{\'\i}z Apell{\'a}niz},
  {Ram{\'\i}rez-Agudelo}, {Taylor}, {Walborn}, {Clark}, {Crowther}, {Herrero},
  {Gieles}, {Langer}, {Lennon}, \& {Vink}}]{2013A&A...550A.107S}
{Sana}, H., {de Koter}, A., {de Mink}, S.~E., {et~al.} 2013, \aap, 550, A107

\bibitem[{{Sana} {et~al.}(2012){Sana}, {de Mink}, {de Koter}, {Langer},
  {Evans}, {Gieles}, {Gosset}, {Izzard}, {Le Bouquin}, \&
  {Schneider}}]{2012Sci...337..444S}
{Sana}, H., {de Mink}, S.~E., {de Koter}, A., {et~al.} 2012, Science, 337, 444

\bibitem[{{Santolaya-Rey} {et~al.}(1997){Santolaya-Rey}, {Puls}, \&
  {Herrero}}]{1997A&A...323..488S}
{Santolaya-Rey}, A.~E., {Puls}, J., \& {Herrero}, A. 1997, \aap, 323, 488

\bibitem[{{Schootemeijer} {et~al.}(2019){Schootemeijer}, {Langer}, {Grin}, \&
  {Wang}}]{2019A&A...625A.132S}
{Schootemeijer}, A., {Langer}, N., {Grin}, N.~J., \& {Wang}, C. 2019, \aap,
  625, A132

\bibitem[{{Shull} {et~al.}(2021){Shull}, {Darling}, \&
  {Danforth}}]{2021ApJ...914...18S}
{Shull}, J.~M., {Darling}, J., \& {Danforth}, C.~W. 2021, \apj, 914, 18

\bibitem[{{Sim{\'o}n-D{\'\i}az} {et~al.}(2017){Sim{\'o}n-D{\'\i}az}, {Godart},
  {Castro}, {Herrero}, {Aerts}, {Puls}, {Telting}, \&
  {Grassitelli}}]{2017A&A...597A..22S}
{Sim{\'o}n-D{\'\i}az}, S., {Godart}, M., {Castro}, N., {et~al.} 2017, \aap,
  597, A22

\bibitem[{{Sim{\'o}n-D{\'\i}az} \& {Herrero}(2014)}]{2014A&A...562A.135S}
{Sim{\'o}n-D{\'\i}az}, S. \& {Herrero}, A. 2014, \aap, 562, A135

\bibitem[{{Sim{\'o}n-D{\'\i}az} {et~al.}(2010){Sim{\'o}n-D{\'\i}az}, {Herrero},
  {Uytterhoeven}, {Castro}, {Aerts}, \& {Puls}}]{2010ApJ...720L.174S}
{Sim{\'o}n-D{\'\i}az}, S., {Herrero}, A., {Uytterhoeven}, K., {et~al.} 2010,
  \apjl, 720, L174

\bibitem[{{Sim{\'o}n-D{\'\i}az} {et~al.}(2020){Sim{\'o}n-D{\'\i}az}, {P{\'e}rez
  Prieto}, {Holgado}, {de Burgos}, \& {Iacob Team}}]{2020sea..confE.187S}
{Sim{\'o}n-D{\'\i}az}, S., {P{\'e}rez Prieto}, J.~A., {Holgado}, G., {de
  Burgos}, A., \& {Iacob Team}. 2020, in XIV.0 Scientific Meeting (virtual) of
  the Spanish Astronomical Society, 187

\bibitem[{Simón-Díaz {et~al.}(2023)Simón-Díaz, de~Burgos, \&
  et~al.}]{simondiaz2023}
Simón-Díaz, S., de~Burgos, A., \& et~al. 2023, \aap, submitted

\bibitem[{{Smartt}(2009)}]{2009ARA&A..47...63S}
{Smartt}, S.~J. 2009, \araa, 47, 63

\bibitem[{{Smith} {et~al.}(2011){Smith}, {Li}, {Silverman}, {Ganeshalingam}, \&
  {Filippenko}}]{2011MNRAS.415..773S}
{Smith}, N., {Li}, W., {Silverman}, J.~M., {Ganeshalingam}, M., \&
  {Filippenko}, A.~V. 2011, \mnras, 415, 773

\bibitem[{{Stothers} \& {Chin}(1975)}]{1975ApJ...198..407S}
{Stothers}, R. \& {Chin}, C.~W. 1975, \apj, 198, 407

\bibitem[{{Telting} {et~al.}(2014){Telting}, {Avila}, {Buchhave}, {Frandsen},
  {Gandolfi}, {Lindberg}, {Stempels}, {Prins}, \& {NOT
  staff}}]{2014AN....335...41T}
{Telting}, J.~H., {Avila}, G., {Buchhave}, L., {et~al.} 2014, Astronomische
  Nachrichten, 335, 41

\bibitem[{{Vanbeveren} {et~al.}(2013){Vanbeveren}, {Mennekens}, {Van
  Rensbergen}, \& {De Loore}}]{2013A&A...552A.105V}
{Vanbeveren}, D., {Mennekens}, N., {Van Rensbergen}, W., \& {De Loore}, C.
  2013, \aap, 552, A105

\bibitem[{{Vink} {et~al.}(2010){Vink}, {Brott}, {Gr{\"a}fener}, {Langer}, {de
  Koter}, \& {Lennon}}]{2010A&A...512L...7V}
{Vink}, J.~S., {Brott}, I., {Gr{\"a}fener}, G., {et~al.} 2010, \aap, 512, L7

\bibitem[{{Vink} {et~al.}(2000){Vink}, {de Koter}, \&
  {Lamers}}]{2000A&A...362..295V}
{Vink}, J.~S., {de Koter}, A., \& {Lamers}, H.~J.~G.~L.~M. 2000, \aap, 362, 295

\bibitem[{{Voss} {et~al.}(2009){Voss}, {Diehl}, {Hartmann}, {Cervi{\~n}o},
  {Vink}, {Meynet}, {Limongi}, \& {Chieffi}}]{2009A&A...504..531V}
{Voss}, R., {Diehl}, R., {Hartmann}, D.~H., {et~al.} 2009, \aap, 504, 531

\bibitem[{{Weis} \& {Bomans}(2020)}]{2020Galax...8...20W}
{Weis}, K. \& {Bomans}, D.~J. 2020, Galaxies, 8, 20

\bibitem[{{Wenger} {et~al.}(2000){Wenger}, {Ochsenbein}, {Egret}, {Dubois},
  {Bonnarel}, {Borde}, {Genova}, {Jasniewicz}, {Lalo{\"e}}, {Lesteven}, \&
  {Monier}}]{2000A&AS..143....9W}
{Wenger}, M., {Ochsenbein}, F., {Egret}, D., {et~al.} 2000, \aaps, 143, 9

\bibitem[{{We{\ss}mayer} {et~al.}(2022){We{\ss}mayer}, {Przybilla}, \&
  {Butler}}]{2022A&A...668A..92W}
{We{\ss}mayer}, D., {Przybilla}, N., \& {Butler}, K. 2022, \aap, 668, A92

\bibitem[{{Woosley} {et~al.}(2002){Woosley}, {Heger}, \&
  {Weaver}}]{2002RvMP...74.1015W}
{Woosley}, S.~E., {Heger}, A., \& {Weaver}, T.~A. 2002, Reviews of Modern
  Physics, 74, 1015

\bibitem[{{Yalyalieva} {et~al.}(2020){Yalyalieva}, {Carraro}, {Vazquez},
  {Rizzo}, {Glushkova}, \& {Costa}}]{2020MNRAS.495.1349Y}
{Yalyalieva}, L., {Carraro}, G., {Vazquez}, R., {et~al.} 2020, \mnras, 495,
  1349

\bibitem[{{Zari} {et~al.}(2018){Zari}, {Hashemi}, {Brown}, {Jardine}, \& {de
  Zeeuw}}]{2018A&A...620A.172Z}
{Zari}, E., {Hashemi}, H., {Brown}, A.~G.~A., {Jardine}, K., \& {de Zeeuw},
  P.~T. 2018, \aap, 620, A172

\end{thebibliography}


\begin{appendix}


\section{pyIACOB package}
\label{apen.pyiacob}
A specific module written in \textit{python 3.8} and named \textit{pyIACOB\footnote{\url{https://github.com/Abelink23/pyIACOB/}}} has been developed for manipulating and analyzing the spectroscopic data obtained from the FIES, HERMES and FEROS echelle spectrographs (see Sect.~\ref{subsection:21_YYYYY}). This module is actually composed of other sub-modules, some of which have been used in this work. In order of use in this work, they are:

\begin{itemize}
    \item The \textit{get\_feros.py} sub-module to retrieve all available public spectra of the FEROS spectrograph from the ESO archive given an input list of starts to search. \smallskip
    
    \item The \textit{db.py} sub-module to: a) Generate tables with the basic information of all the stars available in the IACOB spectroscopic database and additional FEROS spectra. This information includes the coordinates, spectral classifications from \textit{Simbad}, basic photometric ($B_{mag}$, $V_{mag}$), S/N at different wavelength ranges, number of available spectra of each instrument, and best reference file based on the S/N, and b) Generate tables with the astrometric and photometric information from {\em Gaia} DR3, including the parallax zero-point offset. \smallskip
    
    \item The \textit{spec.py} sub-module to: a) review each individual spectra to select the reference one, and b) to visually assign labels to the H$\beta$ and H$\alpha$ line profiles (see Sect.~\ref{subsection:33_XXXXXX}). \smallskip
    
    \item The \textit{measure.py} sub-module to automate the process of cleaning the spectra from cosmic rays and other cosmetic defects, measure and correct the spectra from heliocentric and radial velocity, and finally perform the measurement of \fwhb\ (see Sect.~\ref{subsubsection:312_ZZZZZ}) by fitting the line to a Voigt profile convoluted with a rotational profile and an additional Gaussian profile to account for the cores of the most slow-rotating stars. \smallskip
    
    \item The \textit{IACOBroad.py} sub-module to prepare the input list of spectra and lines to be used in {\tt IACOB-BROAD} for the line-broadening analysis. 

\end{itemize}


\section{Comparison between {\em Gaia} distances}
\label{apen.distances}
\begin{figure}[!t]
\centering
\includegraphics[width=0.45\textwidth]{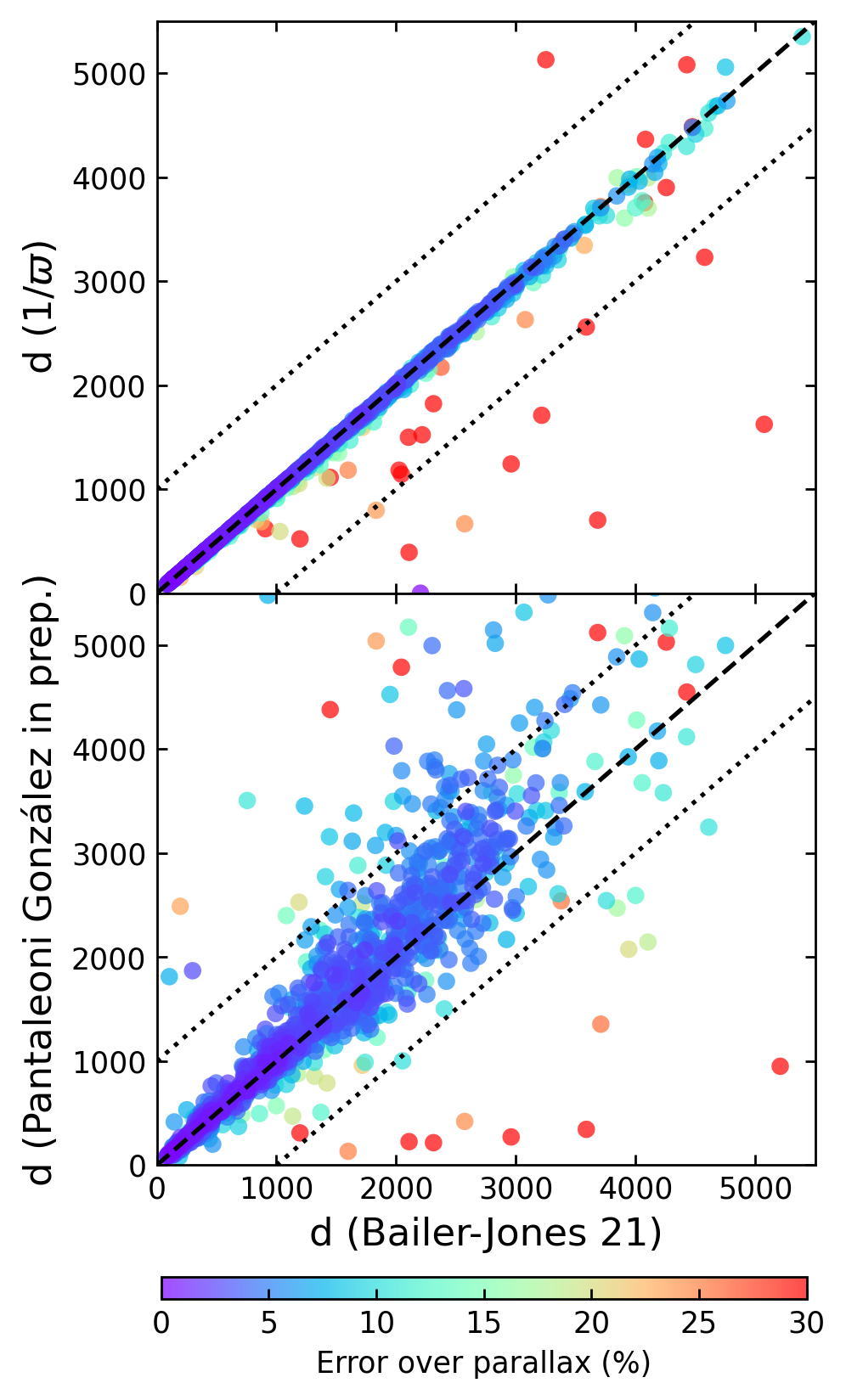}
\caption{Top panel: distances from the inverse of the parallax (corrected from zero-point) against the distances from \citep{2021AJ....161..147B} for the initial full sample of O9\,--\,B9 type stars. Bottom panel: for the same stars, distances from Pantaleoni González, et al. (in prep.) against those from  \citep{2021AJ....161..147B}. The stars are color-coded by their error over parallax from the {\em Gaia DR3} data. In both panels, the two dotter lines indicate a deviation of $\pm$1\,000\,pc.} 
\label{fig:distances}
\end{figure}

As indicated in Sect.~\ref{subsection:22_YYYYY}, the distances used in this work were taken from \citep{2021AJ....161..147B}, who uses a direction-dependent prior to improve distances and their reliability respect to the simple inverse of the parallax. However, we also compared preliminary distances from Pantaleoni González, et al. (in prep.), who uses a different prior in the determination of such distances, optimized for the OB stars. For the initial full sample of stars with available {\em Gaia} data, Fig.~\ref{fig:distances} top and bottom panels compare the distances derived from the inverse of the parallax (corrected from zero-point) and the distances from Pantaleoni González, et al. (in prep.) with those in \citep{2021AJ....161..147B}, respectively. Both panels are color-coded with the error over parallax.

In the top panel, we see that for the vast majority of stars, the difference between both distances is less than 100\,--\,200\,pc up to 5\,000\,pc. We also see in the top panel that the stars with larger error over parallax are those with also larger differences. In particular, from the $\sim$50 stars with $\Delta$(d) > 800\,pc, we find $\sim$20\% to have G\,$<$\,6, and another $\sim$20\% to correspond to SB2+ systems, while many of the rest have LCs I and show strong emission or are identified as stars with pulsations. Independently from the value of $\sigma_{\varpi}/\varpi$, we have found an average difference of only -7\,pc.

The comparison in the bottom panel contrary shows that in this case, differences of less than 500\,pc extend only up to $\sim$1\,500\,pc, with an overall trend towards shorter distances for \citep{2021AJ....161..147B} above that distance. Also above that distance is noticeable the larger scatter, which applies also to stars with very low error over parallax. In this case, stars with larger errors also have larger discrepancies, but we attribute this to the particular cases mentioned above that affect the reliability of the initial data. As a conclusion, the prior used in Pantaleoni González, et al. (in prep.) gives larger distances as a function of the distance itself, populating the sample of stars up to $\sim$4\,000\,pm rather than $\sim$3\,000\,pc as in \citep{2021AJ....161..147B}. If this is correct, it would imply that on the one hand, we might miss less stars than previously shown in Fig~\ref{fig:hist_als} between 1\,000\,--\,3\,000\,pc and therefore increasing our completeness. On the other hand, it would also imply that we may miss additional stars that would move in up in the green are from Fig~\ref{fig:gaia_alsIII}, and therefore likely increasing our number of missing stars.


\section{Examples of SB2+ systems}
\label{apen.sb2}

\begin{figure}[htbp]
    \centering
    \resizebox{\columnwidth}{!}{\includegraphics{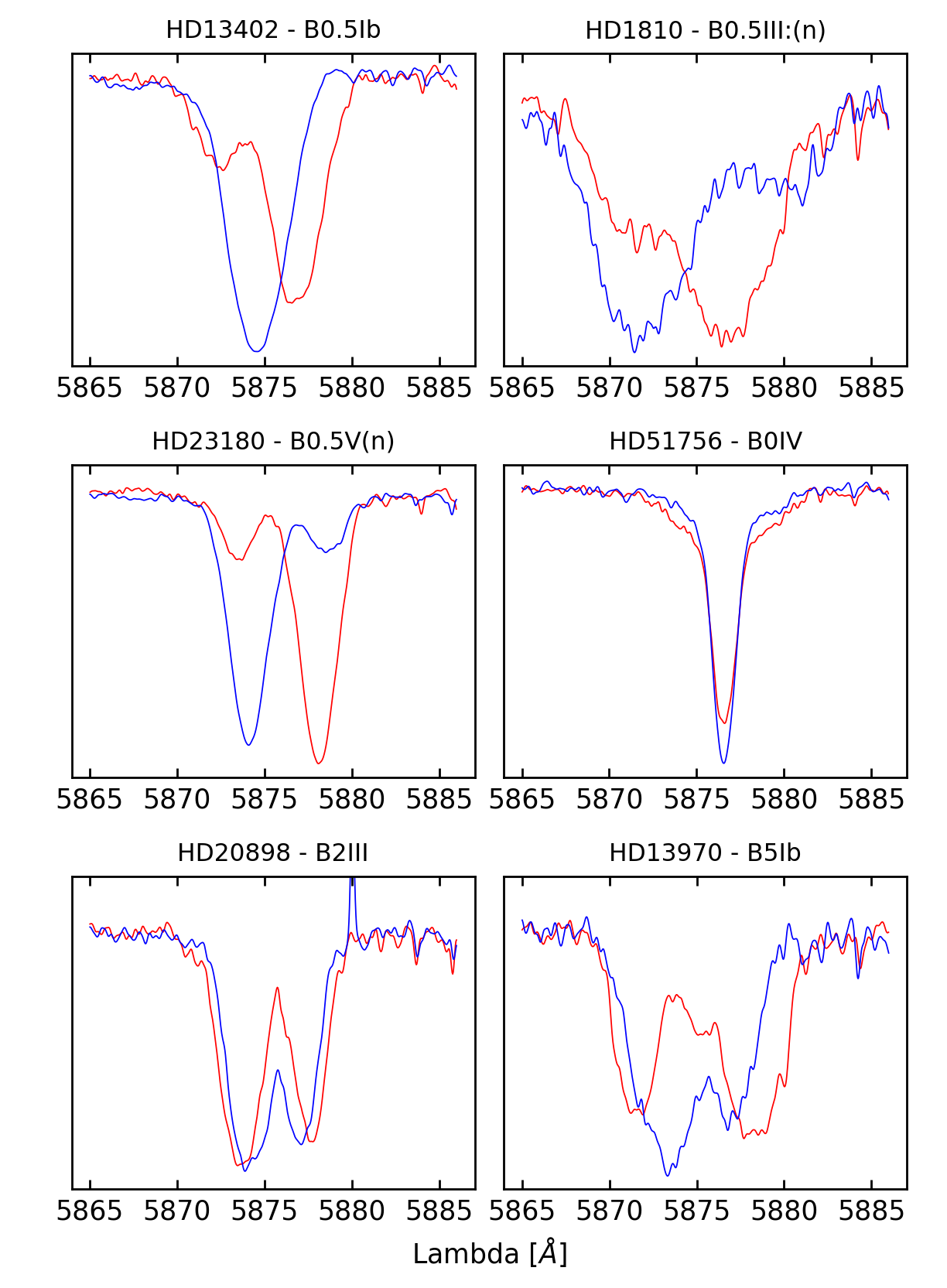}}
    \caption{He~{\sc i} $\lambda$5875.62\,{\AA} line for six SB2 systems identified in this work. In top two figures, two cases of lines blended with short separation. The middle-right figure shows an example of a fast-rotating stars with a normal-rotating star. Bottom figures shows the easiest examples of SB2 where the lines from each star are widely separated.}
    \label{fig:sb2}
\end{figure}

As described in Sect.~\ref{subsubsection:322_YYYYY}, we have looked in the available multi-epoch spectra in search for new or previously identified SB2 or SB3 systems. Fig.~\ref{fig:sb2} include some examples SB2+ systems found within the available spectra.

\FloatBarrier


\section{New spectral classifications}
\label{apen.newclass}

Table~\ref{tab:newclass_BSGs} and Table~\ref{tab:newclass_noLC} include new or reviewed spectral classification based on visual morphological features for two groups of stars. The first table includes new classifications for all LC I stars with \logLs\ < 3.5\,dex from Fig.~\ref{fig:shrd_fw_adb} wrongly classified as such, and independently from the \fwhb. It also includes at the bottom some other LC II stars with \logLs\ < 3.5\,dex for which new classifications have been derived or have been taken from either \citet{1953ApJ...118..318M,1955ApJS....2...41M} or \citet{1968ApJS...17..371L}. The second table includes new classifications for stars without any spectral classification, or spectral sub-type, or without LC in the default \textit{Simbad} query and with \fwhb\ < 7.5\,\AA.


\begin{table}[h]
\caption{New spectral classifications for some stars with LCs I and II and \fwhb\ < 7.5\,\AA\ taken from Fig.~\ref{fig:shrd_fw_adb}.}
\label{tab:newclass_BSGs}
    \centering
    \begin{tabular}{lll}
        ID & SpC$_{\textit{\,Simbad}}$ & SpC$_{\,This\,work}$ \\
        \noalign{\vspace{0.2cm}}\hline\noalign{\smallskip}
        \multicolumn{3}{c}{Stars wrongly classified as supergiants}\\
        \noalign{\smallskip}\hline\noalign{\smallskip\smallskip}\smallskip
        HD~13969   & B0.5~I & B0.7~III \\\smallskip
        BD~+56~528 & B0.5~I & B1~Vn \\\smallskip
        BD~+56~553 & B0.4~I & B1~IV \\\smallskip
        BD~+56~584 & B0.7~I & B1~V \\\smallskip
        HD~14870   & B1~Ib & B1~III \\\smallskip
        HD~51756   & B1/2~Ib & O9.7~IV \\\smallskip
        HD~59882   & B2~Ib/II & B1~V \\\smallskip
        HD~53456   & B1~Ib/II & B0~IV \\\smallskip
        HD~50707   & B1~Ib & B1~IV \\\smallskip
        HD~152685  & B1~Ib & B1.5~III \\\smallskip
        HD~159110  & B4~Ib & B2~III \\\smallskip
        HD~152286  & B1~Ib & B2~III \\\smallskip
        HD~162374  & B6~Ib & B5~IVp \\\smallskip
        HD~164719  & B3~Ib & B4~III \\\smallskip
        HD~164741  & B2~Ib/II & B1~III \\\smallskip
        HD~173502  & B1/2~Ib & B0.5~III \\\smallskip
        HD~165016  & B2~Ib & B0~V \\\smallskip
        HD~165132  & B5/6~Ib & O9.7~V \\\smallskip
        HD~164384  & B1/2~Ib/II & B1~Vnn \\\smallskip
        HD~166922  & B2~Ib & B2~II \\\smallskip
        HD~166965  & B2~Ib & B1.5~II-III \\\smallskip
        HD~167088  & B2~Ib/II & B0.5~V \\\smallskip
        HD~167479  & B1/2~Ib/II & B2.5~III \\\smallskip
        HD~164188  & B1~Ib/II & B0.5~V \\\smallskip
        HD~166803  & B1/2~Ib & B0.5~IV \\\smallskip
        HD~174069  & B2~Ib/II & B1.5~V \\\smallskip
        HD~201638  & B0.5~Ib & B0.5~V \\
        \noalign{\vspace{0.3cm}}\hline\noalign{\smallskip}
        \multicolumn{3}{c}{Stars wrongly classified as bright giants}\\
        \noalign{\smallskip}\hline\noalign{\smallskip\smallskip}\smallskip
        BD~+60~498 & O9.7~II-III & O9~V \\\smallskip
        HD~36591   & B2~II/III & B1~IV \\\smallskip
        HD~42690   & B2~II & B2~V \\\smallskip
        HD~61068   & B2~II & B2~III \\\smallskip
        HD~164002  & B1/2~II & B0.5~V \\\smallskip
        HD~164359  & B1~II & B0~III \\
        \hline
    \end{tabular}
    \tablefoot{Classifications adopted for stars with LC II are either from \citet{1953ApJ...118..318M,1955ApJS....2...41M} or \citet{1968ApJS...17..371L}. The default \textit{Simbad} classification is included for comparison and with the stars ordered by Galactic longitude.}
\end{table}

\begin{table}[h]
\caption{New spectral classifications for the stars with \fwhb\ < 7.5\,\AA\ without any spectral classification, spectral sub-type, or without luminosity class.}
\label{tab:newclass_noLC}
    \centering
    \begin{tabular}{lll}
        ID & SpC$_{\textit{\,Simbad}}$ & SpC$_{\,This\,work}$ \\
        \noalign{\vspace{0.2cm}}\hline\noalign{\smallskip}
        \multicolumn{3}{c}{Stars without full spectral classification}\\
        \noalign{\smallskip}\hline\noalign{\smallskip\smallskip}\smallskip
        HD~218941    & O   & B1~Ib   \\\smallskip
        HD~292164    & OB  & B1.5~III \\\smallskip
        LS~513       & OB  & B0.7~IV \\\smallskip
        LS~516       & OB+ & B0.5~IV \\\smallskip
        CD~-57~3344  & OB  & B1~III  \\\smallskip
        HD~99146     & OBe & B0.5~V  \\
        \noalign{\smallskip}\hline\noalign{\smallskip}
        \multicolumn{3}{c}{Stars without luminosity class}\\
        \noalign{\smallskip}\hline\noalign{\smallskip\smallskip}\smallskip
        HD~239895    & B2 & B8~Ia \\\smallskip
        HD~218325    & B3 & B2.5~IIIe \\\smallskip
        HD~217490    & B0 & B0~II \\\smallskip
        HD~213481    & B8 & B1.5~II-III \\\smallskip
        HD~219287    & B0 & B0.5~Ia \\\smallskip
        HD~219286    & B0 & O8:Ib: \\\smallskip
        HD~240256    & B0 & B3~Ib \\\smallskip
        BD~+60~2525  & B3 & B0.5~III \\\smallskip
        BD~+63~1962  & B5 & B0.7~IV \\\smallskip
        BD~+62~2210  & B0 & B9~Ia \\\smallskip
        BD~+63~1964  & B0 & B0~Iab \\\smallskip
        BD~+60~2615  & B0 & B0.5~Ib(n) \\\smallskip
        BD~+61~2526  & B3 & B1.5~II \\\smallskip
        BD~+61~2529  & B0 & B1~Ib \\\smallskip
        BD~+62~2299  & B2 & O8~II(f) \\\smallskip
        BD~+60~2644  & B1.8 & B1~III \\\smallskip
        HD~236695    & B2 & B1.5~II \\\smallskip
        BD~+00~1617~A& O9 & O9.2~V \\\smallskip
        BD~+00~1617~C& O9 & O9.5~IV \\\smallskip
        HD~292392    & B0 & O8.5~V \\\smallskip
        CPD~-57~3507 & B1 & B0~Iab \\\smallskip
        HD~97400     & B1 & B2.5~Ib \\\smallskip
        HD~306097    & B2 & O9~V \\\smallskip
        HD~332755    & B0 & O7.5~Ib-II \\
        \noalign{\smallskip}\hline
    \end{tabular}
    \tablefoot{The default \textit{Simbad} classification is included for comparison and with the stars ordered by Galactic longitude.}
\end{table}

\FloatBarrier


\section{Further notes about completeness}
\label{apen.completeness}

\begin{figure}[!t]
\centering
\resizebox{\columnwidth}{!}{\includegraphics{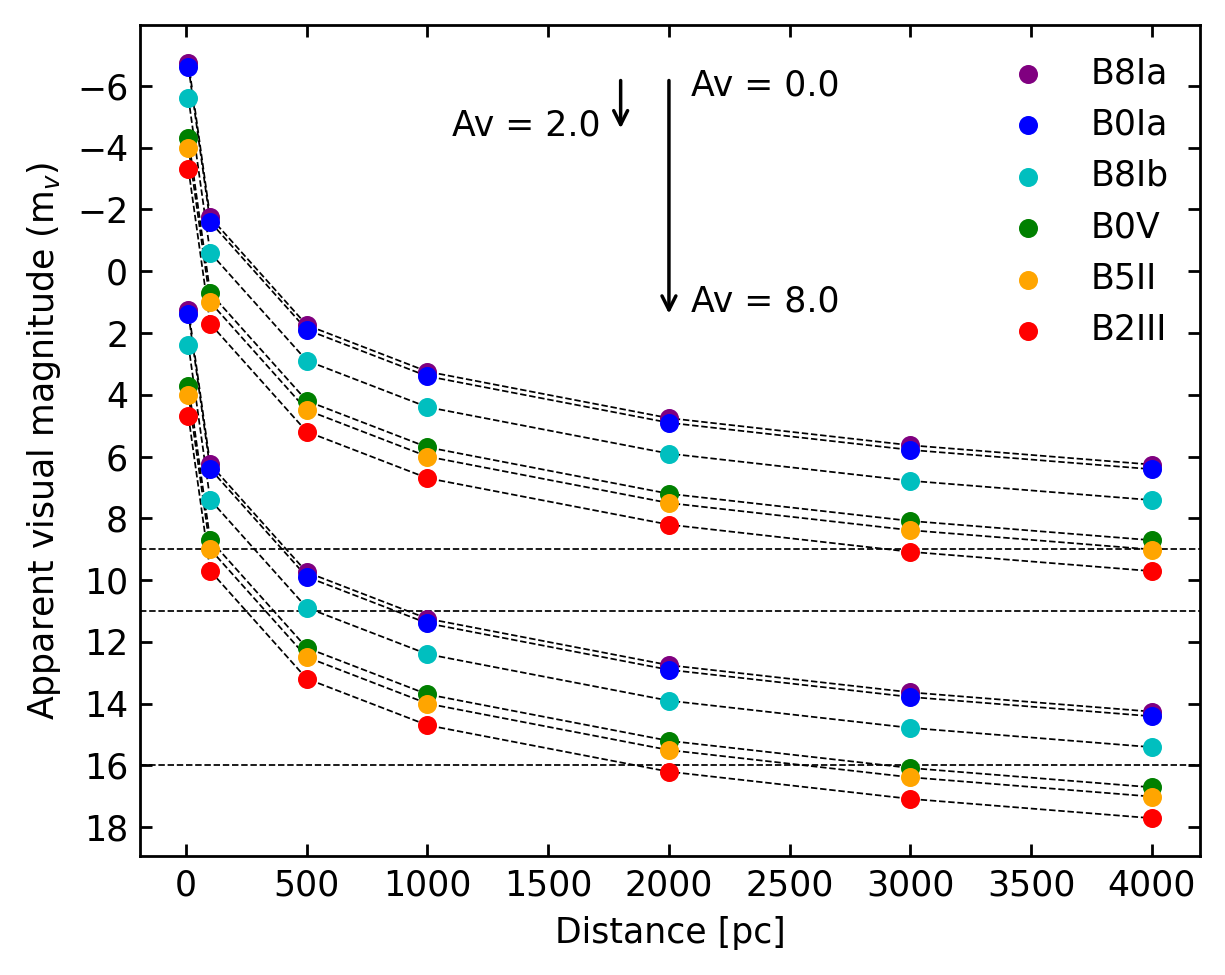}}
\caption{Apparent magnitude against distance for a set of six stars with different SpT and LC and for A${v}$\,=\,0, and A${v}$\,=\,8. Three threshold magnitudes at m${v}$,=,9,,11,,16 are indicated (see Appendix~\ref{apen.completeness}). Two extinction vectors are also included.} 
\label{fig:extinction}
\end{figure}

In Sect.~\ref{subsubsection:423_LLLLL} we provided some information about the usability of the ALS~III as a reference catalog where to check the completeness of our sample of stars up to a given distance. Fig.~\ref{fig:extinction} shows the apparent magnitudes of a set of stars covering the space of SpT and LC in our sample against different distances and for two values of total extinction: A$_{v}$\,=\,0 and A$_{v}$\,=\,8. There, we indicate three threshold magnitudes at m$_{v}$\,=\,9,\,11,\,16. The first and second constraint the magnitudes hardly (but not impossible) accessible from the observing facilities used to build this sample, while the latter corresponding to the magnitude from which point the ALS~III catalog may start missing stars. At m$_{v}$\,=\,16, we can see that in the case of A$_{v}$\,=\,0 (highly ideal) we would be able to observe all stars up to even 4\,000\,pc with an apparent magnitude m$_{v}$\,<\,9. However, as pointed out in Sect.~\ref{subsubsection:423_LLLLL}, the extinction in Milky Way can be severe in certain sight lines (A$_{v}$\,>\,4). If we focus on distances below 2\,000\,pc, still we would be able to say that no stars are missing even at A$_{v}$\,=\,8. However, we refer the reader to Fig.~\ref{fig:shrd_fw_adb} where we can see that both the coldest LC II-III stars above \logLs\,=\,3.5\,dex are actually below the 20\,\MSol\ track, therefore indicating that these stars are likely not evolving from the O-type stars. If we take this into account, we could extend the maximum value of A$_{v}$ up to A$_{v}$\,=\,10\,--\,13 (for LC I) where still we would not expect to miss many BSGs except for for those extremely reddened.

\section{Hypergiant stars}
\label{apen.hypergiants}

As indicated in Sect.~\ref{subsection:41_YYYYY}, we include here the list of hypergiants found within the sample of O9\,--\,B9 stars, independently from whether or not the H$\beta$ with difference was measurable or not.

\begin{table}[h]
\caption{List of hypergiants stars found in the sample of O9\,--\,B9 stars. The table includes the identifier of the star together with the \textit{Simbad} default classifications and those adopted in this work.}
\label{tab:hypergiants}
    \centering
    \begin{tabular}{lllll}
        ID & SpC$_{\textit{\,Simbad}}$ & SpC$_{\textit{\,reviewed}}$ \\
        \noalign{\smallskip}\hline\noalign{\smallskip}\smallskip
        HD\,173010     & O9.7\,Ia   & O9.7\,Ia+ \\\smallskip 
        HD\,13256      & B1\,Ia     & B1\,Ia(+) \\\smallskip 
        HD\,169454     & B1\,Ia     & B1\,Ia$^+$ (1) \\\smallskip
        HD\,152236     & B1\,Ia-0ek & B1.5\,Ia$^+$ (1) \\\smallskip
        BD\,-14\,5037  & B1.5\,Ia   & B1.5\,Ia$^+$ \\\smallskip 
        HD\,190603     & B1.5\,Ia   & B1.5\,Ia$^+$ (1) \\\smallskip
        HD\,80077      & B2\,Ia+e   & B2.5\,Ia$^+$ (1) \\\smallskip
        HD\,50064      & B6\,Ie     & B5\,Ia$^+$ \\\smallskip 
        HD\,183143     & B6\,Ia     & B7\,Iae (1) \\\smallskip
        HD\,303143     & A1\,Iae    & B8\,Ia$^+$ \\\smallskip 
        HD\,199478     & B8\,Ia     & B9\,Iae (1) \\\smallskip
        HD\,168607     & B9\,Iaep   & B9\,Ia$^+$ (1) \\
        \noalign{\smallskip}\hline
    \end{tabular}
    \tablefoot{Adopted classifications from \citet[][]{2012A&A...541A.145C}, if available. The stars are ordered by the spectral type of the adopted classifications.}
\end{table}

The selection has been made by filtering the full sample by stars with the ``PCy++" label in the H$\beta$ line profile, according to the morphological classification described in Sect.~\ref{subsubsection:425_YYYYY}. Their classification have been revised and are included together with the default classification from \textit{Simbad}.

\FloatBarrier


\section{Comparison of line broadening}
\label{apen.vsini}

\begin{figure}[!t]
\centering
\resizebox{\columnwidth}{!}{\includegraphics{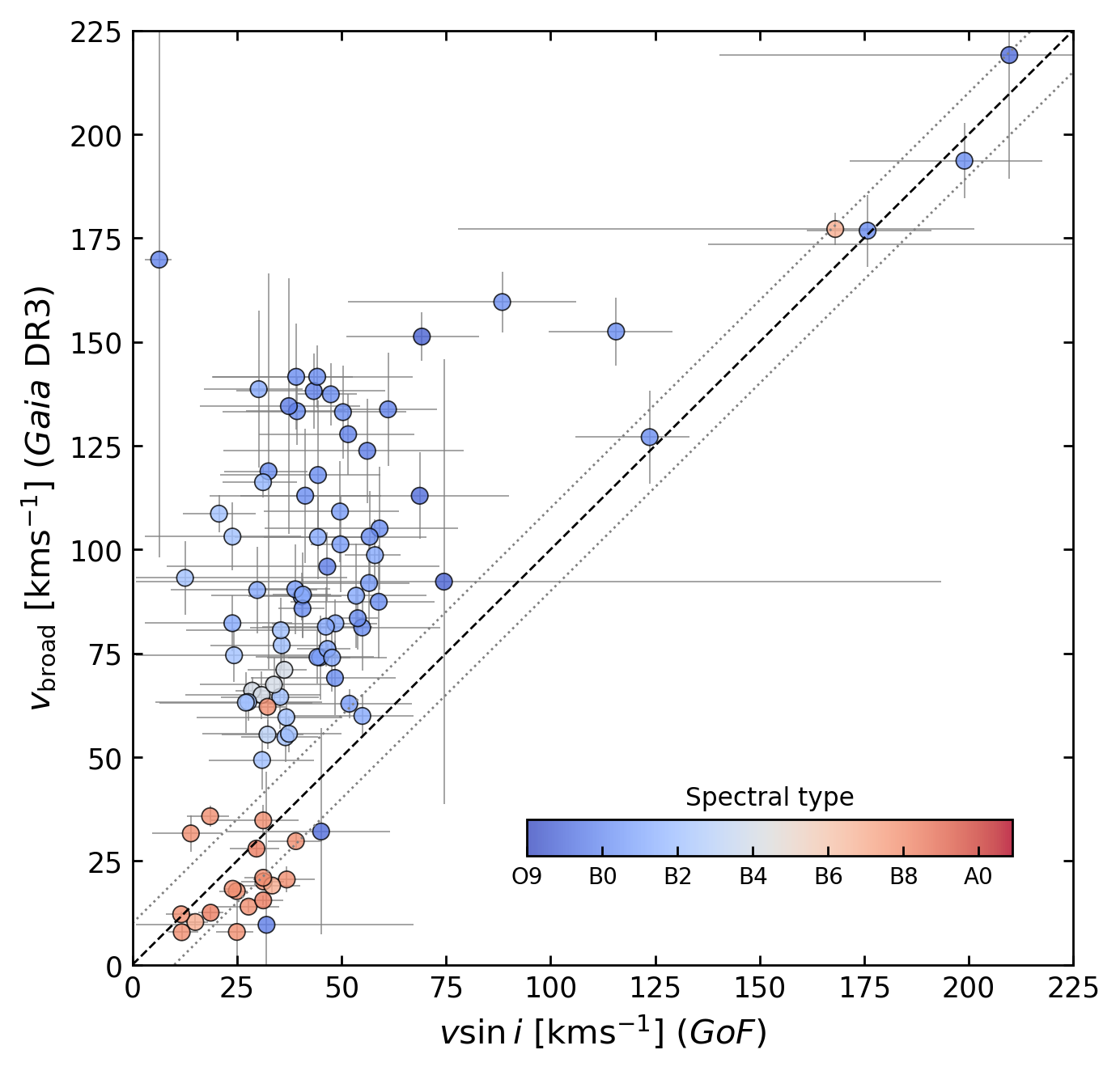}}
\caption{Comparison of the projected rotational velocity derived from the Goodness of Fit method (\vsini) and the line broadening parameter (\textit{Vbroad}) from {\em Gaia}, for the sample of O9\,--\,B9 stars with a FW3414(H$\beta$) < 7.5\,\AA. The stars are color-coded with the spectral type. Three dashed lines indicate the 1:1 line for comparison and additional two lines shifted $\pm$10\,\kms.} 
\label{fig:vbroad}
\end{figure}

We have compared our results for the \vsini\ and those obtained with the Radial Velocity Spectrometer ($R\approx$\,11\,500) onboard {\em Gaia} to account for the line broadening (\textit{Vbroad}). As pointed out in \citet{2022arXiv220610986F}, the parameter \textit{Vbroad} does not account for other mechanisms affecting the line profiles, neither the macroturbulence \citep[e.g.][]{2009A&A...508..409A,2010ApJ...720L.174S}, which is assumed to be 0\,\kms, or the microturbulence \citep[e.g.][]{2009A&A...499..279C,2015ApJ...808L..31G} as it is in our case. Moreover, hottest stars for which radial velocities are derived in the DR3 reaches only 14\,500\,K \citep{2022arXiv220605486B}, and therefore we did not expected to obtain accurate measurements \textit{Vbroad} but only a rough estimate. Fig.~\ref{fig:vbroad} shows for the comparison of both measurements for the sample of O9\,--\,B9 stars with a FW3414(H$\beta$) < 7.5\,\AA, excluding those with poor determination of \vsini\ (see Sect.~\ref{subsection:43_TTTTTTT}) and those SB2+. We note that only $\sim$15\% of the stars had values of \textit{Vbroad}. From the comparison we can see that, on the one hand, only a very few stars had values for \textit{Vbroad} above \vsini\,\gs\,70\,\kms\ compared to the number of stars with \vsini\ above that value (see Fig.~\ref{fig:hist_vsini}), but for those with \vsini\,\gs\,120\,\kms\ the agreement is less than 10\,\kms. On the other hand, those O9\,--\,B4 stars with \vsini\,\ls\,70\,\kms\ have systematic larger values of \vsini\ with differences \gs\,120\,\kms\ in several cases, thus indicating that in this domain, the values for \textit{Vbroad} are not reliable. The only case where the results from \textit{Vbroad} are similar to ours correspond to those late-B stars with \vsini\,\ls\,40\,\kms. 

\section{Long tables}
\onecolumn
\small{
\begin{landscape}
\begin{longtable}{lclllcclllc}
\label{tab:observations}\\ %
\caption[]{Summary of information for the sample of stars.}\\ %

 \hline\hline
 \noalign{\smallskip}

 ID & B$_{mag}$ & SpT$^{a}$ & LC$^{a}$ & Ref.\,file & S/N & \fwhb$^{e}$ & SB2+$^{f}$ & H$\alpha$ & H$\beta$ & \vsini\ \\
 &  &  &  &  &  & [\AA] &  &  &  & [\kms] \\

 \noalign{\smallskip}
 \hline
 \noalign{\smallskip}		
 \endfirsthead

\caption[]{Summary of information for the sample of stars.}\\ %
 \hline\hline
 \noalign{\smallskip}		

 ID & B$_{mag}$ & SpT$^{a}$ & LC$^{a}$ & Ref.\,file & S/N & \fwhb$^{e}$ & SB2+$^{f}$ & H$\alpha$ & H$\beta$ & \vsini\ \\
 &  &  &  &  &  & [\AA] &  &  &  & [\kms] \\

 \noalign{\smallskip}
 \hline
 \noalign{\smallskip}
 \endhead

 \noalign{\smallskip}
 \hline
 \endfoot
 \noalign{\smallskip}
HD~164032 & 7.53 & B1/2 & Ib & HD164032\_20220830\_201330\_M\_V85000\_log & 106 & 4.7 & -- & DP+ & DP & 108 \\
\noalign{\smallskip}
HD~164019 & 9.49 & O9.5 & IVp & HD164019\_20080608\_075144\_F\_V48000 & 183 & 4.7 & -- & Ab & Ab & 78 \\
\noalign{\smallskip}
HD~163613 & 8.88 & B1 & I/II & HD163613\_20200804\_213522\_N\_V25000 & 96 & 4.1 & -- & DP & DP & 71 \\
\noalign{\smallskip}
HD~160430 & 8.93 & B2 & II & HD160430\_20200802\_220513\_N\_V25000 & 96 & 6.1 & -- & Ab & Ab & 20 \\
\noalign{\smallskip}
HD~162717 & 9.81 & B2 & II & HD162717\_20180731\_215140\_M\_V85000 & 36 & 6.1 & -- & DP+ & DP & 264 \\
\noalign{\smallskip}
HD~164741 & 9.48 & (B2) & (Ib/II) & HD164741\_20180731\_223907\_M\_V85000 & 38 & 7.1 & -- & Ab & Ab & 20 \\
\noalign{\smallskip}
HD~173502 & 9.61 & (B1/2) & (Ib) & HD173502\_20190710\_003102\_N\_V25000 & 93 & 6.4 & LPV/SB2? & Ab & Ab & 131 \\
\noalign{\smallskip}
HD~156779 & 9.57 & B2 & II & HD156779\_20210621\_224538\_M\_V85000\_log & 80 & 7.2 & -- & Ab & Ab & 97 \\
\end{longtable}

\tablefoot{
\tablefoottext{a}{Stars with the SpT and LC in parenthesis have their spectral classification reviewed (see Appendix~\ref{apen.newclass}).}
\tablefoottext{b}{B-type stars wrongly classified as A-type in the default \textit{Simbad} classification and included in this work.}
\tablefoottext{c}{For HD~13970 the classification has been taken from \citet{2020A&A...643A.116D}.}
\tablefoottext{d}{Stars for which the default \textit{Simbad} classification includes SpT and LC for a second component but we have only considered the first one listed.}
\tablefoottext{e}{We adopt a global uncertainty for \fwhb\ of 0.1\,\AA.}
\tablefoottext{f}{Stars marked with ``*" in the SB column are new identifications.}
\tablefoottext{g}{Stars are ordered by Galactic longitude.}
}
\end{landscape}
}

\onecolumn
\small{
\begin{longtable}{lcccccccccc}
\label{tab:phot_kinem_Gaia}\\ %
\caption{Photometric and kinematic data from {\em Gaia}.}\\ %

\hline\hline
\noalign{\smallskip}

ID  & l     & b     & G & G$_{BP}$ & G$_{RP}$ & $\varpi$$^{a,b}$ & $\mu_{\alpha}\cos{\delta}$ & $\mu_{\delta}$   & RUWE & Distance (B-J) \\
    & [deg] & [deg] &   &          &          &  [mas]         & [mas\,yr$^{-1}$]           & [mas\,yr$^{-1}$] &      & [pc] \\

 \noalign{\smallskip}
 \hline
 \noalign{\smallskip}		
 \endfirsthead

\caption[]{Photometric and kinematic data from {\em Gaia}.}\\ %
 \hline\hline
 \noalign{\smallskip}		

ID & l     & b     & G & G$_{BP}$ & G$_{RP}$ & $\varpi$$^{a,b}$ & $\mu_{\alpha}\cos{\delta}$ & $\mu_{\delta}$   & RUWE & Distance (B-J) \\
   & [deg] & [deg] &   &          &          &  [mas]         & [mas\,yr$^{-1}$]           & [mas\,yr$^{-1}$] &      & [pc] \\

 \noalign{\smallskip}
 \hline
 \noalign{\smallskip}
 \endhead

 \noalign{\smallskip}
 \hline
 \endfoot
 \noalign{\smallskip}
HD~164032 & 0.8767 & -3.237 & 7.39 & 7.43 & 7.25 & 0.627$\pm$0.035 & 1.432$\pm$0.036 & 0.171$\pm$0.022 & 0.62 & 1623~$_{-84}^{120}$ \\
\noalign{\smallskip}
HD~164019 & 1.9099 & -2.6166 & 9.19 & 9.29 & 8.96 & 0.396$\pm$0.028 & 1.592$\pm$0.027 & 0.060$\pm$0.018 & 0.55 & 2578~$_{-148}^{178}$ \\
\noalign{\smallskip}
HD~163613 & 2.0881 & -1.9638 & 8.41 & 8.58 & 8.07 & 0.508$\pm$0.030 & 0.439$\pm$0.030 & -1.827$\pm$0.019 & 0.84 & 1991~$_{-100}^{107}$ \\
\noalign{\smallskip}
HD~160430 & 3.7823 & 3.5878 & 7.80 & 7.98 & 7.43 & 0.830$\pm$0.034 & 1.290$\pm$0.035 & -1.363$\pm$0.020 & 0.82 & 1223~$_{-40}^{56}$ \\
\noalign{\smallskip}
HD~162717 & 4.9039 & 0.9037 & 9.19 & 9.46 & 8.73 & 0.668$\pm$0.022 & 0.915$\pm$0.027 & -1.654$\pm$0.016 & 1.06 & 1499~$_{-53}^{57}$ \\
\noalign{\smallskip}
HD~164741 & 5.164 & -1.6449 & 8.90 & 9.08 & 8.56 & 0.741$\pm$0.019 & -0.492$\pm$0.023 & -1.031$\pm$0.017 & 0.74 & 1356~$_{-34}^{33}$ \\
\noalign{\smallskip}
HD~173502 & 5.3641 & -12.2722 & 9.70 & 9.64 & 9.79 & 0.277$\pm$0.044 & -0.351$\pm$0.045 & -3.536$\pm$0.039 & 0.97 & 3908~$_{-585}^{633}$ \\
\noalign{\smallskip}
HD~156779 & 5.4256 & 10.3249 & 9.23 & 9.31 & 9.03 & 0.355$\pm$0.020 & 1.409$\pm$0.025 & -6.854$\pm$0.016 & 1.00 & 2858~$_{-129}^{160}$ \\
\end{longtable}
\tablefoot{
\tablefoottext{a}{Parallaxes are corrected from zero-point offset.}
\tablefoottext{b}{Negative parallaxes have been masked.}
}
}

\onecolumn
\small{
\begin{longtable}{lccccccc}
\label{tab:phot_kinem_Hipp}\\ %
\caption[]{Photometric and kinematic data from {\em Hipparcos}.}\\ %

 \hline\hline
 \noalign{\smallskip}

ID & l     & b     & B-V & $\varpi$ & $\mu_{\alpha}\cos{\delta}$ & $\mu_{\delta}$   & Distance  \\
   & [deg] & [deg] &     & [mas]    & [mas\,yr$^{-1}$]           & [mas\,yr$^{-1}$] & [pc]      \\
   
 \noalign{\smallskip}
 \hline
 \noalign{\smallskip}		
 \endfirsthead

\caption[]{Photometric and kinematic data from {\em Hipparcos}.}\\ %
 \hline\hline
 \noalign{\smallskip}		

ID & l     & b     & B-V & $\varpi$ & $\mu_{\alpha}\cos{\delta}$ & $\mu_{\delta}$   & Distance  \\
   & [deg] & [deg] &     & [mas]    & [mas\,yr$^{-1}$]           & [mas\,yr$^{-1}$] & [pc]      \\

 \noalign{\smallskip}
 \hline
 \noalign{\smallskip}
 \endhead

 \noalign{\smallskip}
 \hline
 \endfoot
 \noalign{\smallskip}
HD~166937 & 10.0033 & -1.6005 & 0.195 & 0.09$\pm$0.28 & 0.30 & -0.48 & ... \\
\noalign{\smallskip}
HD~202214 & 98.5202 & 7.9852 & 0.110 & 0.76$\pm$0.42 & -2.39 & -3.17 & ... \\
\noalign{\smallskip}
BD~+55~2770 & 104.751 & -1.4029 & 0.350 & 0.46$\pm$1.37 & -1.76 & -4.27 & ... \\
\noalign{\smallskip}
HD~36486 & 203.8559 & -17.7397 & -0.175 & 4.71$\pm$0.58 & 0.64 & -0.69 & ... \\
\noalign{\smallskip}
HD~37128 & 205.2121 & -17.2417 & -0.184 & 1.65$\pm$0.45 & 1.44 & -0.78 & ... \\
\noalign{\smallskip}
HD~37742 & 206.4522 & -16.5852 & -0.199 & 4.43$\pm$0.64 & 3.19 & 2.03 & ... \\
\noalign{\smallskip}
HD~37743 & 206.4522 & -16.5852 & -0.199 & 4.43$\pm$0.64 & 3.19 & 2.03 & ... \\
\noalign{\smallskip}
HD~34085 & 209.2412 & -25.2454 & -0.030 & 3.78$\pm$0.34 & 1.31 & 0.50 & 265$\pm$24 \\
\end{longtable}

\tablefoot{
\tablefoottext{a}{Distances for stars with $\sigma_{\varpi}/\varpi$ > 0.1 are not included.}
\tablefoottext{b}{HD\,51756 has a negative parallax and therefore the value and the corresponding distance are not included in the table.}
}}

\end{appendix}

\end{document}